\documentclass[twocolumn,final]{svjour3}
\usepackage{graphicx}
\usepackage{hyperref}
\usepackage{epsfig}
\usepackage{amsmath}
\usepackage{amssymb}
\usepackage{hyperref}
\usepackage{lineno}
\usepackage{enumerate}

\usepackage{cite}

\usepackage{color}

\title{Impact of atrial fibrillation on the cardiovascular system through a lumped-parameter approach}

\author{S. Scarsoglio \and A. Guala \and C. Camporeale \and L. Ridolfi}

\institute{S. Scarsoglio
\at Department of Mechanical and Aerospace Engineering, Politecnico di Torino, Torino, Italy\\\email{stefania.scarsoglio@polito.it}
\and
A. Guala \and C. Camporeale \and L. Ridolfi
\at Department of Environment, Land and Infrastructure Engineering, Politecnico di Torino, Torino, Italy\\
}

\date{}

\begin{document}

\maketitle

\begin{abstract}
Atrial fibrillation (AF) is the most common arrhythmia affecting millions of people in the Western countries and, due to the widespread impact on the population and its medical relevance, is largely investigated in both clinical and bioengineering sciences. However, some important feedback mechanisms are still not clearly established. The present study aims at understanding the global response of the cardiovascular system during paroxysmal AF through a lumped-parameter approach, which is here performed paying particular attention to the stochastic modeling of the irregular heartbeats and the reduced contractility of the heart. AF can be here analyzed by means of a wide number of hemodynamic parameters and avoiding the presence of other pathologies, which usually accompany AF.
Reduced cardiac output with correlated drop of ejection fraction and decreased amount of energy converted to work by the heart during blood pumping, as well as higher left atrial volumes and pressures are some of the most representative results aligned with the existing clinical literature and here emerging during acute AF.
The present modeling, providing new insights on cardiovascular variables which are difficult to measure and rarely reported in literature, turns out to be an efficient and powerful tool for a deeper comprehension and prediction of the arrythmia impact on the whole cardiovascular system.
\end{abstract}


\section{Introduction}

Atrial fibrillation (AF) occurs when the electrical activity of the atria, governed by the sinoatrial node, is disorganized, causing irregular and rapid heartbeats \cite{Fuster}. AF can lead to disabling symptoms, such as palpitations, chest discomfort, anxiety, fall in blood pressure, decreased exercise tolerance, pulmonary congestion, which are all related to rapid heart rate and inefficient cardiac output. Moreover, the persistency of fibrillated conditions can enhance heart failure and stroke, being AF responsible for $15$ to $20$ percent of ischemic strokes, and increasing the risk of suffering an ischemic stroke by five times \cite{Lloyd-Jones}.

\noindent AF incidence gets higher with age: $2.3\%$ of people older than $40$ years are affected, up to more than $8\%$ of people older than 80 years \cite{Krahn}, with a prevalence which is markedly amplifying in industrialized countries \cite{Alpert}. AF currently affects almost 7 million people in the USA and Europe, but because of the rise of life expectancy in Western countries, incidence is expected to double within the next forty years \cite{Lloyd-Jones}. Nevertheless, AF is responsible for substantial morbidity and mortality in the general population \cite{Benjamin}.

For the above reasons, AF is a subject of broad interest under several aspects. Some examples are, among many others, statistical analyses on the heartbeat distributions \cite{Tateno,Hayano}, risk factors \cite{Tsang,Kannel}, correlation with other cardiac pathologies \cite{Verdecchia}.

However, several key points on the consequences of AF are not completely understood \cite{Magnani}. Literature data, as we will show later on, reveal for example contrasting trends regarding pulmonary and systemic arterial pressures during AF: hypotension, normotension and hypertension seem to be equally probable when AF emerges. Invasive measures could, for instance, better clarify how AF acts on arterial systemic and pulmonary pressures. But, these and other non-invasive measures are often difficult to be performed especially during atrial fibrillation, since the heart rate variability causes problems to oscillometric instruments and the clinical framework often requires immediate medical treatment. Moreover, the anatomical and structural complexity of some cardiac regions (e.g., right ventricle) makes estimates not always feasible and accurate \cite{Haddad}. This leads to a substantial absence of well-established information regarding the behavior of the right ventricle and central venous pressure during AF. Nevertheless, AF usually occurs in presence of other pathologies (such as hypertension, atrial dilatation, coronary heart disease, mitral stenosis), therefore the specific role of atrial fibrillation on the whole cardiovascular system is not easily detectable and distinguishable. It is still nowadays debatable whether atrial dilatation acts mainly as cause or effect of AF events \cite{Sanfilippo,Osranek}, while the role of hypertension as causative agent and predictor makes it difficult to discern whether it is also a consequence of AF \cite{Verdecchia}.

The present work arises in this scenario and aims at being a first attempt to quantify, through a stochastic modeling approach, the impact of acute AF on the cardiovascular variables with respect to the normal rhythm. Structural remodeling effects due to persistent or chronic AF are not taken into account. 
Our goal of understanding the global response of the cardiovascular system during fibrillated paroxysmal events can be achieved by means of a lumped-parameter approach \cite{Korakianitis-a,Korakianitis-b}, which is here carried out paying particular attention to the stochastic modeling of the irregular heartbeats and the reduced contractility function of the heart. In particular, the beating model accounts for a time-correlated regular beat typical of the normal rhythm, while the irregular fibrillated beating is composed by two differently correlated random distributions.

\noindent The proposed approach has a double advantage. First, AF conditions can be analyzed without the presence of other side pathologies. Therefore, the outcomes should be read as purely consequent of a fibrillated cardiac status in a healthy young adult. To this end, the AF parametrization through the modeling is presented to highlight single cause-effect relations, trying to address from a mechanistic point of view the cardiovascular feedbacks which are currently poorly understood. Second, the main cardiac variables and hemodynamics parameters can all be obtained at the same time, while clinical studies usually focus only on a few of them at a time. The present global response is indeed compared with more than thirty works in literature (reported for convenience in Online Resource 2), which describe the variations of different hemodynamic features between normal sinus rhythm and atrial fibrillation conditions. The overall agreement with clinical state-of-the-art measures is rather good. Moreover, additional information related to the statistical properties of the cardiovascular variables as well as hemodynamic parameters (such as right ventricle data) which are difficult to measure and almost never offered in literature, are here provided. An accurate statistical analysis of the cardiovascular dynamics, yielding the main values and the probability distributions, is not easily accomplished by in vivo measurements, while it is here carried out thanks to the statistically stationarity of the performed simulations.


\section{Materials and Methods}

\begin{figure*}
\centering
\includegraphics[width=1.5\columnwidth]{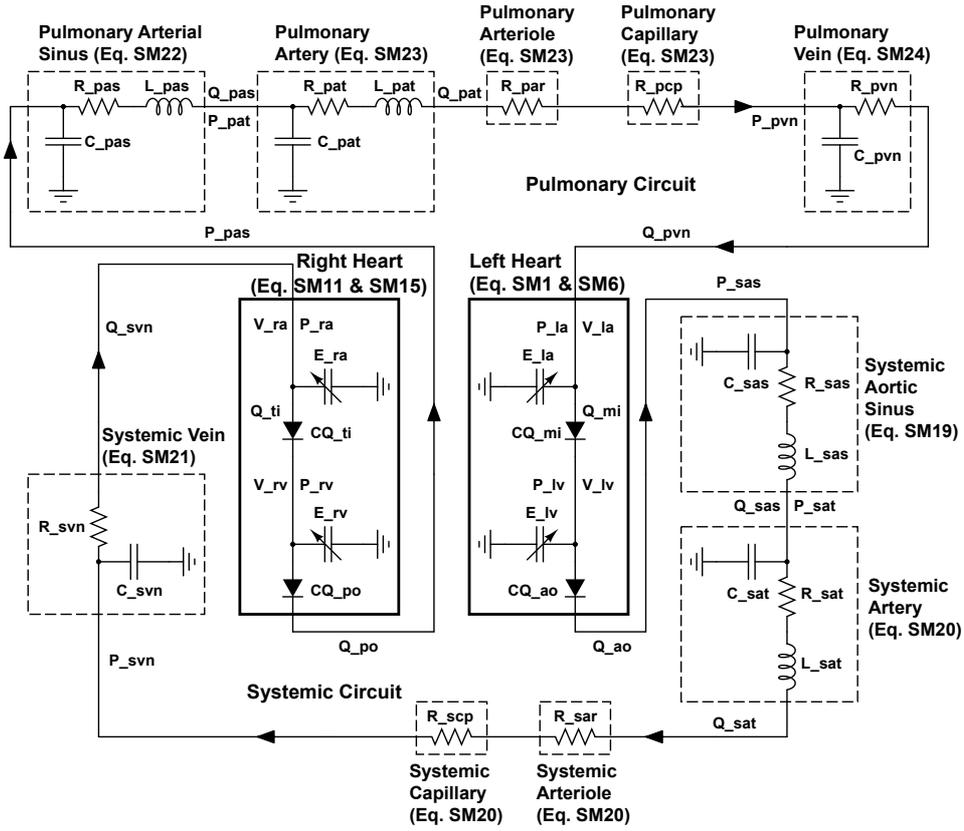}
\caption{Scheme of the model: blood flow $Q$, volume $V$, pressure $P$. Equations are listed in Online Resource 1}
\label{scheme}
\end{figure*}

The present lumped model, proposed by Korakianitis and Shi \cite{Korakianitis-a,Korakianitis-b}, extends the windkessel approach to the whole (arterial and venous) circulation system, and is combined to an active atrial and ventricular representation, through four time-varying elastance functions, for both the left and right heart chambers. The cardiovascular model is composed by a network of compliances, $C$, resistances, $R$, and inductances, $L$, representing the pumping heart coupled to the systemic and pulmonary systems. The viscous effects are taken into account by the resistances, $R$ [mmHg s/ml], the inertial terms are considered by the inductances, $L$ [mmHg s$^2$/ml], while the elastic vessel properties are described by the compliances, $C$ [ml/mmHg]. Three cardiovascular variables are involved at each section: the blood flow, $Q$ [ml/s], the volume, $V$ [ml], the pressure, $P$ [mmHg]. A schematic representation of the cardiovascular system is shown in Fig. \ref{scheme}.

All the four chambers of the pumping heart are described. The pulsatility properties are included by means of two pairs of time-varying elastance functions, one for the atria and one for the ventricles, which are then used in the constitutive equations (relating pressure, $P$, and volume, $V$). 
For the four heart valves, the basic pressure-flow relation is described by an orifice model. The valve motion mechanisms are deeply analyzed and account for the following blood-flow effects: pressure difference across the valve, frictional effects from neighboring tissue resistance, the dynamic motion effect of the blood acting on the valve leaflet, the action of the vortex downstream of the valve. The introduction of a time-varying elastance for the atria accounts for the atrial contraction, while the heart dynamics description provides to model valvular regurgitation and the dicrotic notch mechanism. An equation for the mass conservation (accounting for the volume variation, $dV/dt$) concludes the description of each chamber.

The systemic and pulmonary systems are divided into 5 parts (see Fig. \ref{scheme}). The systemic arteries are described by four parts (aortic sinus, artery, arteriole and capillary), while the systemic venous circulation is characterized by a unique compartment. The pulmonary circuit follows the same architecture. Each section of the systemic and pulmonary circuits may contain three components (the viscous term, $R$, the inertial term, $L$, and the elastic term, $C$), and is characterized by three equations: an equation of motion (accounting for the flow variation, $dQ/dt$), an equation for the conservation of mass (expressed in terms of pressure variations, $dP/dt$), and a linear state equation between pressure and volume.

The resulting differential system is numerically solved by means of a multistep adaptative scheme based on the numerical differentiation formulas (NDFs). Details of the modeling and the computational method are given in Online Resource 1.

\subsection{Cardiac cycle simulation: physiologic and fibrillated beating}

The main aim of the present work is to compare the cardiovascular outcomes of the physiologic case (normal sinus rhythm, NSR) with those during acute AF events. We recall that $RR$ [s] is the temporal range between two consecutive heart beats, while the heart rate, $HR$, is the number of heartbeats per minute.

The normal heart-beating is an example of pink noise, which means that the electrocardiogram has an approximately $1/f$ power spectrum \cite{Kobayashi,Hayano}, where $f$ is the beat frequency. The presence of a pink noise induces a temporal correlation, differently from the white noise, which is instead uncorrelated. During NSR the $RR$ interval is usually Gaussian distributed \cite{Hennig,Pikkujamsa}, with a coefficient of variation $cv$=$\sigma/\mu$ ($\sigma$ is the standard deviation, $\mu$ is the mean value of the $RR$ distribution), which is between $0.05$ and $0.13$ \cite{Pikkujamsa,Sosnowski}. 
Thus, we here extract the $RR$ intervals from a pink Gaussian distribution with an average value for normal adults, $\mu$=$0.8$ s, and choosing $cv$=$0.07$. Pink noise is generated as follows. An uncorrelated temporal signal is created extracting values from a white Gaussian distribution. A Fourier transform of the signal is carried out and the spectrum is multiplied by a filter so that the resulting spectral density is proportional to $1/f$. In the end, an inverse Fourier transform is performed to recover the filtered signal in time.

\begin{figure*}
\begin{minipage}[]{0.8\columnwidth}
\includegraphics[width=\columnwidth]{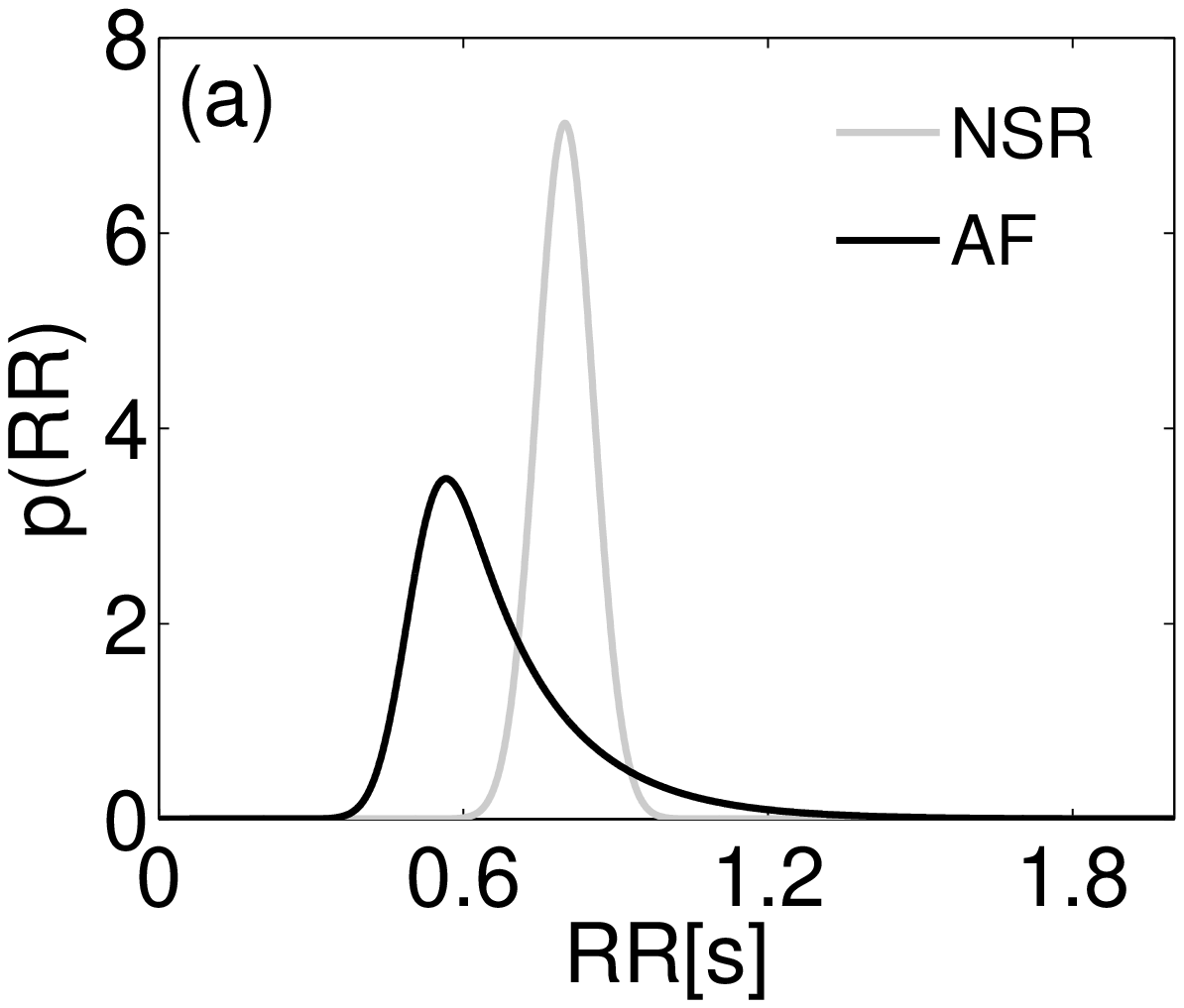}
\end{minipage}
\begin{minipage}[]{0.8\columnwidth}
\includegraphics[width=\columnwidth]{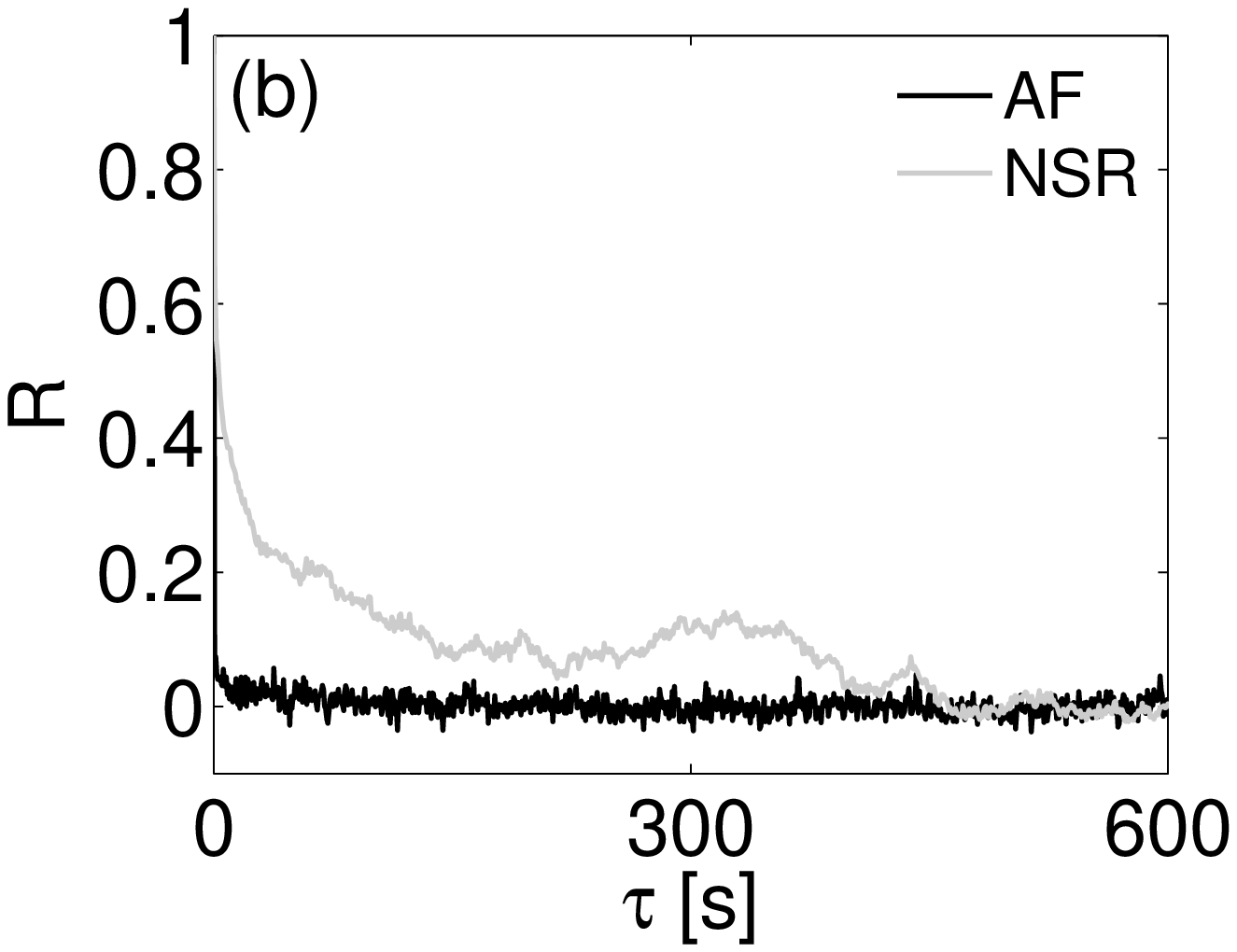}
\end{minipage}\\
\begin{minipage}[]{1.5\columnwidth}
\includegraphics[width=\columnwidth]{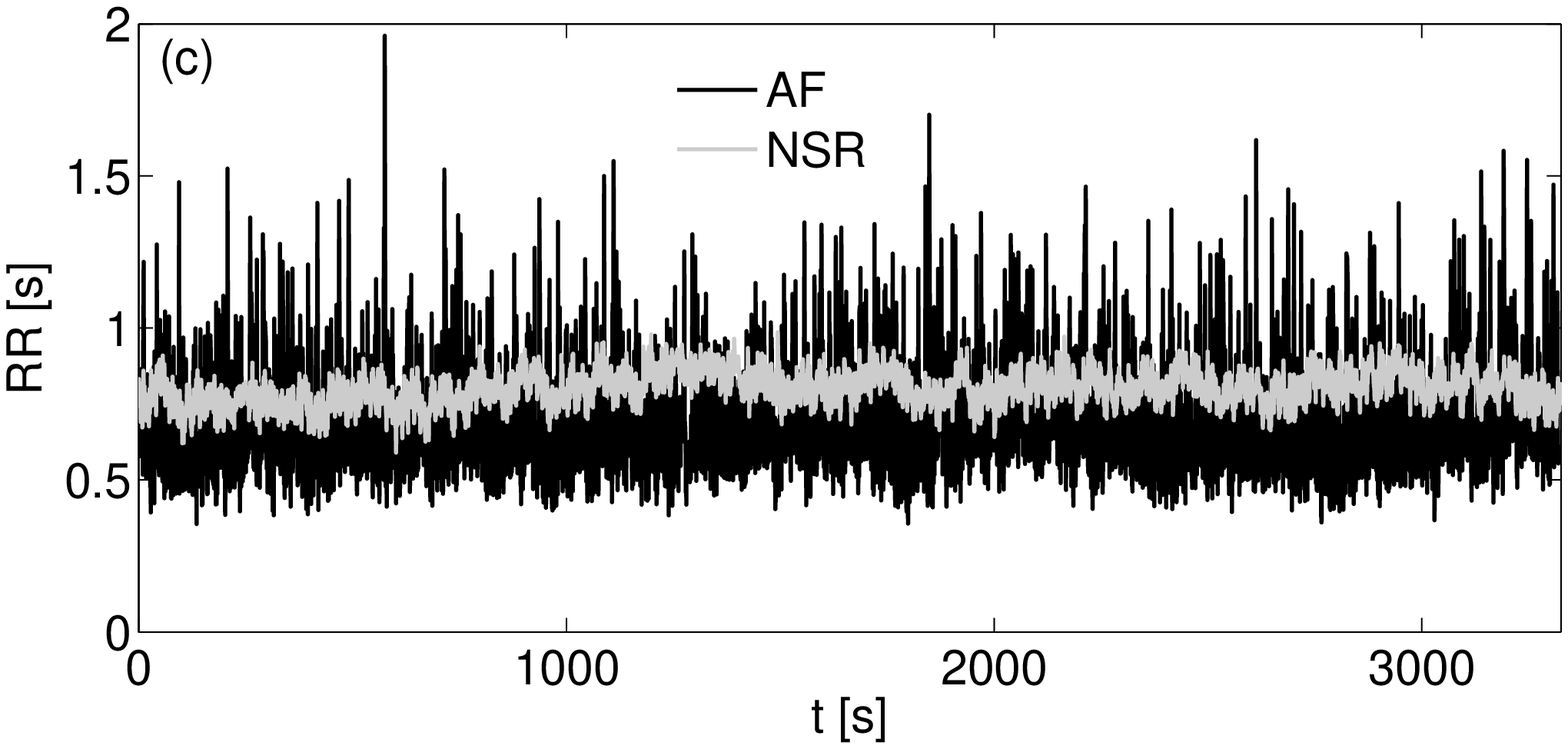}
\end{minipage}\\
\begin{minipage}[]{0.8\columnwidth}
\includegraphics[width=\columnwidth]{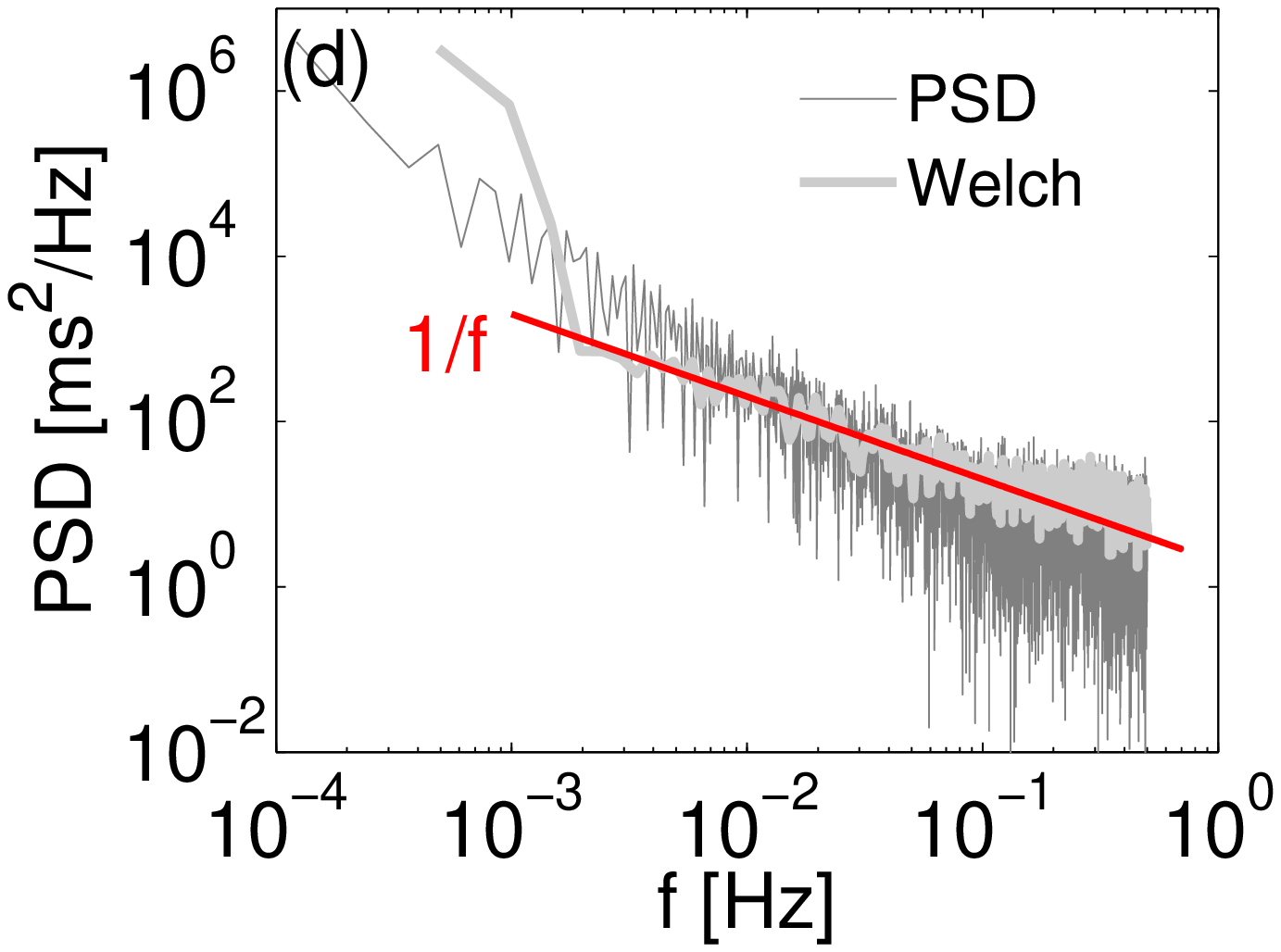}
\end{minipage}
\begin{minipage}[]{0.8\columnwidth}
\includegraphics[width=\columnwidth]{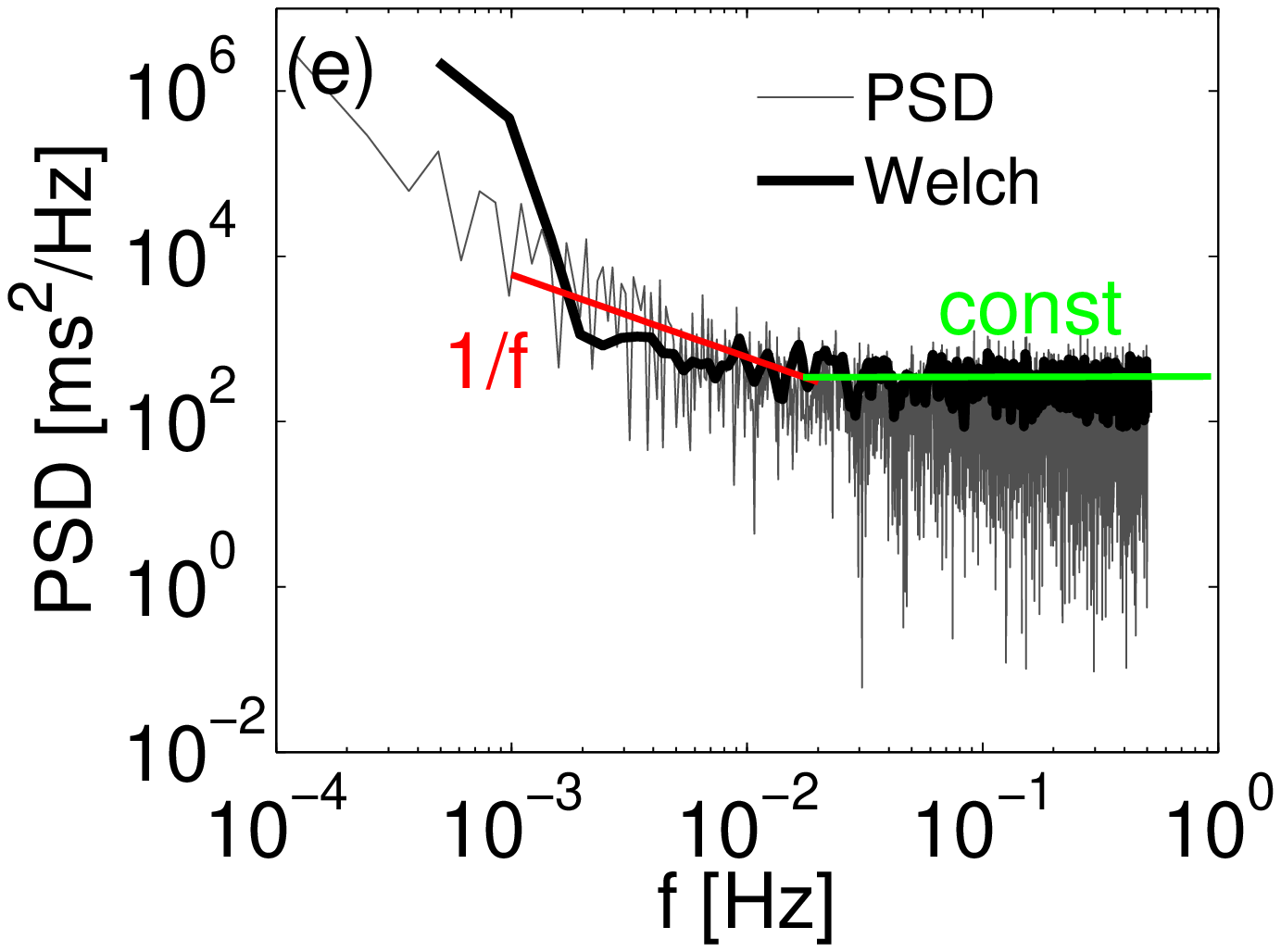}
\end{minipage}
\caption{(a) $RR$ distributions: NSR ($\mu$=$0.8$ s, $\sigma$=$0.06$ s), AF ($\mu_G$=$0.5$ s, $\sigma_G$=$0.05$ s, $\gamma$=$6$ Hz). (b) Autocorrelation functions, R. (c) $RR$ temporal series. (d)-(e) Power spectrum density, PSD: (d) NSR, (e) AF. Light: NSR, dark: AF}
\label{RR_distributions_series}
\end{figure*}

During AF, the $RR$ distribution can be unimodal ($61\%$ of the cases), bimodal ($32\%$), trimodal ($5\%$), or multimodal ($2\%$) \cite{Weismuller}. The presence of multiple modes is often attributed to a modification of the atrio-ventricular node conduction properties \cite{Rokas,Weismuller}. Here we focus on the unimodal distribution, affecting the large part of AF patients, which can be fully described by the superposition of two statistically independent times \cite{Hennig,Hayano}, $RR$=$\tau+\eta$. $\tau$ is extracted from a correlated pink Gaussian distribution (mean $\mu_G$, standard deviation $\sigma_G$), 
%
while $\eta$ is drawn from an uncorrelated exponential distribution (rate parameter $\gamma$). 
%
%
The resulting $RR$ distribution is an exponentially modified Gaussian (EMG) distribution, with mean $\mu$=$\mu_G+\gamma^{-1}$, standard deviation $\sigma$=$\sqrt{\sigma_G^2+\gamma^{-2}}$, and probability density function:

\begin{eqnarray}
p(RR; \mu_G,\sigma_G, \gamma) &=& \frac{\gamma}{2} e^{\frac{\gamma}{2}(2 \mu_G + \gamma \sigma_G^2 - 2 RR)} \\
&\times& \textmd{erfc}\left(\frac{\mu_G + \gamma \sigma_G^2 - RR}{\sqrt{2} \sigma_G}\right), \nonumber
\end{eqnarray}

\noindent where $\textmd{erfc}(\cdot)$ is the complementary error function. The parameters are suggested by the fibrillated $RR$ data available \cite{Hennig,Sosnowski}, and by considering that the coefficient of variation, $cv$, is around $0.24$ during AF \cite{Tateno}.

\noindent In Fig. \ref{RR_distributions_series}a, the fibrillated and physiologic distributions, from which $RR$ intervals are extracted, are shown for comparison.

Two other important features of the AF are introduced into the model. First, there is a complete loss of atrial contraction [B8,B25], which can be taken into account, for both atria, by considering a constant atrial elastance. In so doing, we exclude the \emph{atrial kick}, that is the atrial contribution to the ventricular filling during late diastole. Second, a reduced systolic left ventricular function is experienced during AF \cite{Cha,Tanabe} and this aspect is included by decreasing the maximum left ventricular elastance, $E_{lv,max}$ (see Online Resource 1), to a value depending on the preceding ($RR1$) and pre-preceding ($RR2$) heartbeats \cite{Tanabe}:

\begin{equation}
E_{lv,max}=0.59 \frac{RR1}{RR2} + 0.91 \,\,\, \textmd{mmHg/ml}.
\label{eq_tanabe}
\end{equation}

\noindent A reduced contractile capacity is also observed for the right ventricle \cite{Haddad}. However, since the partial disfunction of the right ventricle cannot be easily traduced into elastance terms, here we prefer not to change the $E_{rv,max}$ value with respect to the normal case.

\subsection{Runs}

The duration of AF episodes can greatly vary, depending on which kind of fibrillation is considered \cite{Fuster}. For example, during paroxysmal atrial fibrillation recurrent episodes usually occur and self-terminate in less than $7$ days. Permanent fibrillation instead induces an ongoing long-term episode, which can last up to a year or more.

\noindent The model equations are integrated in time until the main statistics (mean and variance) of the cardiac variables remain insensitive to further extensions of the computational domain size, that is the results are statistically stationary. The first $20$ periods are left aside since they reflect the transient dynamics of the system. In fact, the NSR configuration shows periodic solutions after 5-6 cycles (also refer to \cite{Korakianitis-a}), therefore we can assume that for both NSR and AF simulations the transient dynamics are completely extinguished after 20 periods. Once the system exceeds the transient and reaches a statistically stationary state, $5000$ cycles are then simulated for the normal and the fibrillated cases. Our choice of computing 5000 heartbeats can be thought of as a representative acute fibrillation episode of about one hour length, affecting a healthy young adult without pre-existing and structural pathologies.

\noindent The two simulated cases are listed below:

\begin{enumerate}[(a)]
\item Normal sinus rhythm (NSR)
 \begin{itemize}
 \item $RR$ extracted from a correlated pink Gaussian distribution: $\mu$=$0.8$ s, $\sigma$=$0.06$ s;
     \item Time varying (right and left) atrial elastance;
     \item Full left ventricular contractility: $E_{lv,max}$=$2.5$ mmHg/ml.
 \end{itemize}
 \item Atrial fibrillation (AF)
 \begin{itemize}
\item $RR$ extracted from an EMG distribution: $\mu_G$=$0.5$ s, $\sigma_G$=$0.05$ s, $\gamma$=$6$ Hz; $\mu$=$0.67$ s, $\sigma$=$0.17$ s, $cv$=$0.26$;
     \item Constant (right and left) atrial elastance;
     \item Reduced left ventricular contractility, Eq. (\ref{eq_tanabe}).
 \end{itemize}
\end{enumerate}

\noindent For both NSR and AF we report the autocorrelation functions, $R$, (Fig. \ref{RR_distributions_series}b), the $RR$ time series (Fig. \ref{RR_distributions_series}c), and the power spectrum density, PSD (Fig. \ref{RR_distributions_series}, panels d and e), computed directly from the $RR$ series (thin curves), and estimated through the Welch method (thick curves) to smooth the oscillations. The NSR spectrum shows the typical $1/f$ power-law scaling observed in healthy subjects, while the AF spectrum displays two distinct power-law scalings which is a common feature of a real fibrillated beating \cite{Hayano,Hennig}. As for the autocorrelation function, $R$, during AF it immediately drops to zero, revealing furthermore its uncorrelated nature with respect to the NSR.

\noindent A sensitivity analysis of the present results is discussed in Online Resource 1.

\section{Results}


We focus on the most relevant variables of the cardiovascular model, in terms of pressures, $P$, and volumes, $V$. The temporal series, the probability density function (PDF), and the main statistics (computed over 5000 periods) are reported, trying to highlight significant trends emerging with a fibrillated heartbeat. The ranges where the temporal series are visualized are therefore chosen as representative of the general behaviours.

We also evaluate pressures and volumes at particular instants of the heartbeat: end-systole (ES) and end-diastole (ED). Left (or right) end-systole is the instant defined by the closure of the aortic (or pulmonary) valve, while left (or right) end-diastole corresponds to the closure of the mitral (or tricuspid) valve.

\begin{table}
\centering
\begin{tabular}{|c|c|c|}
  \hline
  Variable & Present & Literature \\
           & Model   &  Data \\
  \hline
  $P_{la}$ & + & + [B9,B29] \\
  $P_{laed}$ & + & / \\
  $P_{laes}$ & + & / \\
  \hline
  $P_{lv}$ & - &  / \\
  $P_{lved}$ & + & - [B2] \\
  $P_{lves}$ & - & / \\
  \hline
  $P_{ra}$ & = & + [B2,B6,B7,B29] \\
  $P_{raed}$ & = & / \\
  $P_{raes}$ & - & / \\
  \hline
  $P_{rv}$ & = &  / \\
  $P_{rved}$ & + & - [B2] \\
  $P_{rves}$ & - & / \\
  \hline
  $V_{la}$ & + & = [B24], - [B6] \\
  $V_{laed}$ & + &  + [B1,B30,B31,B32,B34,B35] \\
  $V_{laes}$ & + &  + [B1,B30,B31,B32,B34,B35] \\
  \hline
  $V_{lv}$ & + & / \\
  $V_{lved}$ & + & + [B3,B32], = [B24] \\
  $V_{lves}$ & + & + [B3,B32], = [B24] \\
  \hline
  $V_{ra}$ & - & / \\
  $V_{raed}$ & = & + [B32]* \\
  $V_{raes}$ & - & + [B32]* \\
  \hline
  $V_{rv}$ & - & / \\
  $V_{rved}$ & - & + [B32]* \\
  $V_{rves}$ & = & + [B32]* \\
  \hline
  $P_{sas}$ & - & - [B8,B25], + [B14], = [B2,B7,B21] \\
  $P_{sas,s}$ & - & + [B14,B19], = [B2,B7,B21] \\
  $P_{sas,d}$ & - & + [B14,B19], = [B2,B7,B21] \\
  $P_{p}$ & - &  - [B19] \\
  \hline
  $P_{pas}$ & = & = [B26,B29], + [B2,B7] \\
  $P_{pas,s}$ & - & + [B6,B7], = [B2] \\
  $P_{pas,d}$ & + & + [B2,B6,B7] \\
  \hline
  $P_{svn}$ & - & / \\
  \hline
  $P_{pvn}$ & + & + [B2,B6,B7,B11], = [B24] \\
  \hline
  $SV$ & - & - [B2,B9,B14,B18,B28], = [B21] \\
  \hline
  $EF$ & - & - [B3,B6,B13,B32,B34], = [B35] \\
  \hline
  $SW$ & - & - [B20,B24] \\
  \hline
  $CO$ & - & - [B6,B8,B10,B11,B16,B18,B21] \\
       &   & - [B7,B9,B20,B22,B24,B25,B28,B33] \\
       &   & = [B2,B14,B26,B29] \\
  \hline
\end{tabular}
\caption{Present outcomes (II column) and literature data (III column): $+$ increase during AF, $-$ decrease during AF, $=$ no significant variations during AF, $/$ no data available. All the variables of the present model are intended as averaged over 5000 cycles. * Slight increase which is significant only after 6 months. Literature references can be found in Online Resource 2.}
\label{results_table}
\end{table}

Other important hemodynamics parameters, referred to the left ventricle, are evaluated: the stroke volume ($SV$), the ejection fraction ($EF$), the stroke work ($SW$), and the cardiac output ($CO$).

When in vivo fibrillated data are available, they are compared with the present outcomes (literature references are reported for convenience in Online Resource 2). To provide a quick overview of the comparison, the current results and the available literature data are summarized in Table \ref{results_table}, where the variations during AF with respect to NSR are synthesized as follows: $+$ increase, $-$ decrease, $=$ no substantial variation, $/$ no data available. It should be reminded that measurements come from very different types of arrythmia, i.e. persistent, chronic, paroxysmal, induced, benign fibrillations. Moreover, AF is usually present along with other pathologies, which can themselves substantially affect the cardiac response. Nevertheless, depending on the clinical case, different techniques (cardioversion, catheter ablation, drugs treatment, ...) are used to restore the normal sinus rhythm, as well as differing follow-up periods (from 24 hours to 6 months) are analyzed. The invasiveness of some medical procedures, together with the anatomical and structural heart complexity, can contribute to the incompleteness of data regarding certain hemodynamics variables (e.g., right ventricle measurements).

\noindent Keeping in mind these aspects, the literature data should not be taken for a comparison on the specific values (which can largely vary from case to case), but as indicators of general trends during AF. 

\begin{table*}
\centering
\begin{tabular}{|c|c|c|c|}
  \hline
  Variable & NSR & AF & AF with normal  \\
           &     &    & LV contractility \\
  \hline
  $P_{la}$ & $\mu=9.12$, $\sigma=0.74$  &  $\mu=10.29$, $\sigma=0.70$ & $\mu=9.16$, $\sigma=0.75$ \\
  $P_{laed}$ & $\mu=8.71$, $\sigma=0.05$  &  $\mu=10.73$, $\sigma=0.12$ & $\mu=9.59$, $\sigma=0.08$ \\
  $P_{laes}$ & $\mu=10.06$, $\sigma=0.10$  &  $\mu=11.11$, $\sigma=0.15$ & $\mu=10.08$, $\sigma=0.15$ \\
  \hline
  $P_{lv}$ & $\mu=43.23$, $\sigma=45.18$  &  $\mu=40.64$, $\sigma=38.98$ & $\mu=45.21$, $\sigma=45.37$ \\
  $P_{lved}$ & $\mu=11.25$, $\sigma=0.40$  &  $\mu=17.16$, $\sigma=0.82$ & $\mu=17.23$, $\sigma=0.94$ \\
  $P_{lves}$ & $\mu=100.18$, $\sigma=0.78$  &  $\mu=86.45$, $\sigma=2.99$ & $\mu=96.76$, $\sigma=2.83$ \\
  \hline
  $P_{ra}$ & $\mu=7.90$, $\sigma=1.56$  &  $\mu=7.50$, $\sigma=1.55$ & $\mu=8.06$, $\sigma=1.71$ \\
  $P_{raed}$ & $\mu=7.70$, $\sigma=0.10$  &  $\mu=7.72$, $\sigma=0.33$ & $\mu=8.29$, $\sigma=0.54$ \\
  $P_{raes}$ & $\mu=10.41$, $\sigma=0.09$  &  $\mu=9.73$, $\sigma=0.28$ & $\mu=10.56$, $\sigma=0.32$ \\
  \hline
  $P_{rv}$ & $\mu=13.33$, $\sigma=8.84$  &  $\mu=13.06$, $\sigma=8.65$ & $\mu=13.53$, $\sigma=8.78$ \\
  $P_{rved}$ & $\mu=10.48$, $\sigma=0.13$  &  $\mu=11.81$, $\sigma=0.55$ & $\mu=12.66$, $\sigma=1.21$ \\
  $P_{rves}$ & $\mu=19.38$, $\sigma=0.32$  &  $\mu=17.67$, $\sigma=0.89$ & $\mu=18.00$, $\sigma=0.89$ \\
  \hline
  $V_{la}$ & $\mu=56.53$, $\sigma=6.25$  &  $\mu=65.95$, $\sigma=4.64$ & $\mu=58.41$, $\sigma=5.03$ \\
  $V_{laed}$ & $\mu=55.37$, $\sigma=0.35$  &  $\mu=68.84$, $\sigma=0.80$ & $\mu=61.26$, $\sigma=0.54$ \\
  $V_{laes}$ & $\mu=64.41$, $\sigma=0.67$  &  $\mu=71.41$, $\sigma=1.00$ & $\mu=64.53$, $\sigma=1.03$ \\
  \hline
  $V_{lv}$ & $\mu=93.19$, $\sigma=27.72$  &  $\mu=108.76$, $\sigma=23.43$ & $\mu=88.62$, $\sigma=25.80$ \\
  $V_{lved}$ & $\mu=119.79$, $\sigma=1.68$  &  $\mu=126.93$, $\sigma=5.03$ & $\mu=110.30$, $\sigma=5.28$ \\
  $V_{lves}$ & $\mu=55.95$, $\sigma=0.97$  &  $\mu=79.72$, $\sigma=7.34$ & $\mu=56.69$, $\sigma=2.19$ \\
  \hline
  $V_{ra}$ & $\mu=48.70$, $\sigma=11.17$  &  $\mu=47.36$, $\sigma=10.34$ & $\mu=51.10$, $\sigma=11.38$ \\
  $V_{raed}$ & $\mu=48.72$, $\sigma=0.64$  &  $\mu=48.77$, $\sigma=2.20$ & $\mu=52.62$, $\sigma=3.59$ \\
  $V_{raes}$ & $\mu=66.71$, $\sigma=0.63$  &  $\mu=62.22$, $\sigma=1.88$ & $\mu=67.72$, $\sigma=2.17$ \\
  \hline
  $V_{rv}$ & $\mu=70.36$, $\sigma=25.95$  &  $\mu=62.90$, $\sigma=22.42$ & $\mu=66.87$, $\sigma=25.03$ \\
  $V_{rved}$ & $\mu=98.25$, $\sigma=2.98$  &  $\mu=81.90$, $\sigma=8.27$ & $\mu=87.75$, $\sigma=11.16$ \\
  $V_{rves}$ & $\mu=34.41$, $\sigma=0.33$  &  $\mu=34.69$, $\sigma=8.27$ & $\mu=35.09$, $\sigma=1.68$ \\
  \hline
  $P_{sas}$ & $\mu=99.52$, $\sigma=10.12$  &  $\mu=89.12$, $\sigma=9.10$ & $\mu=99.47$, $\sigma=10.20$ \\
  $P_{sas,s}$ & $\mu=116.22$, $\sigma=1.32$  &  $\mu=103.66$, $\sigma=3.00$ & $\mu=114.92$, $\sigma=2.98$ \\
  $P_{sas,d}$ & $\mu=83.24$, $\sigma=2.42$  &  $\mu=77.24$, $\sigma=5.18$ & $\mu=86.23$, $\sigma=6.11$ \\
  $P_{p}$ & $\mu=32.99$, $\sigma=1.13$  &  $\mu=26.42$, $\sigma=6.20$ & $\mu=28.69$, $\sigma=3.54$ \\
  \hline
  $P_{pas}$ & $\mu=20.14$, $\sigma=4.00$  &  $\mu=20.10$, $\sigma=3.37$ & $\mu=20.15$, $\sigma=3.78$ \\
  $P_{pas,s}$ & $\mu=26.73$, $\sigma=0.22$  &  $\mu=25.38$, $\sigma=0.74$ & $\mu=26.03$, $\sigma=0.84$ \\
  $P_{pas,d}$ & $\mu=14.27$, $\sigma=0.39$  &  $\mu=15.68$, $\sigma=1.21$ & $\mu=15.21$, $\sigma=1.41$ \\
  \hline
  $P_{svn}$ & $\mu=13.89$, $\sigma=0.16$  &  $\mu=12.84$, $\sigma=0.16$ & $\mu=14.04$, $\sigma=0.19$ \\
  \hline
  $P_{pvn}$ & $\mu=9.60$, $\sigma=0.46$  &  $\mu=10.72$, $\sigma=0.45$ & $\mu=9.64$, $\sigma=0.49$ \\
  \hline
  $SV$ & $\mu=63.84$, $\sigma=2.63$  &  $\mu=47.21$, $\sigma=8.32$ & $\mu=53.61$, $\sigma=7.26$ \\
  \hline
  $EF$ & $\mu=53.27$, $\sigma=1.46$  &  $\mu=37.12$, $\sigma=6.01$ & $\mu=48.40$, $\sigma=4.29$ \\
  \hline
  $SW$ & $\mu=0.87$, $\sigma=0.02$  &  $\mu=0.57$, $\sigma=0.14$ & $\mu=0.74$, $\sigma=0.06$ \\
  \hline
  $CO$ & $\mu=4.80$, $\sigma=0.17$  &  $\mu=4.38$, $\sigma=0.67$ & $\mu=4.92$, $\sigma=0.55$ \\
  \hline
\end{tabular}
\caption{Mean ($\mu$) and standard deviation ($\sigma$) values for the analyzed parameters. II column: NSR, III column: AF, IV column: AF without the reduced and variable LV elastance (i.e. $E_{lv,max} = 2.5$ mmHg/ml). The pressures $P$ are expressed in [mmHg], the volumes $V$ in [ml], the stroke volume $SV$ in [ml], the ejection fraction $EF$ in percentage, the stroke work $SW$ in [J], the cardiac output $CO$ in [l/min].}
\label{value_table}
\end{table*}

The pressures $P$ are expressed in terms of [mmHg], the volumes $V$ in [ml], the stroke volume $SV$ in [ml], the ejection fraction $EF$ in percentage, the stroke work $SW$ in [J], the cardiac output $CO$ in [l/min], the temporal variable $t$ in [s]. Mean, $\mu$, and standard deviation, $\sigma$, values of the present simulations are summarized in Table \ref{value_table} for all the variables here considered. Second column refers to the NSR, third column to the AF case, while fourth column refers to AF without the forced decrease of the LV elastance ($E_{lv,max}$ is constant and equal to 2.5 mmHg/ml).

\subsection{Left heart pressures}

\begin{figure}
\begin{minipage}[]{0.5\columnwidth}
\includegraphics[width=\columnwidth]{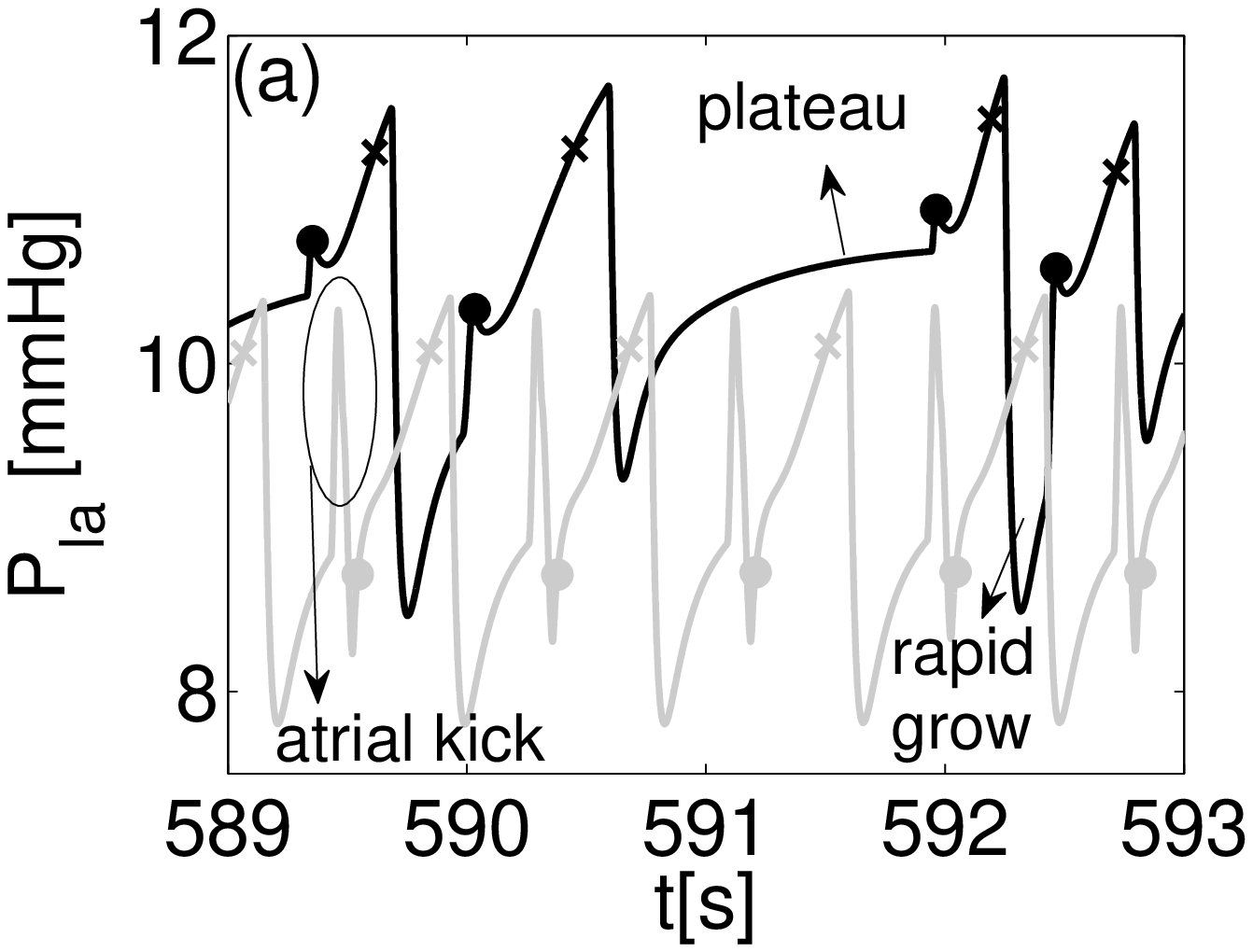}
\end{minipage}
\hspace{-0.3cm}
\begin{minipage}[]{0.5\columnwidth}
\includegraphics[width=\columnwidth]{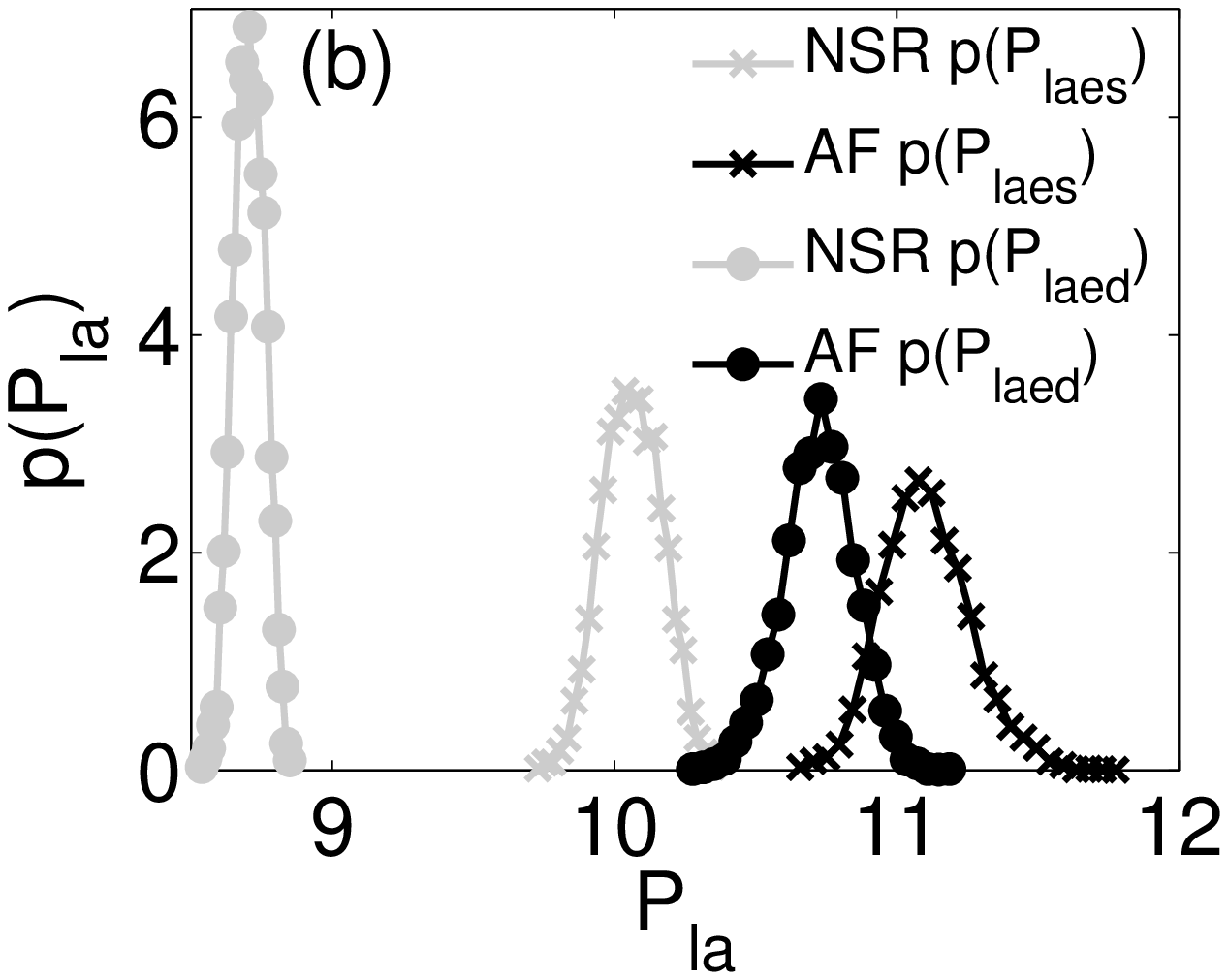}
\end{minipage}
\hspace{-0.5cm}
\begin{minipage}[]{0.5\columnwidth}
\includegraphics[width=\columnwidth]{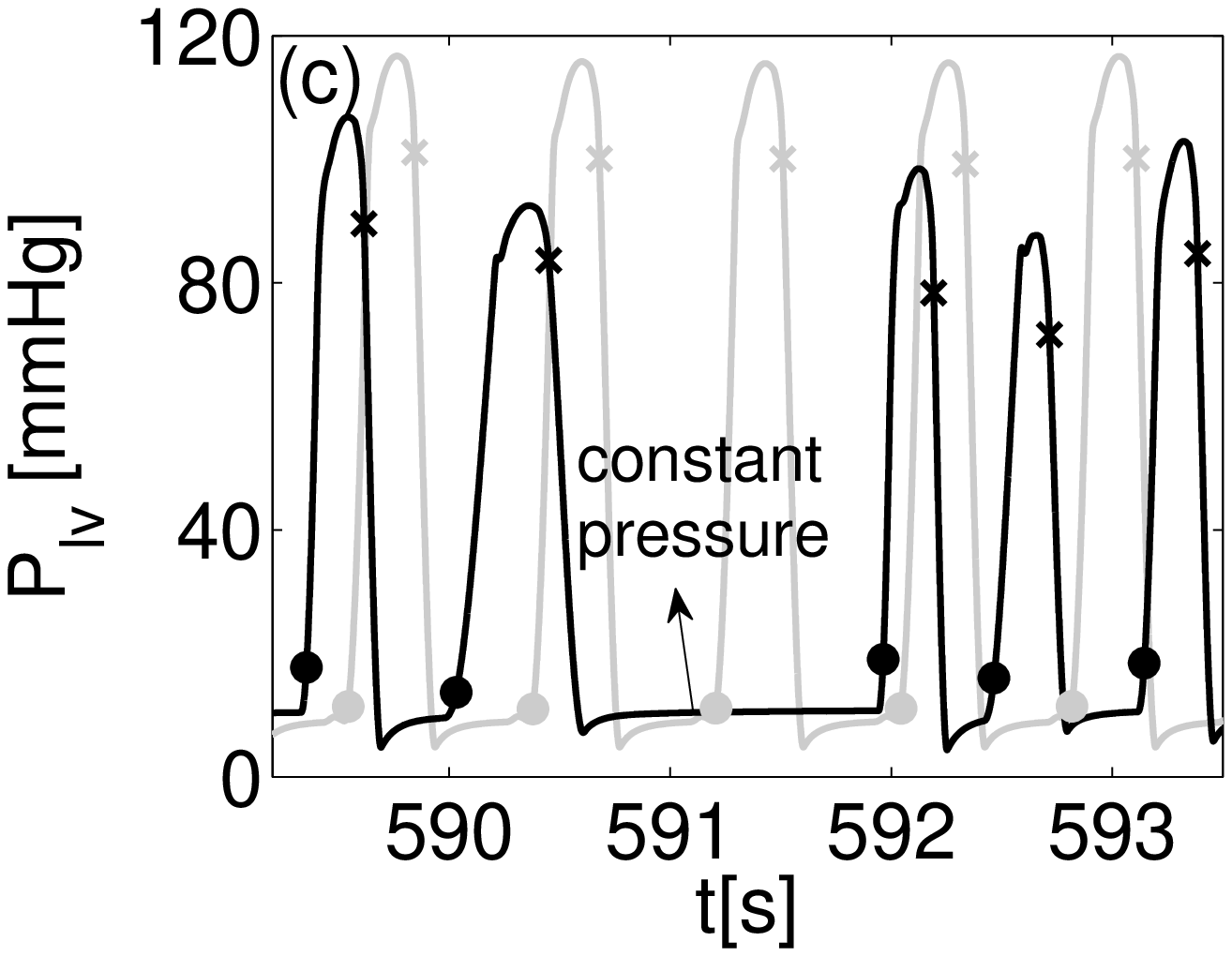}
\end{minipage}
\hspace{-0.3cm}
\begin{minipage}[]{0.5\columnwidth}
\includegraphics[width=\columnwidth]{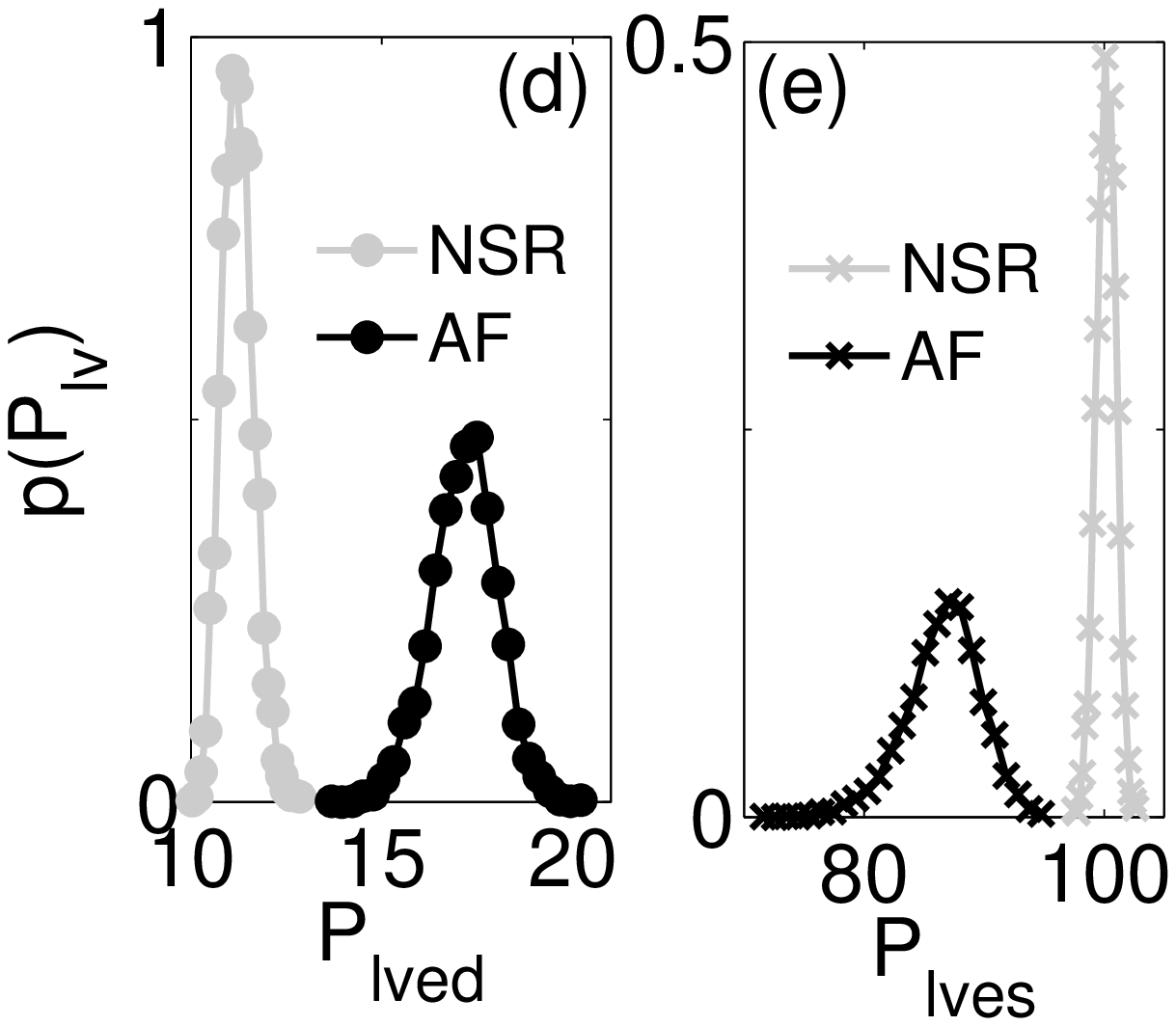}
\end{minipage}
\caption{Left heart pressures: (a) left atrial pressure series, (b) PDFs of left atrial end-systolic and end-diastolic pressures. (c) Left ventricular pressure series, (d)-(e) PDFs of left ventricular end-diastolic and end-systolic pressures. Light: NSR, dark: AF. Symbols indicate end-systole ($\times$) and end-diastole ($\bullet$) values.}
\label{pl}
\end{figure}

The left atrial pressure increases during atrial fibrillation, by shifting its mean value from $9.12$ mmHg (NSR) to $10.29$ mmHg (AF). The range of values reached during AF is a bit less wide than in the normal case (AF $\sigma$=$0.70$, NSR $\sigma$=$0.74$). The PDFs of end-diastolic and end-systolic left atrial pressures (see Fig. \ref{pl}b) also reveal that, during AF, end-diastolic and end-systolic pressure values are by far closer to each other than in the normal rhythm. The differences emerging in the temporal series of normal and fibrillated cases (see Fig. \ref{pl}a) are mainly due to the passive atrial role. When the heartbeat is longer, $P_{la}$ slowly increases until it almost reaches a plateau value. When instead the heartbeat is shorter, the pressure rapidly grows and no plateau region is observable. 

\noindent A decrease of the left atrial pressure is found [B9,B29] after cardioversion in patients with AF, a result which is in agreement with the present observations.

During AF the left ventricular pressure reduces its mean value, from $43.23$ mmHg (NSR) to $40.64$ mmHg (AF), and the standard deviation value as well (NSR $\sigma$=$45.18$, AF $\sigma$=$38.98$). Lower maxima values in the fibrillated case are in general found (see Fig. \ref{pl}c), while the longer the heartbeat, the higher is the next maximum value with respect to the previous one. For a long heartbeat the pressure remains constant for more than 50\% of the $RR$ interval, while for a short heartbeat a phase of constant pressure almost disappears. 

\noindent The PDFs (see Fig. \ref{pl}d, e) of end-diastolic and end-systolic pressure values show, during AF, an increase and a decrease, respectively. As for left atrial pressure, the fibrillated sequence tends to get the two values closer. 
Although there is no clear evidence of the effects of AF on the left ventricular pressure ([B2] only showed a decrease of the end-diastolic pressure during induced AF), an increase of the end-diastolic left ventricular pressure is in general a symptom of heart failure risk and ventricle dysfunction \cite{Mielniczuk}.

\subsection{Right heart pressures}

\begin{figure}
\begin{minipage}[]{0.5\columnwidth}
\includegraphics[width=\columnwidth]{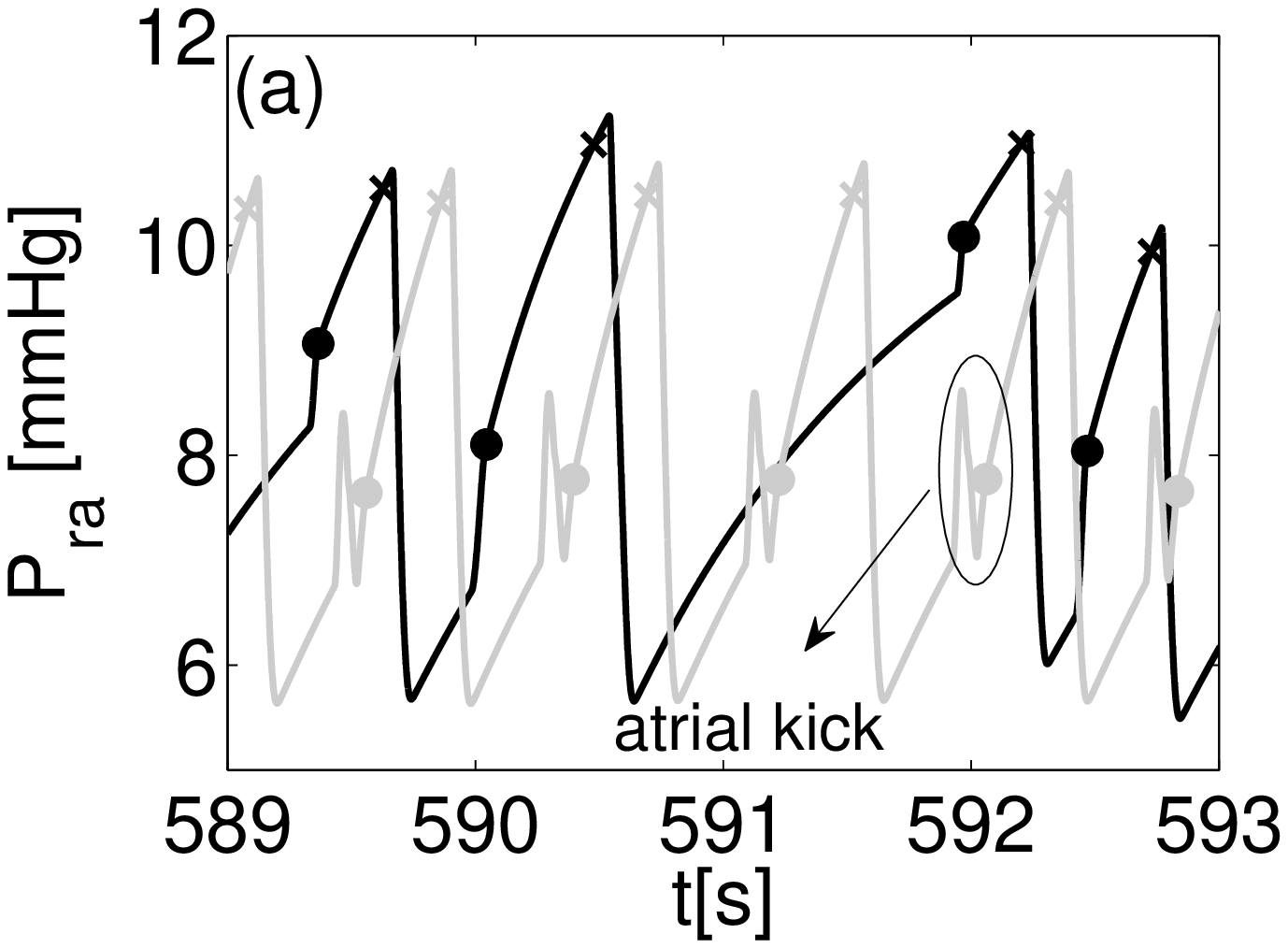}
\end{minipage}
\hspace{-0.3cm}
\begin{minipage}[]{0.5\columnwidth}
\includegraphics[width=\columnwidth]{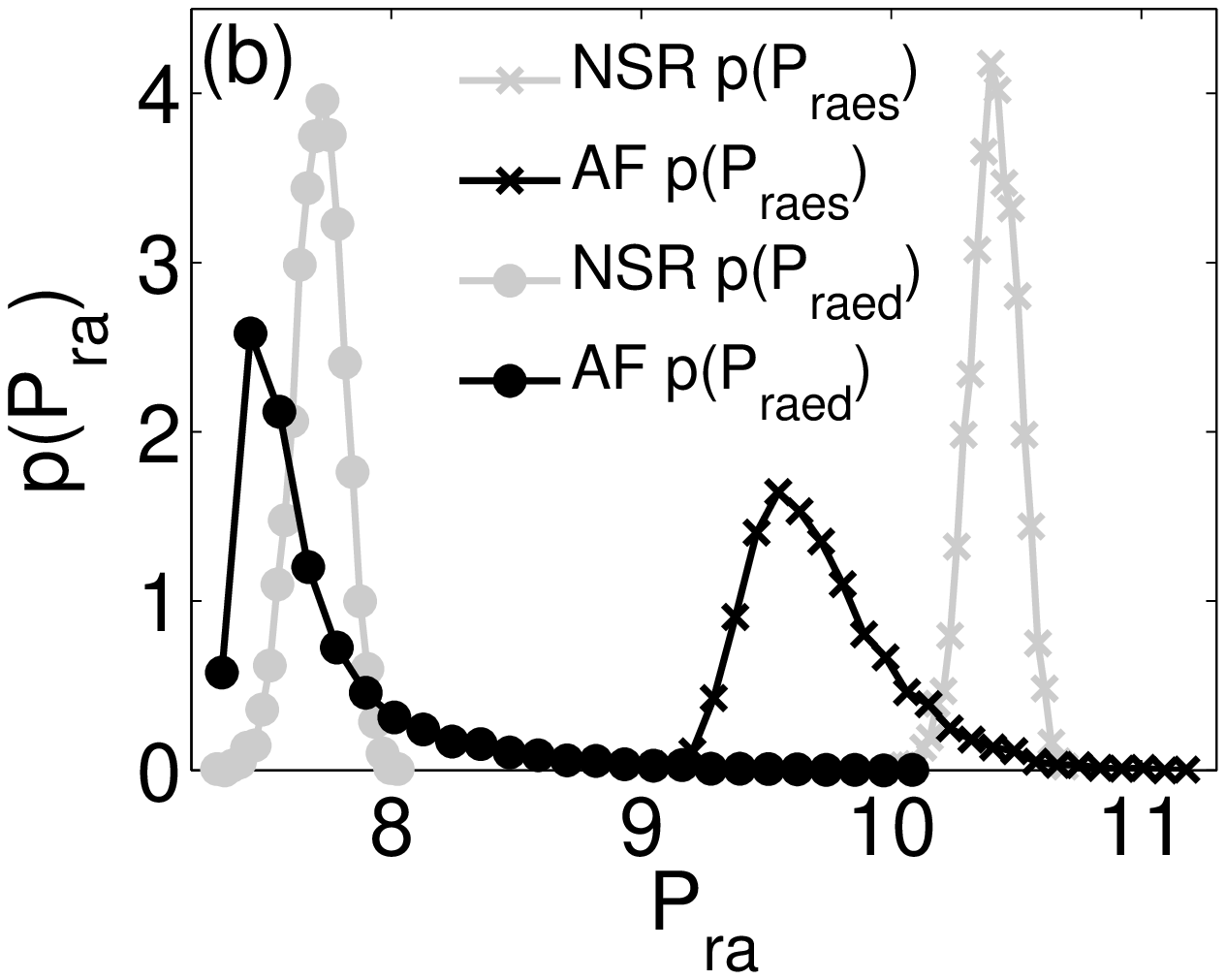}
\end{minipage}
\begin{minipage}[]{0.5\columnwidth}
\includegraphics[width=\columnwidth]{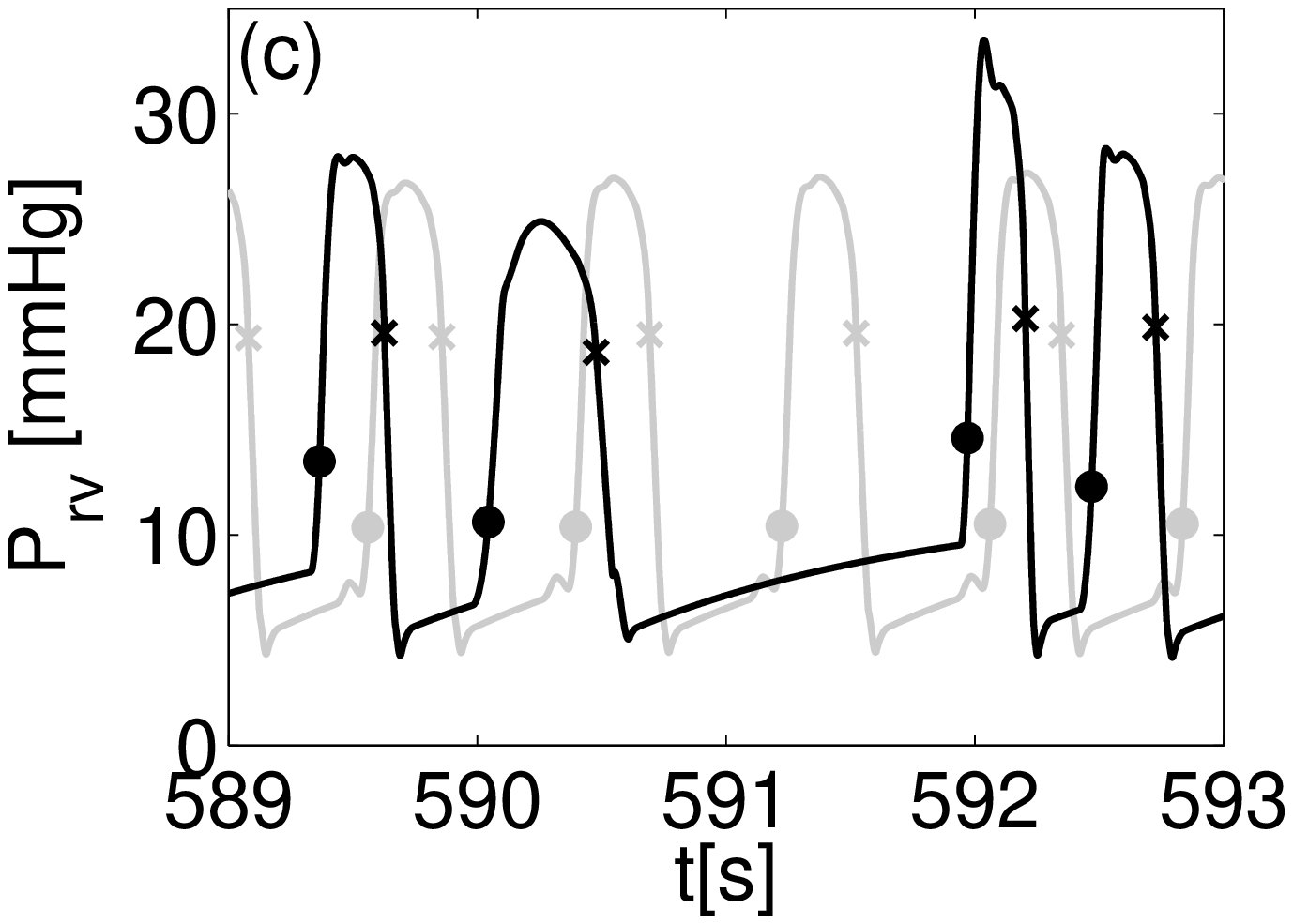}
\end{minipage}
\hspace{-0.3cm}
\begin{minipage}[]{0.5\columnwidth}
\includegraphics[width=\columnwidth]{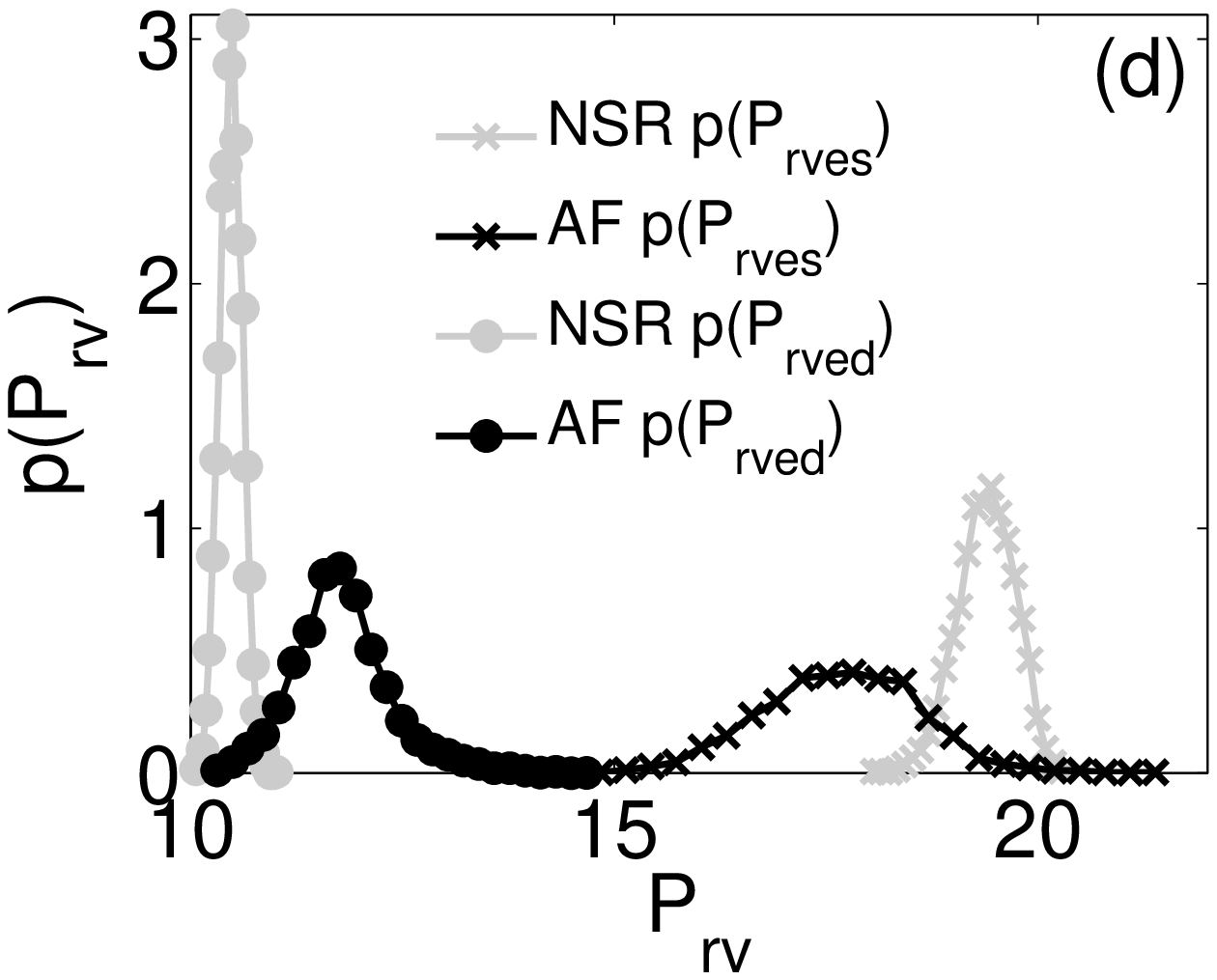}
\end{minipage}
\caption{Right heart pressures: (a) right atrial pressure series, (b) PDFs of end-diastolic and end-systolic right atrial pressures. (c) Right ventricular pressure series, (d) PDFs of end-diastolic and end-systolic right ventricular pressures. Light: NSR, dark: AF. Symbols indicate end-systole ($\times$) and end-diastole ($\bullet$) values.}
\label{pr}
\end{figure}

Even though the mean right atrial pressure is frequently reported to increase its mean value during atrial fibrillation [B2,B6,B7,B29], here we do not identify substantial differences with respect to the physiologic case, in terms of mean value (NSR $\mu$=$7.90$, AF $\mu$=$7.50$) and standard deviation (NSR $\sigma$=$1.56$, AF $\sigma$=$1.55$). The comparison of the time series in Fig. \ref{pr}a mainly highlights the lack of the atrial kick in the fibrillated case. The PDFs of end-diastolic pressures confirm no strong differences in the mean values, while the mean end-systolic pressure is shifted towards lower values during AF (see Fig. \ref{pr}b). Both end-systolic and end-diastolic pressure PDFs present pronounced right tails during AF, meaning that higher maxima values are possible.

Probably due to the geometrical and structural complexity of the right ventricle \cite{Haddad} and to the difficulty of invasive measures, effects of AF in terms of right ventricular pressure are not easily available \cite{Haddad}. Although there are no significant variations of the two fundamental statistical measures (NSR: $\mu$=$13.33$, $\sigma$=$8.84$; AF: $\mu$=$13.06$, $\sigma$=$8.65$), an inspection of the temporal series (Fig. \ref{pr}c) reveals that higher values are possible during AF. As for the left ventricle, higher maxima are more probable after a long heartbeat, while after a short heartbeat the maximum pressure value is usually lower than the previous one. When considering the end-diastolic pressure (Fig. \ref{pr}d), an average increase is experienced during atrial fibrillation (NSR $\mu$=$10.48$, AF $\mu$=$11.81$), with values that are much more spread out (NSR $\sigma$=$0.13$, AF $\sigma$=$0.55$). On the contrary, end-systolic pressure decreases during AF (NSR $\mu$=$19.38$, AF $\mu$=$17.67$), thereby getting closer to the end-diastolic mean value. The only available result come from induced AF [B2], showing a decrease of the end-diastolic pressure.

\subsection{Atrial Volumes}

Atrial dilatation, together with systemic and pulmonary hypertension, is by itself one of the most important risk factors for AF \cite{Kannel,Tsang}. For this reason, when looking at the available data in literature it is not easy to discern whether atrial enlargement is a cause or a consequence of atrial fibrillation.

\noindent In the present work, at least for the left atrium, we observe an average increase of the atrial volume during atrial fibrillation (NSR: $\mu$=$56.53$, $\sigma$=$6.25$; AF: $\mu$=$65.95$, $\sigma$=$4.64$), that is therefore here considered as a consequence of AF. This average enlargement is confirmed by higher average end-diastolic and end-systolic volumes (end-diastolic volumes: NSR $\mu$=$55.37$, AF $\mu$=$68.84$; end-systolic volumes: NSR $\mu$=$64.41$, AF $\mu$=$71.41$). Several results [B1,B30,B31,B32,B34,B35] agree with an increase of the left atrial (end-diastolic and end-systolic) volumes during AF. Only two works are not aligned, by remarking no substantial differences [B24] or a decrease [B6] of the left atrial volume when a fibrillated heartbeat is present.

Concerning the volumes of the right atrium, during atrial fibrillation we observe a small decrease of the mean (NSR: $\mu$=$48.70$, AF: $\mu$=$47.36$) and end-systolic (NSR: $\mu$=$66.71$, AF $\mu$=$62.22$) values, while the end-diastolic volume remains unvaried (NSR: $\mu$=$48.72$, AF: $\mu$=$48.77$). Right atrial volume is rarely measured during atrial fibrillation and sometimes only to assess the effects of varying volume loading conditions and heart rate [B35]. A statistically significant decrease of right atrial end-diastolic and end-systolic volumes after cardioversion was found only after 6 months [B32].

\noindent As a consequence of the atrial elastance modeling, the atrial volumes qualitatively resemble the atrial pressure behaviors and therefore are not shown here. 

\subsection{Ventricular Volumes}

End-systolic and end-diastolic left ventricle volume data are not often offered in the atrial fibrillation literature, maybe because the related ejection fraction better synthesizes a similar kind of information. The available results show an increase of both end-diastolic and end-systolic volumes during atrial fibrillation [B3,B32], while in one case no significant variations are encountered [B24]. Here we find an increase of the mean values of end-diastolic (EDV) and end-systolic (ESV) volumes during atrial fibrillation (EDV: NSR $\mu$=$119.79$, AF $\mu$=$126.93$; ESV: NSR $\mu$=$55.95$, AF $\mu$=$79.72$), as well as an increase of the mean left ventricle volume (NSR $\mu$=$93.19$, AF $\mu$=$108.76$).

\begin{figure}
\begin{minipage}[]{0.5\columnwidth}
\includegraphics[width=\columnwidth]{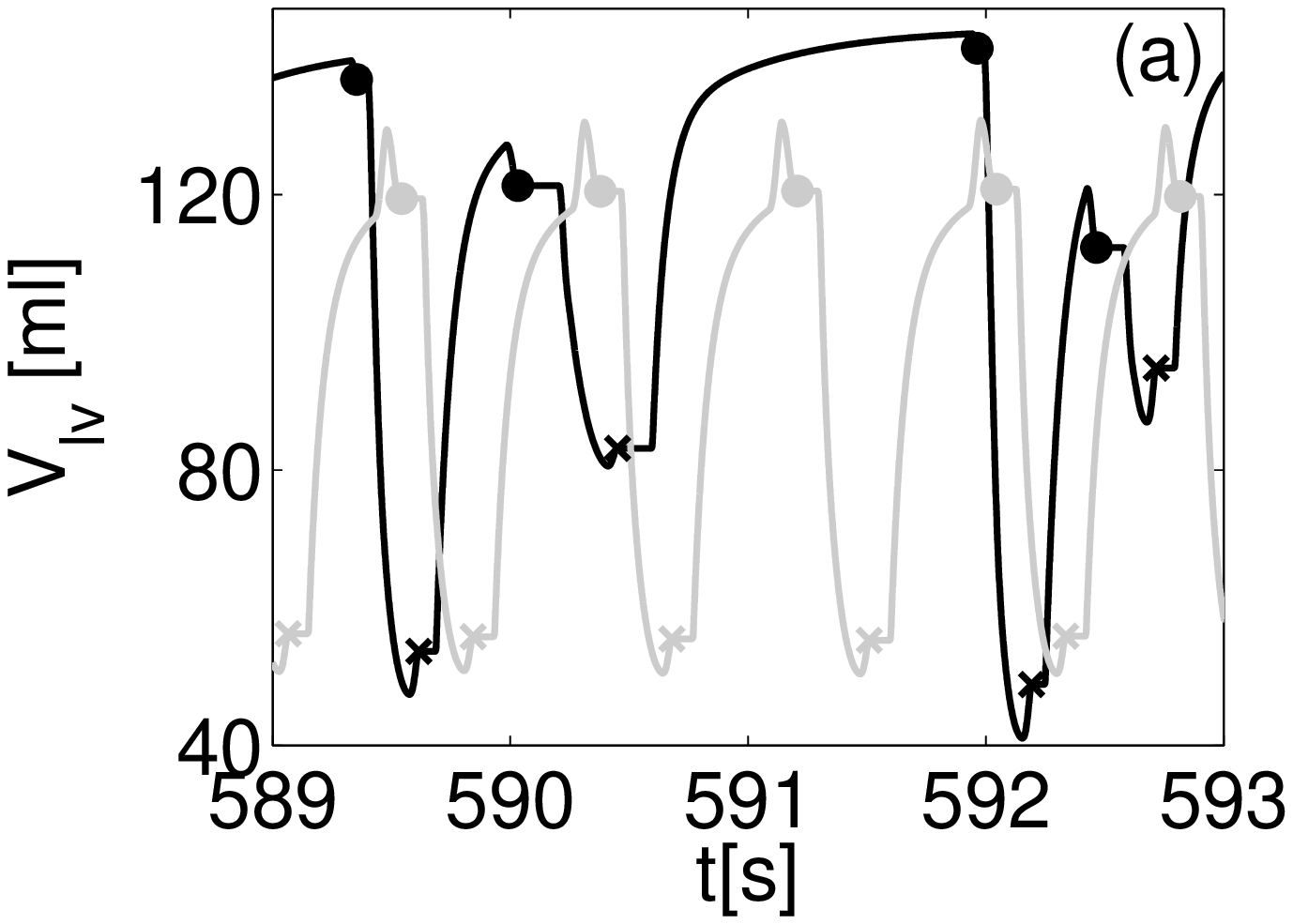}
\end{minipage}
\hspace{-0.3cm}
\begin{minipage}[]{0.5\columnwidth}
\includegraphics[width=\columnwidth]{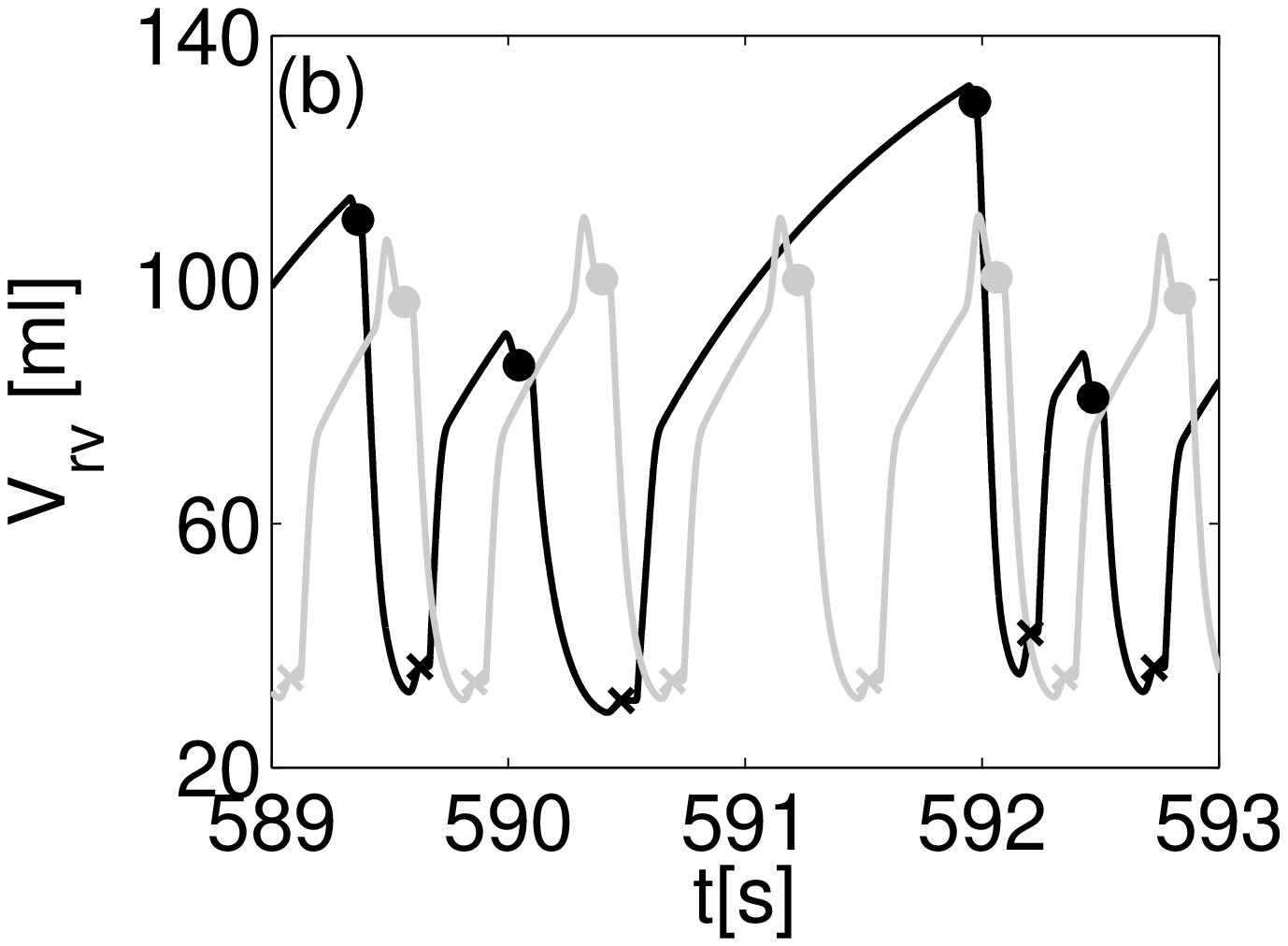}
\end{minipage}
\begin{minipage}[]{0.5\columnwidth}
\includegraphics[width=\columnwidth]{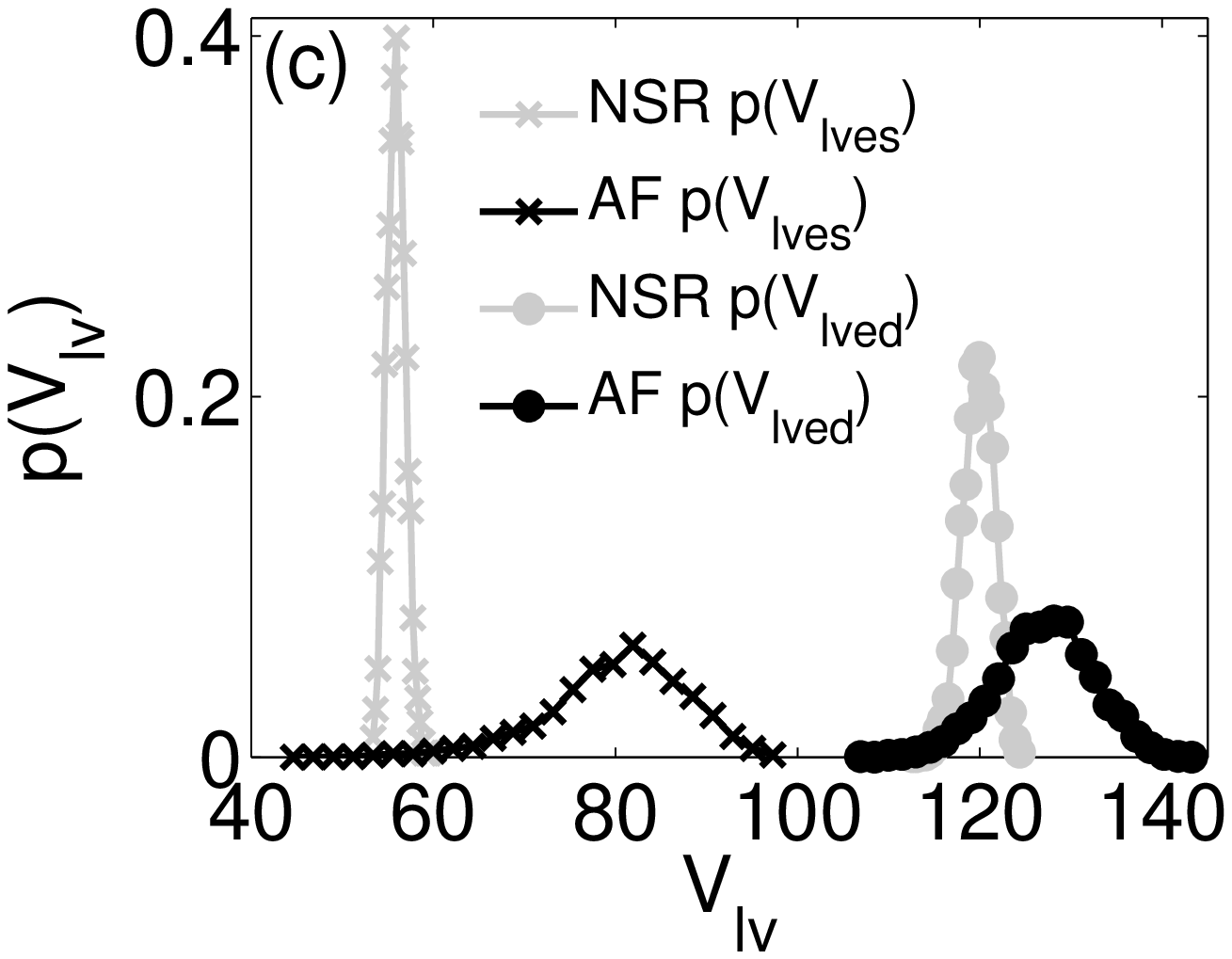}
\end{minipage}
\hspace{-0.3cm}
\begin{minipage}[]{0.5\columnwidth}
\includegraphics[width=\columnwidth]{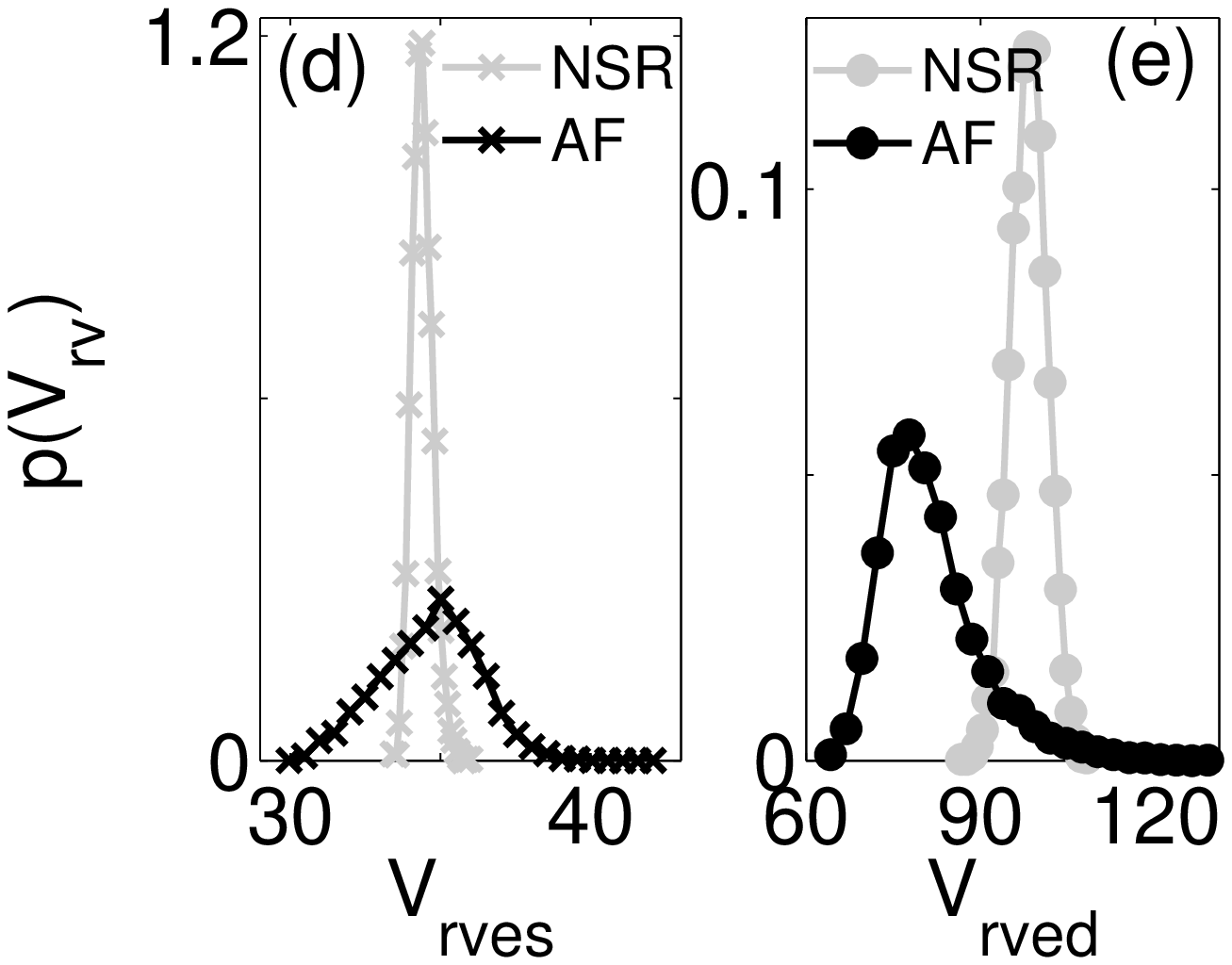}
\end{minipage}
\begin{minipage}[]{0.5\columnwidth}
\includegraphics[width=\columnwidth]{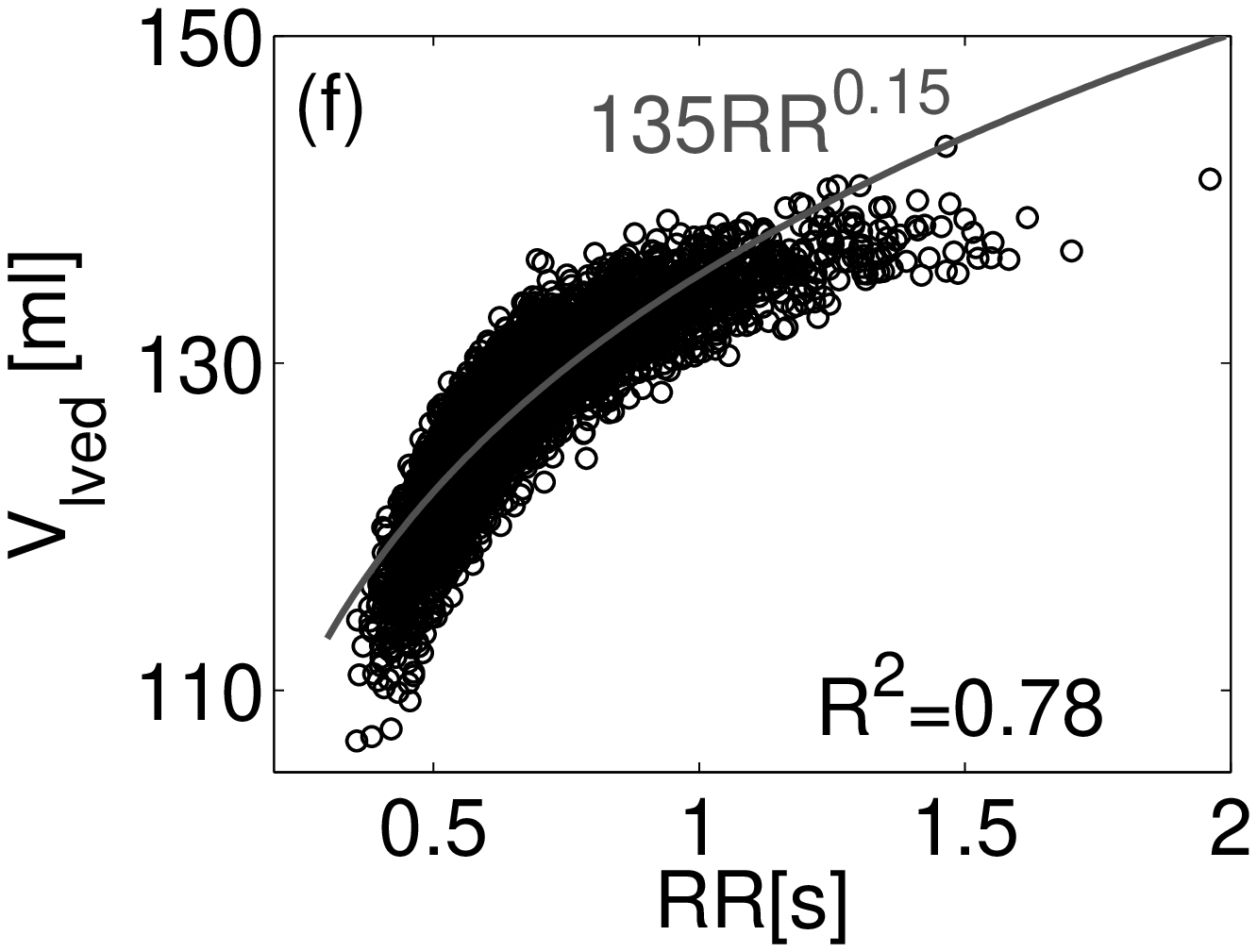}
\end{minipage}
\hspace{-0.3cm}
\begin{minipage}[]{0.5\columnwidth}
\includegraphics[width=\columnwidth]{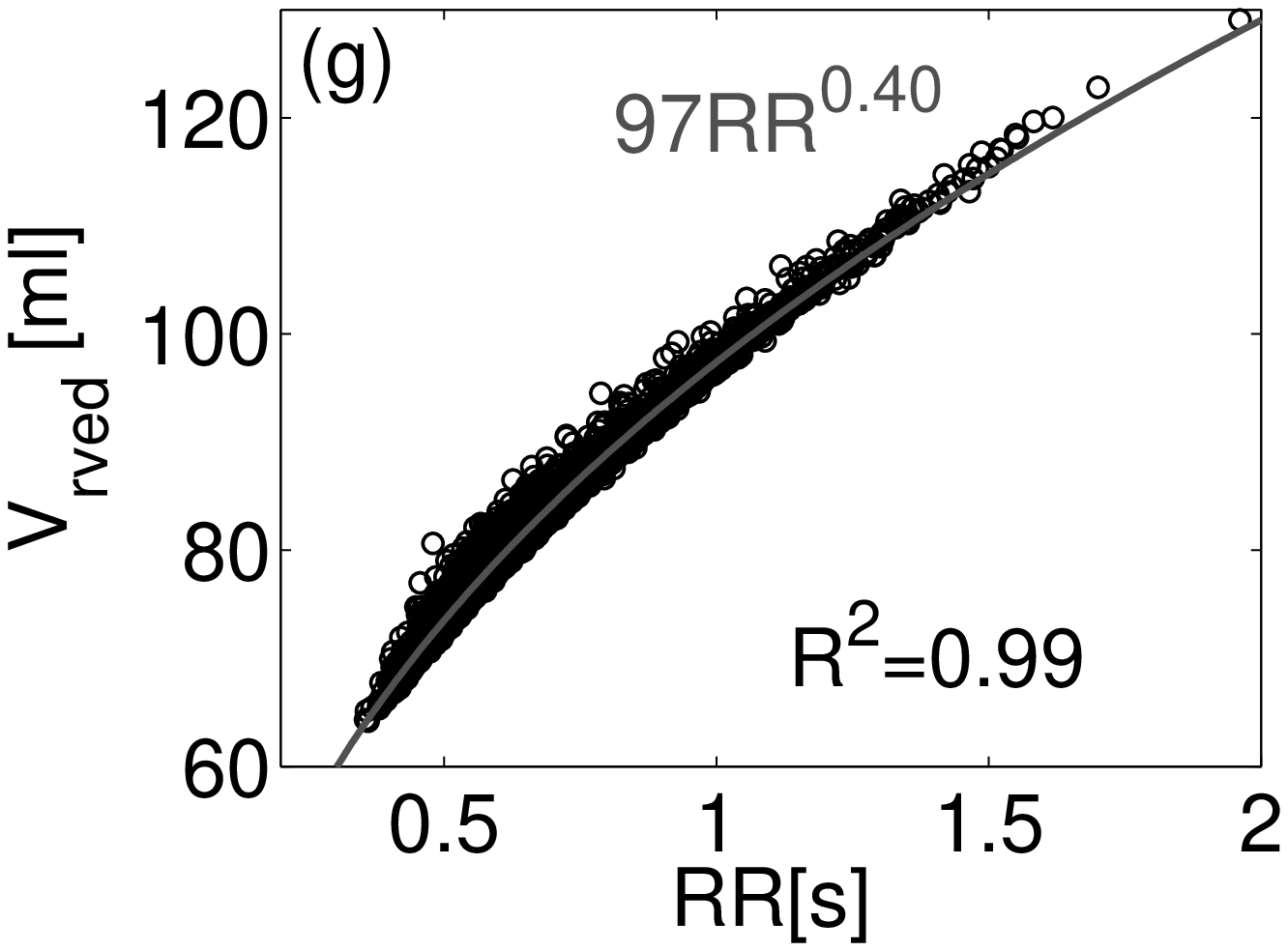}
\end{minipage}
\caption{(a),(c),(f) Left ventricular volume: (a) temporal series, (c) PDFs of end-systolic and end-diastolic left ventricular volumes. (f) AF: left ventricle end-diastolic volume ($V_{lved}$) as function of $RR$. (b),(d)-(e),(f) Right ventricular volume: (b) temporal series, (d)-(e) PDFs of end-systolic and end-diastolic right ventricular volumes. (g) AF: right ventricle end-diastolic volume ($V_{rved}$) as function of $RR$. (f),(g) Power-law fittings of the data and coefficients of determination, $R^2$, are introduced. Light: NSR, dark: AF. Symbols indicate end-systole ($\times$) and end-diastole ($\bullet$) values.}
\label{Vv}
\end{figure}

\noindent The shifts of the EDV and ESV towards higher values during atrial fibrillation are clearly detectable by the PDFs reported in Fig. \ref{Vv}c. Braunwald et al. [B4] found a direct proportion between the end-diastolic volume and the heartbeat length ($RR$). This relation can be verified by considering the fibrillated temporal series in Fig. \ref{Vv}a. The left ventricle is overfilled when the beat is long, while is underfilled when the beat is short. Figure \ref{Vv}f shows the relation $V_{lved}(RR)$, confirming the positive correlation between $RR$ and $V_{lved}$.

During atrial fibrillation the right ventricle experiences, with less data dispersion (coefficient of determination, $R^2$=$0.99$), a filling dynamics similar to the left ventricle: overfilling occurs for long beats, while underfilling follows a short heartbeat (Fig. \ref{Vv}b, g). The PDFs reveal that end-systolic values are more spread out maintaining the mean value of the NSR, while end-diastolic volume decreases even if during atrial fibrillation important overfilling are possible (see Fig. \ref{Vv}d, e). The mean RV volume is lower than in normal conditions (NSR $\mu$=$70.36$, $\sigma$=$25.95$; AF $\mu$=$62.90$, $\sigma$=$22.42$). Although not included as a common effect of atrial fibrillation, the combined action of irregular heartbeat and overfilling can compromise the right ventricle function \cite{Haddad}. Due to the complex geometrical shape, the right ventricle volume is not easily estimated. Recent magnetic resonance images [B32] show, after cardioversion, a slight decrease of right ventricle (end-systolic and end-diastolic) volumes which is significant only considering a six-month follow-up.

\subsection{Systemic and Pulmonary Arterial Pressures}

Systemic and pulmonary arterial pressures are perhaps the most controversial hemodynamics variables regarding the effects of atrial fibrillation. The first reason is that systemic hypertension is among the most common risk factors inducing atrial fibrillation \cite{Kannel,Verdecchia}. More recently, AF has been reported in patients with pulmonary hypertension [B27]. Therefore, it is not straightforward to identify the specific role of AF on systemic and pulmonary pressure levels. Secondly, systemic arterial pressure is in general variable during AF, and often difficult to estimate as the irregular heartbeat causes problems for non-invasive blood pressure measurements. 

\noindent The variability of systemic arterial pressure is present in the literature of fibrillated data. During AF no substantial differences in terms of systolic, diastolic and mean pressures are displayed [B2,B7], in particular when patients are not affected by other pathologies [B21]. An increase of these quantities is instead registered by Giglioli et al. [B14]. Furthermore, [B19] showed an increase of diastolic and systolic pressure, along with a decrease of the pulsatile pressure.

\begin{figure}
\begin{minipage}[]{0.5\columnwidth}
\includegraphics[width=\columnwidth]{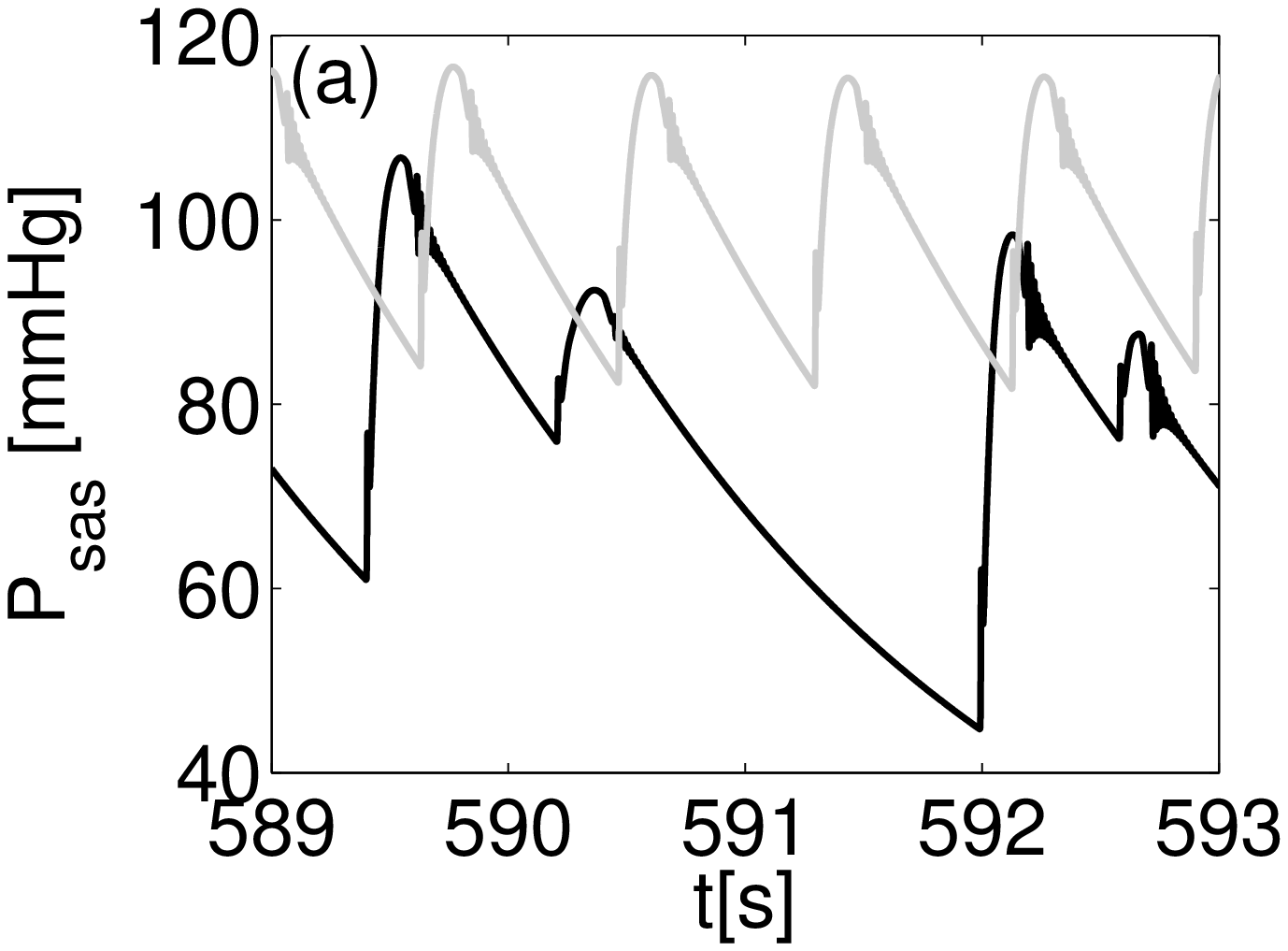}
\end{minipage}
\hspace{-0.3cm}
\begin{minipage}[]{0.5\columnwidth}
\includegraphics[width=\columnwidth]{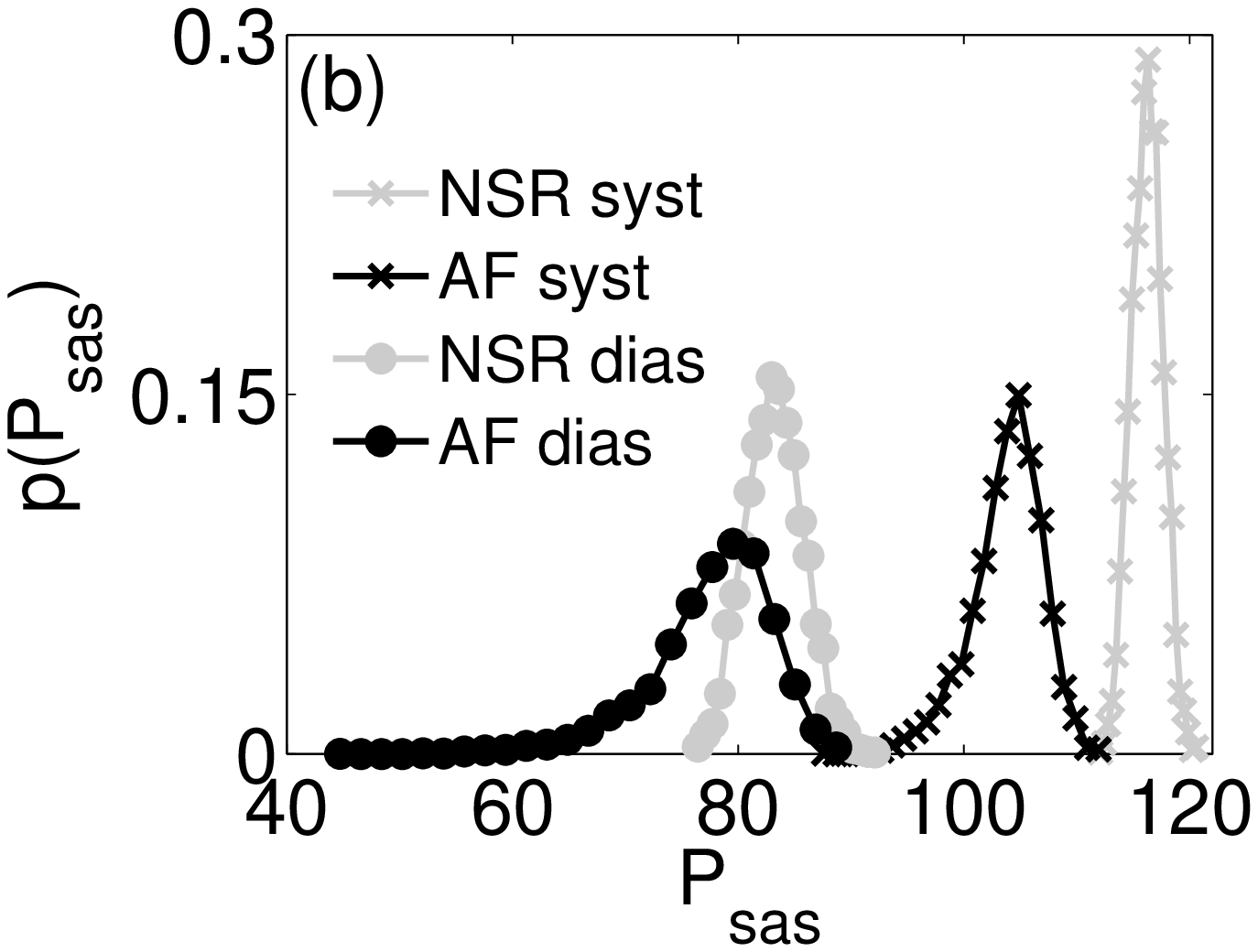}
\end{minipage}
\begin{minipage}[]{0.5\columnwidth}
\includegraphics[width=\columnwidth]{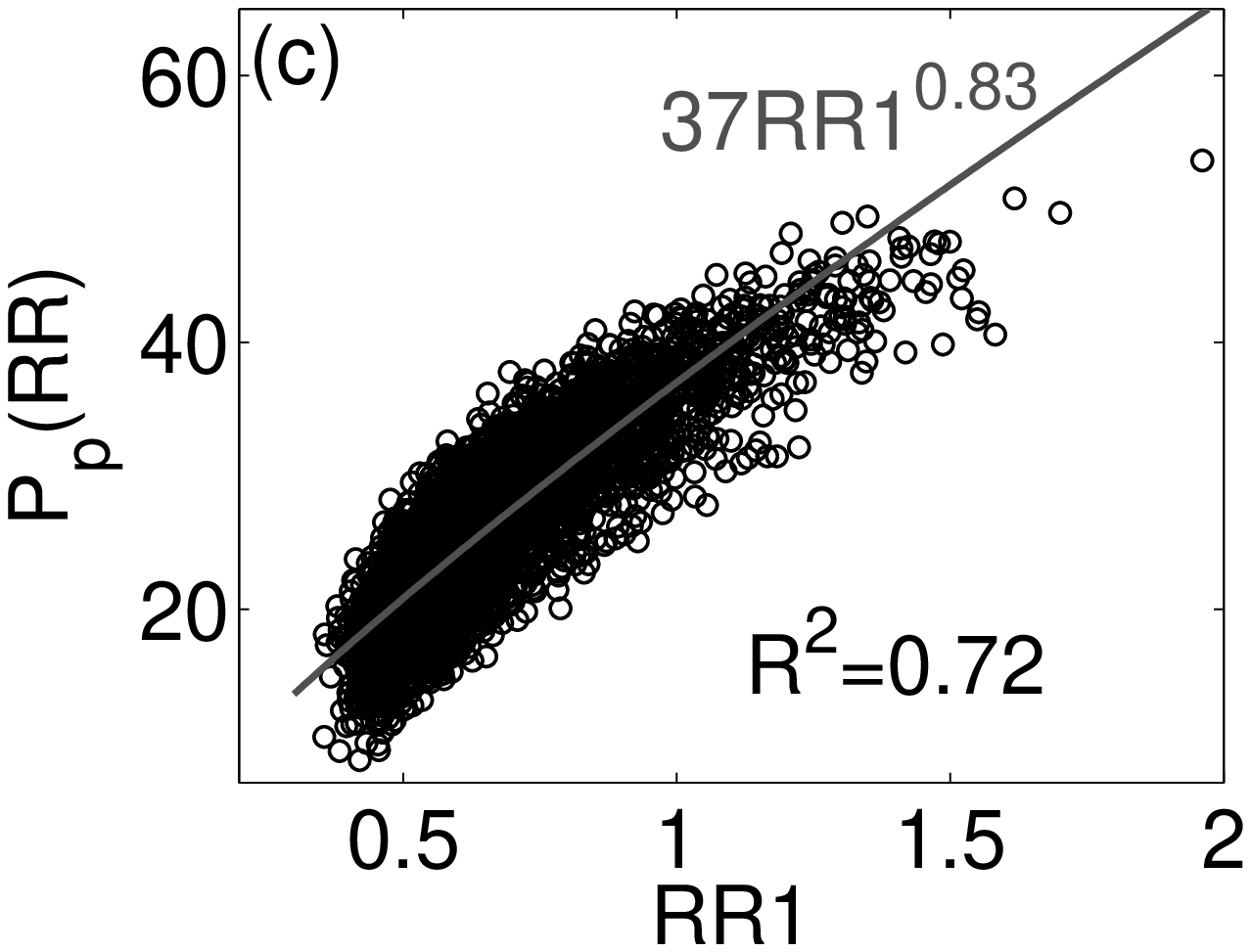}
\end{minipage}
\hspace{-0.3cm}
\begin{minipage}[]{0.5\columnwidth}
\includegraphics[width=\columnwidth]{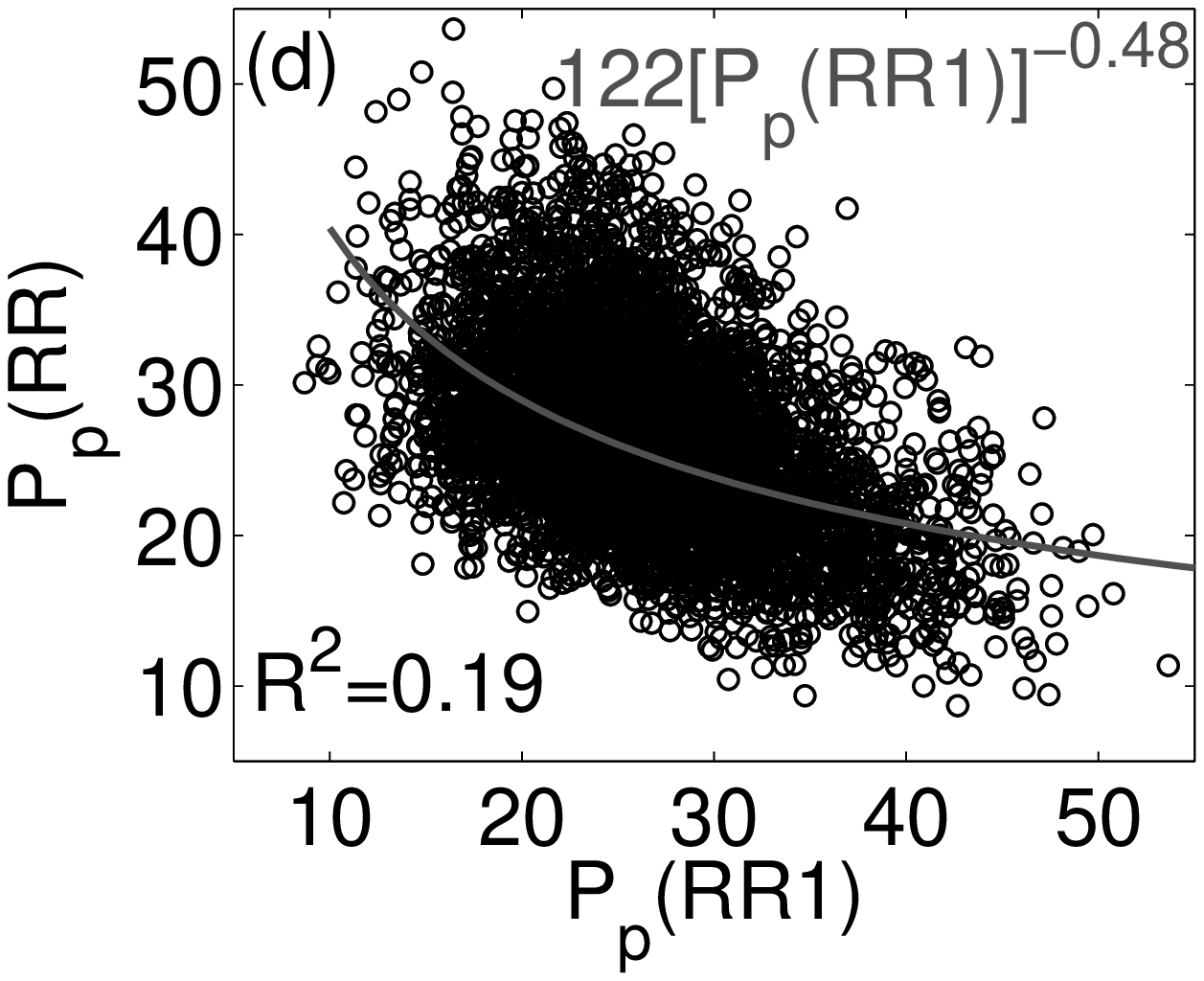}
\end{minipage}
\begin{minipage}[]{0.5\columnwidth}
\includegraphics[width=\columnwidth]{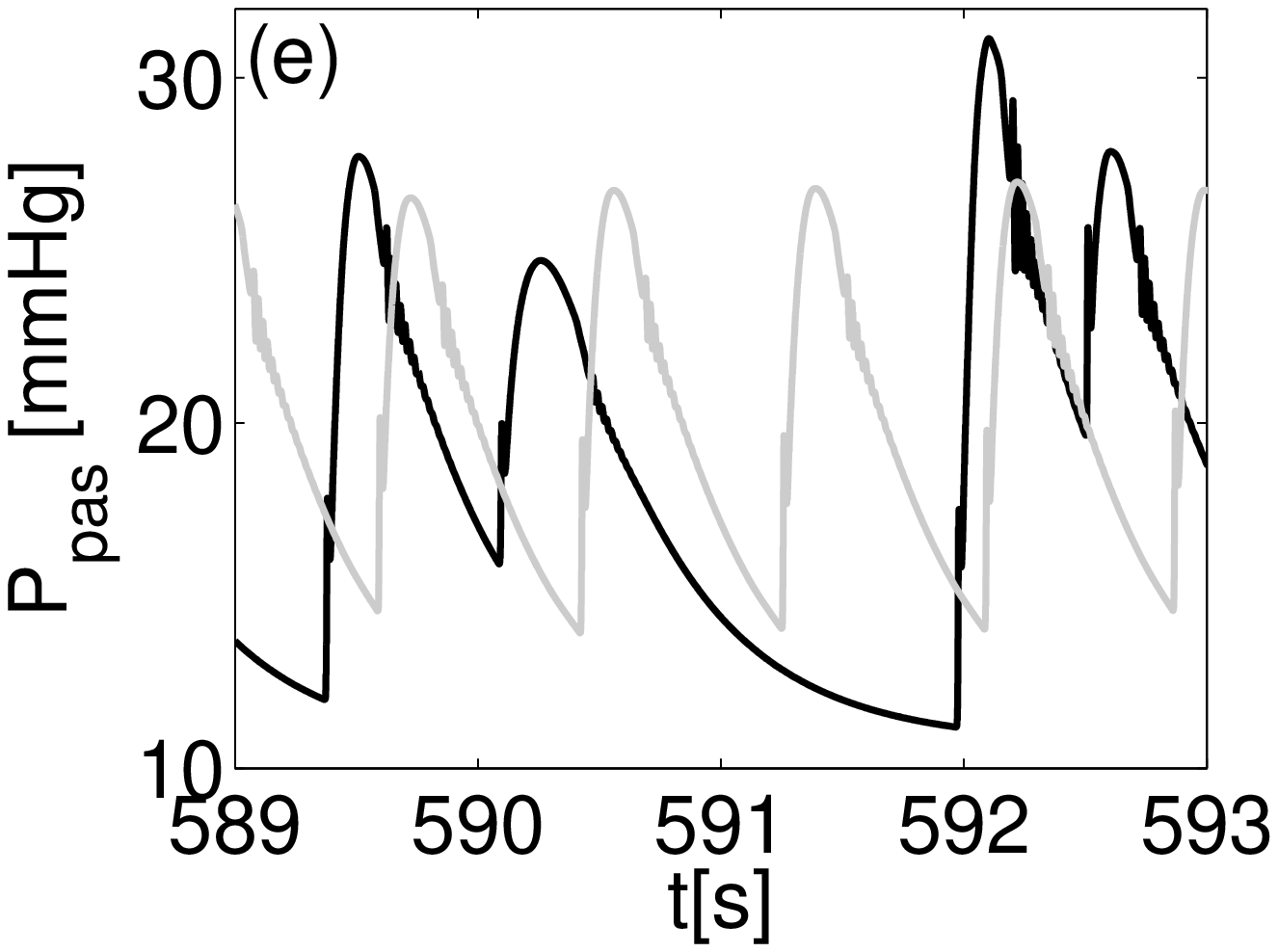}
\end{minipage}
\hspace{-0.3cm}
\begin{minipage}[]{0.5\columnwidth}
\includegraphics[width=\columnwidth]{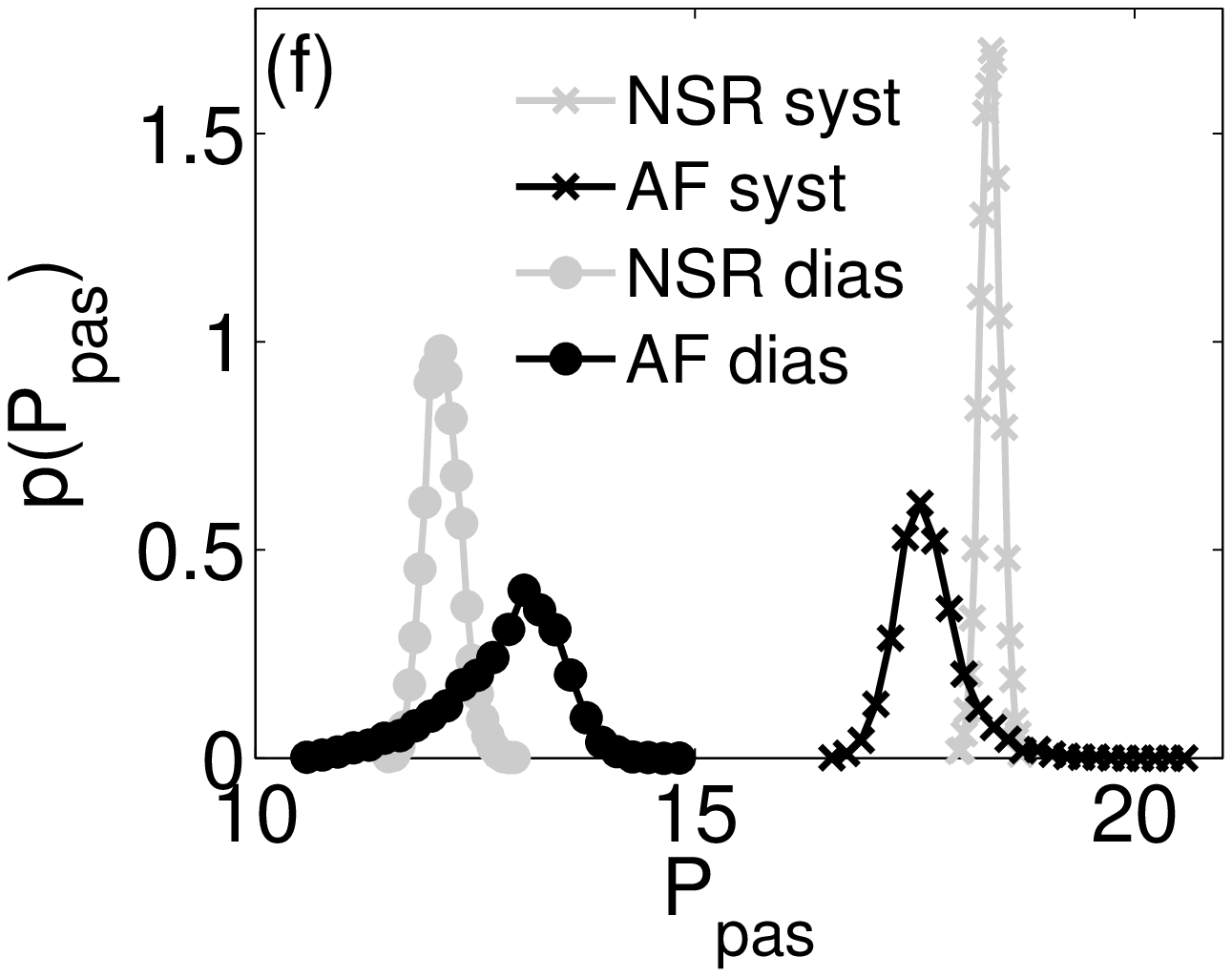}
\end{minipage}
\caption{(Top) Systemic arterial pressure: (a) temporal series; (b) PDFs of systolic and diastolic pressures. (Middle) AF pulsatile arterial pressure, power-law fittings of the data and coefficients of determination, $R^2$, are provided: (c) amplitude, $P_p(RR)$, as a function of the preceding heartbeat, $RR1$; (d) amplitude, $P_p(RR)$, as a function of the pulsatile pressure of the preceding beat, $P_p(RR1)$. (Bottom) Pulmonary arterial pressure: (a) temporal series, (b) PDFs of systolic and diastolic pressures. Light: NSR, dark: AF.}
\label{sas_pas}
\end{figure}

\noindent We notice a decrease of the mean and standard deviation values of the systemic arterial pressure during atrial fibrillation (NSR: $\mu$=$99.52$, $\sigma$=$10.12$; AF: $\mu$=$89.12$, $\sigma$=$9.10$), which is in agreement with the common sign of hypotension caused by a decreased cardiac output [B8,B25]. The average decrease is accompanied by a decrease of the mean systolic (NSR $\mu$=$116.22$, AF $\mu$=$103.66$, see Fig. \ref{sas_pas}b), diastolic (NSR $\mu$=$83.24$, AF $\mu$=$77.24$, see Fig. \ref{sas_pas}b), and pulsatile (NSR $\mu$=$32.99$, AF $\mu$=$26.42$) pressures. The pulsatile arterial pressure, $P_{p}$=$P_{syst}-P_{dias}$, is the variable which is mostly affected by the variability of the heartbeat (NSR $\sigma$=$1.13$, AF $\sigma$=$6.20$), reaching values that oscillate from $10$ to $50$ mmHg (in the normal case, $P_p$ varies from $29$ to $36$ mmHg). The analysis of the fibrillated time series (Fig. \ref{sas_pas}a) highlights that the pulsatile pressure amplitude, $P_p(RR)$, increases with the length of the preceding heartbeat, $RR1$, while decreases with the increase of the pressure amplitude of the preceding beat, $P_p(RR1)$. These findings are summarized in Fig. \ref{sas_pas} (panels c and d), and confirm what observed by Dodge et al. [B12].

As the systemic pressure, the pulmonary arterial pressure suffers from the same variability of registered data during fibrillation events. The mean value remains unchanged during atrial fibrillation (NSR: $\mu$=$20.14$, AF $\mu$=$20.10$), in line with the findings of [B26,B29]. An increase is instead displayed by the results in [B2,B7]. Systolic pressure decreases (NSR $\mu$=$26.73$, AF $\mu$=$25.38$, see Fig. \ref{sas_pas}f), [B6,B7] register an increase, while [B2] do not measure significant variations. Diastolic pressure increases during atrial fibrillation (NSR $\mu$=$14.27$, AF $\mu$=$15.68$, see Fig. \ref{sas_pas}f), and this result aligns with all the other available data [B2,B6,B7]. Both systolic and diastolic pressure values are more spread out during AF (see Fig. \ref{sas_pas}f). The temporal dynamics is analogous to the arterial systemic pressure (Fig. \ref{sas_pas}e): a large pulsatile pressure amplitude is due to a long preceding heartbeat and is followed by a smaller pressure amplitude.

\subsection{Pulmonary and Systemic Vein Pressures}

The pulmonary capillary wedge pressure (or pulmonary vein pressure) is an indirect estimate of the left atrium pressure. 
During AF an increase of the pulmonary vein pressure is observed by [B2,B6,B7], while no particular differences emerge from [B11,B24]. An increase of the mean pulmonary vein pressure is confirmed by the present data (NSR $\mu$=$9.60$, AF $\mu$=$10.72$), and is in agreement with an increase of the left atrium pressure (see Section 3.1).

The systemic vein pressure (also referred to as central venous pressure) estimates the pressure in the thoracic vena cava and approximates the right atrial pressure. As for this latter variable, the systemic vein pressure decreases during atrial fibrillation (NSR $\mu$=$13.89$, AF $\mu$=$12.84$). To the best of our knowledge, no data treat the effects of atrial fibrillation on the systemic vein pressure.


\subsection{Stroke Volume and Ejection Fraction}

The stroke volume, $SV$, is the difference between end-diastolic and end-systolic ventricular volumes and here is used to represent the volume of blood pumped from the left ventricle with each beat, $SV$=$V_{lved}-V_{lves}$. $SV$ coincides with the integral of the flow across the aortic valve.

\noindent There is almost unanimity among the collected data in literature saying that atrial fibrillation reduces the stroke volume [B2,B9,B14,B18,B28]. The relative increase of $SV$ when passing from AF to NSR is up $40\%$ [B9,B14,B18] and in one case is around $30\%$ [B28]. Only Killip et al. [B21] found no substantial differences with respect to the normal conditions.

\begin{figure}
\begin{minipage}[]{0.5\columnwidth}
\includegraphics[width=\columnwidth]{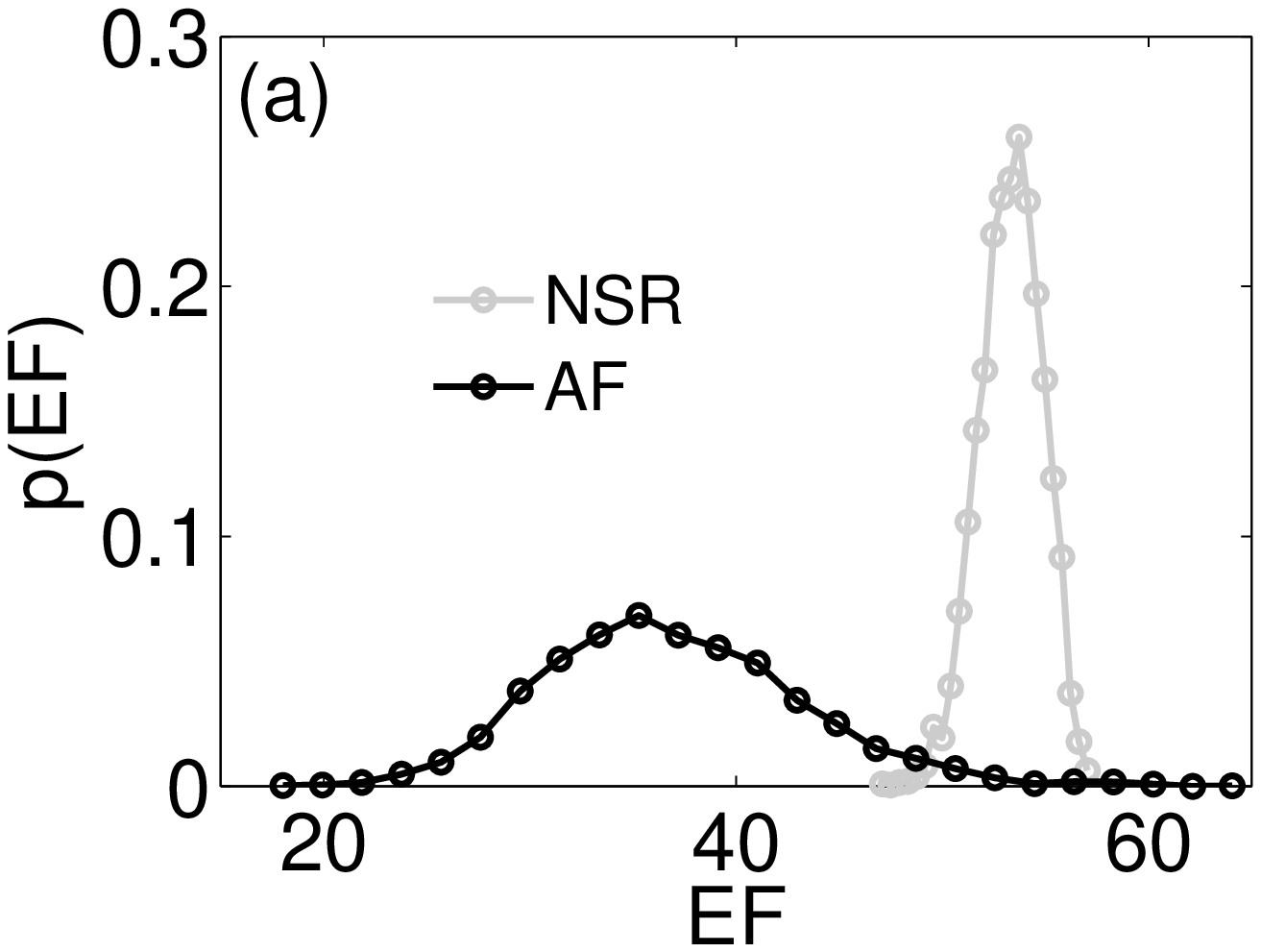}
\end{minipage}
\hspace{-0.3cm}
\begin{minipage}[]{0.5\columnwidth}
\includegraphics[width=\columnwidth]{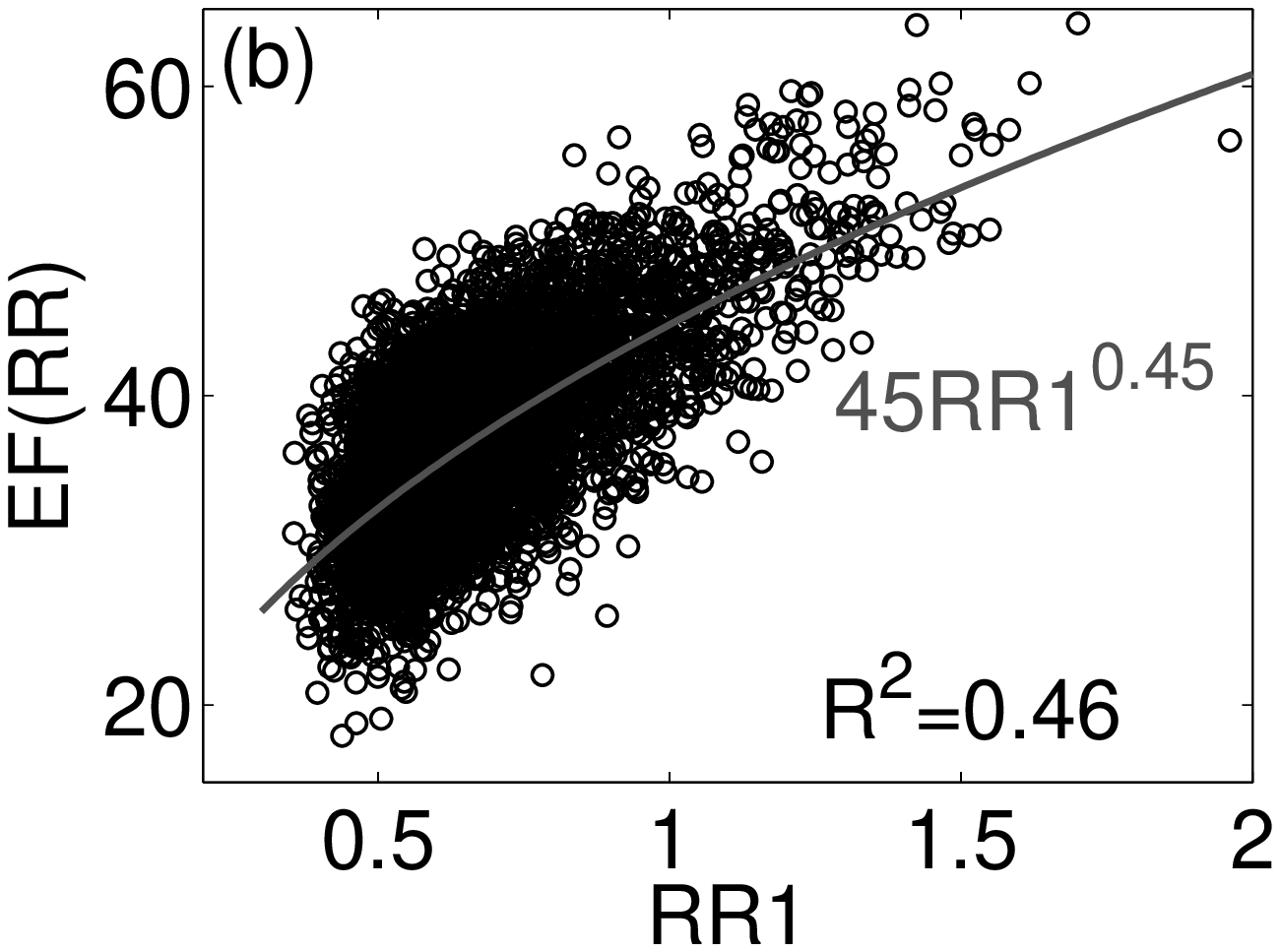}
\end{minipage}
\begin{minipage}[]{0.5\columnwidth}
\includegraphics[width=\columnwidth]{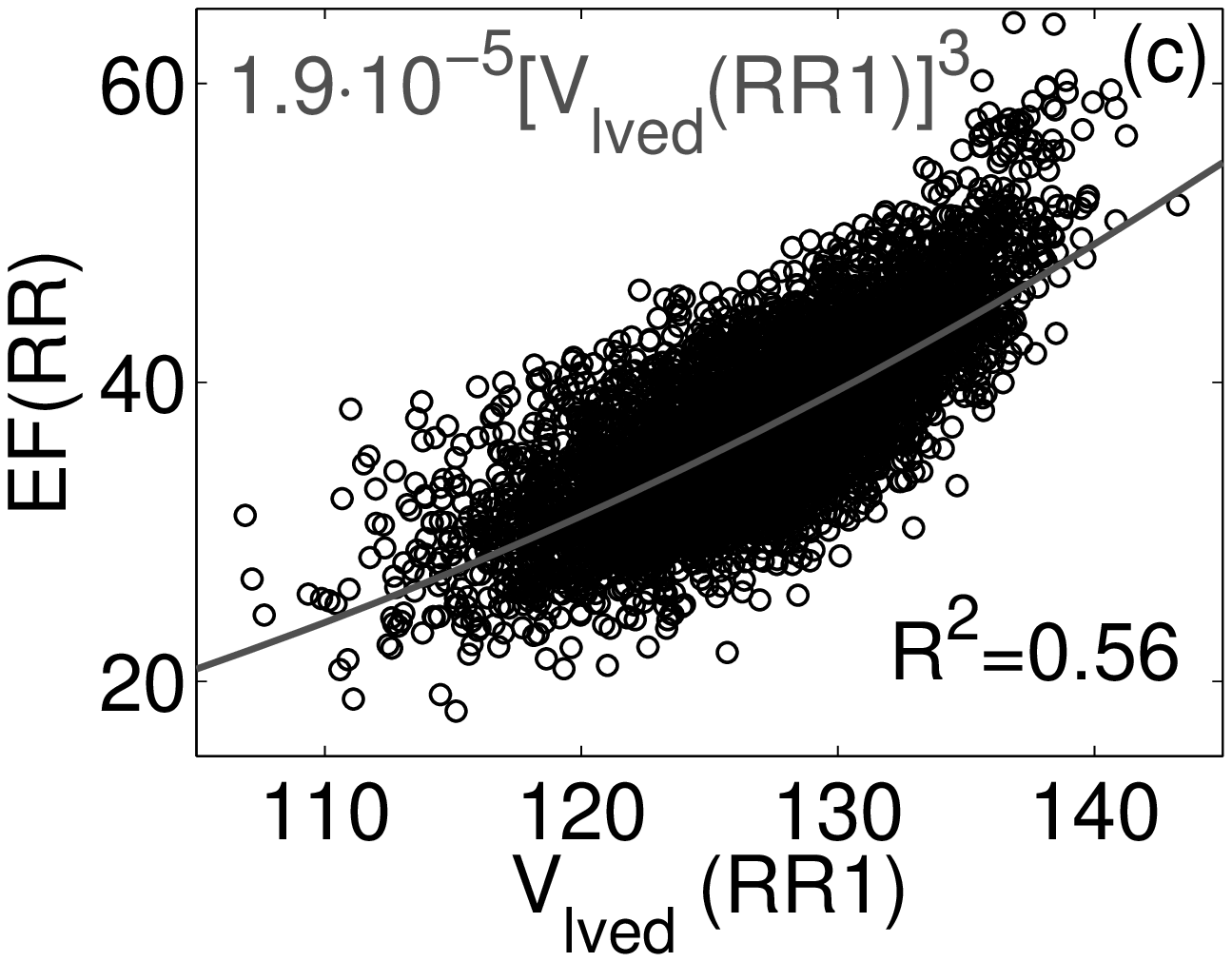}
\end{minipage}
\hspace{-0.3cm}
\begin{minipage}[]{0.5\columnwidth}
\includegraphics[width=\columnwidth]{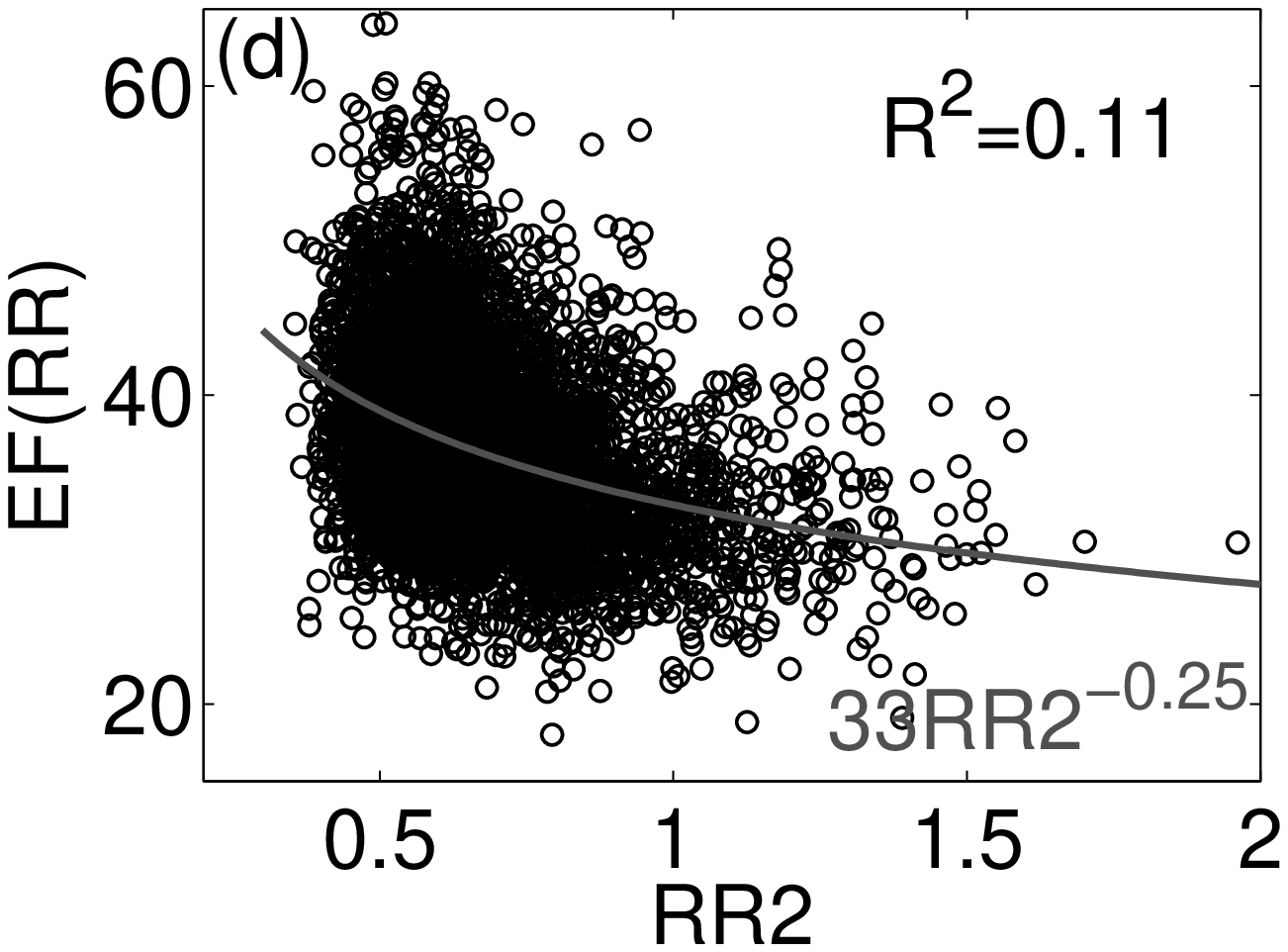}
\end{minipage}
\caption{Ejection fraction. Power-law fittings of the data and coefficients of determination, $R^2$, are added in panels (b), (c), (d). (a) Ejection fraction PDFs. (b) AF: ejection fraction, $EF(RR)$, as a function of the preceding beat, $RR1$. (c) AF: ejection fraction, $EF(RR)$, as a function of the end-diastolic volume of the preceding beat, $V_{lved}(RR1)$. (d) AF: ejection fraction, $EF(RR)$, as a function of the pre-preceding beat, $RR2$. Light: NSR, dark: AF.}
\label{SV_EF}
\end{figure}

The present results fully agree with the measured data in literature: from a fibrillated to a normal condition there is a relative increase of $SV$ of about $35\%$ (average $SV$: NSR $\mu$=$63.84$, AF $\mu$=$47.21$). The standard deviation of the stroke volume is much higher during atrial fibrillation (NSR $\sigma$=$2.63$, AF $\sigma$=$8.32$), due to the ejection variability introduced by the irregular heartbeat. 
Greenfield et al. [B17] discovered that an inverse relation between stroke volume, $SV(RR)$, and the heart rate of the preceding beat, $HR(RR1)$, exists, a trend which is positively verified by our results.

The same concordance found for the stroke volume is encountered in literature regarding the ejection fraction. We recall that the ejection fraction is the fraction of blood ejected into the systemic circulation by the left ventricle relative to its end-diastolic volume, $EF$=$(SV/V_{lved})/100$. Apart from one case  [B35] where no variation is found for the $EF$, all the other results [B3,B6,B13,B32,B34] agree in observing a decrease of the $EF$ during atrial fibrillation.

\noindent We are in agreement with the literature data, by noting a decrease of mean $EF$ during atrial fibrillation, from $EF$=$53.27\%$ (normal case) to $EF$=$37.12\%$ (fibrillation), with values that are much more spread out (NSR: $\sigma$=$1.46$, AF $\sigma$=$6.01$, see also the PDF reported in Fig. \ref{SV_EF}a). Gosselink et al. [B15] determine an ejection fraction of about $34\%$ during atrial fibrillation, evidencing a positive correlation between $EF(RR)$ and $RR1$ (length of the preceding beat), and between $EF(RR)$ and the end-diastolic volume of the preceding beat, $V_{lved}(RR1)$. The same correlations are obtained by Mun-tinga et al. [B23], and here furthermore confirmed (see Fig. \ref{SV_EF}b and \ref{SV_EF}c). Gosselink et al. [B15] also recognize a negative correlation between $EF(RR)$ and the pre-preceding beat, RR2, a trend which is in qualitative accordance with the present outcomes (Fig. \ref{SV_EF}d), despite their sparsity as highlighted by a low coefficient of determination, $R^2$=$0.11$.

\subsection{Stroke Work, Pressure-Volume Loop and Cardiac Output}

Stroke work, $SW$, represents the work done by the left ventricle to eject a volume of blood (i.e., stroke volume) into the aorta, and quantifies the amount of energy converted to work by the heart during each heartbeat. 
The left ventricle stroke work is computed at every heartbeat as the area within the left ventricle pressure-volume loop.

\begin{figure}
\begin{minipage}[]{0.5\columnwidth}
\includegraphics[width=\columnwidth]{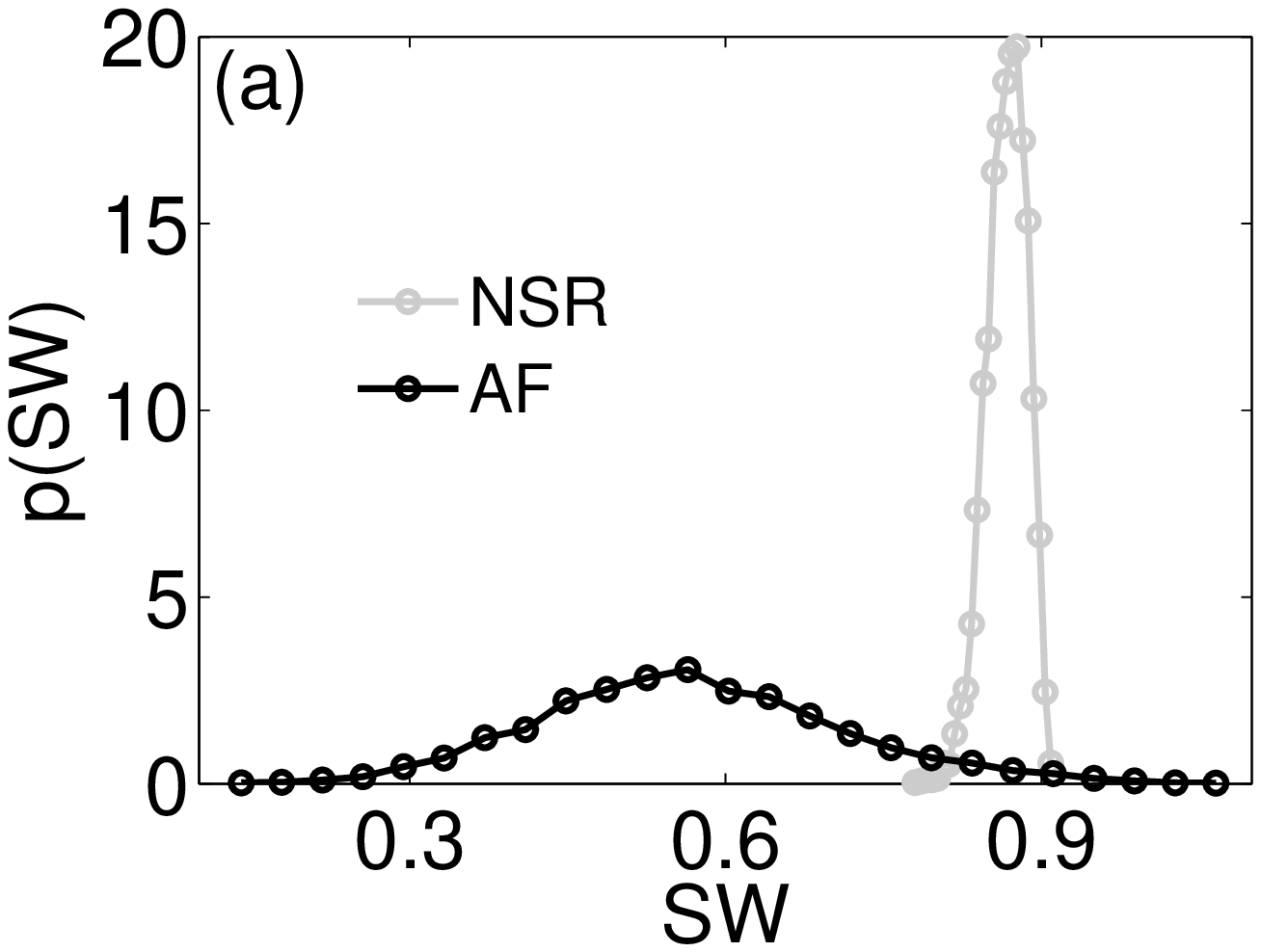}
\end{minipage}
\hspace{-0.3cm}
\begin{minipage}[]{0.5\columnwidth}
\includegraphics[width=\columnwidth]{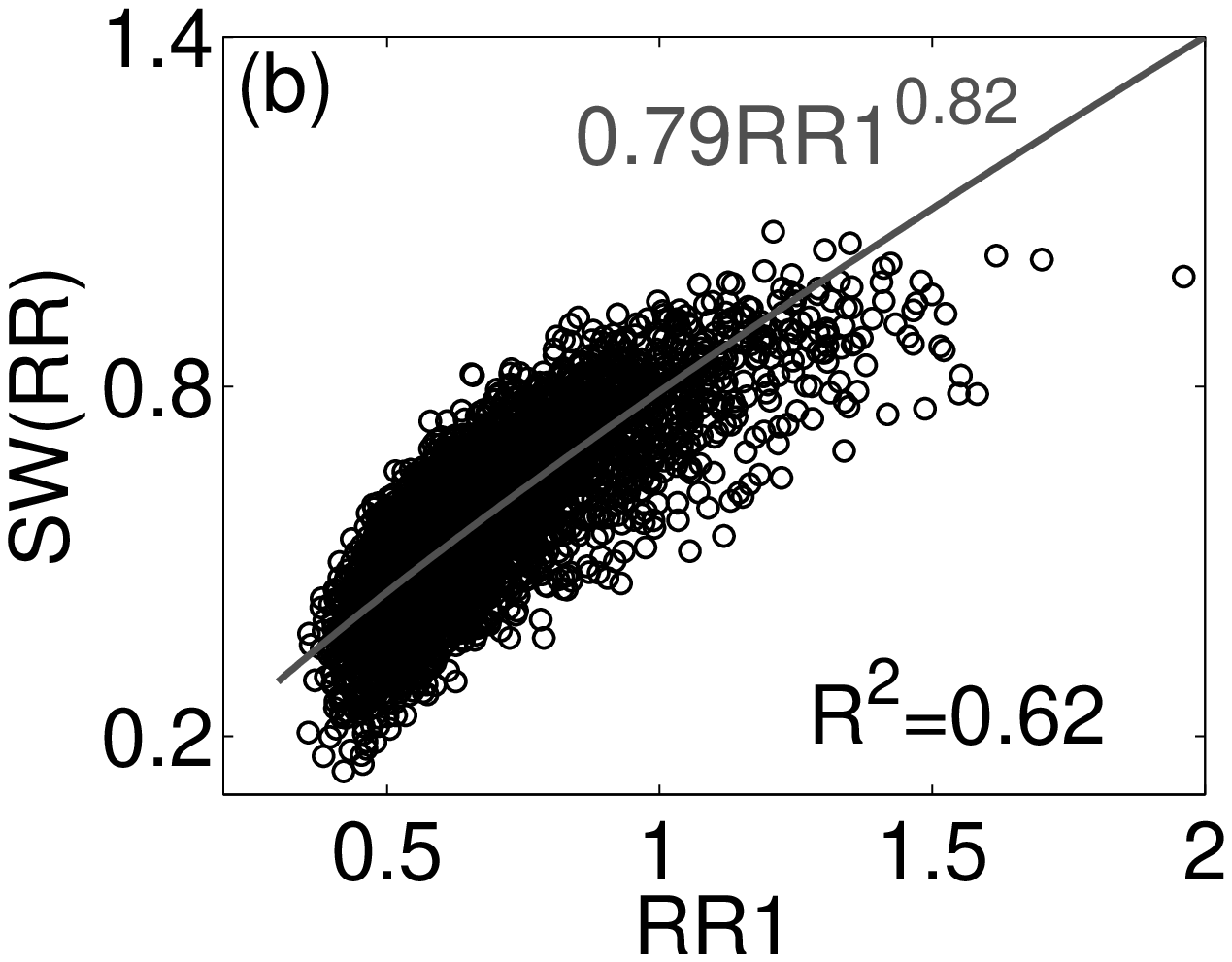}
\end{minipage}
\begin{minipage}[]{0.5\columnwidth}
\includegraphics[width=\columnwidth]{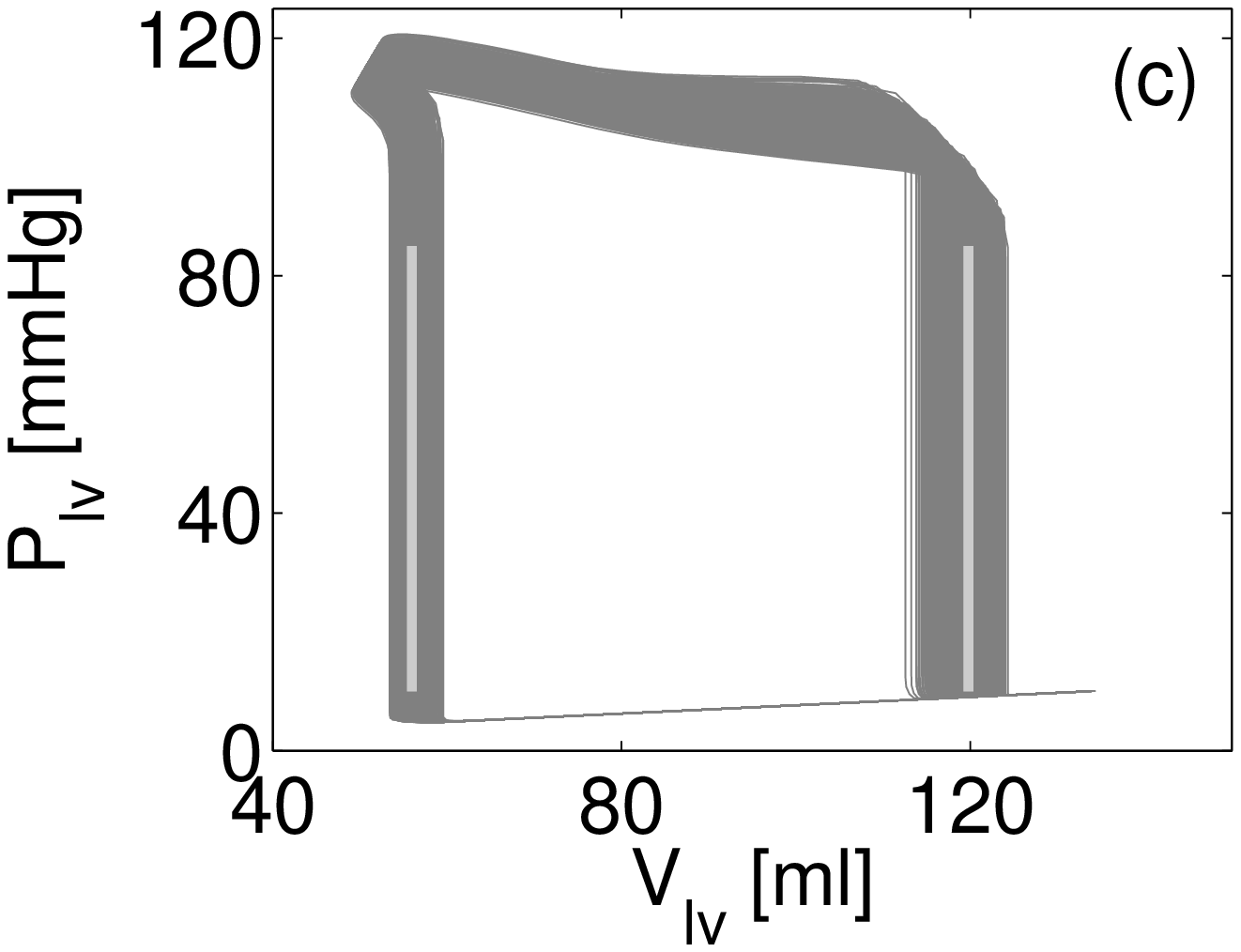}
\end{minipage}
\hspace{-0.3cm}
\begin{minipage}[]{0.5\columnwidth}
\includegraphics[width=\columnwidth]{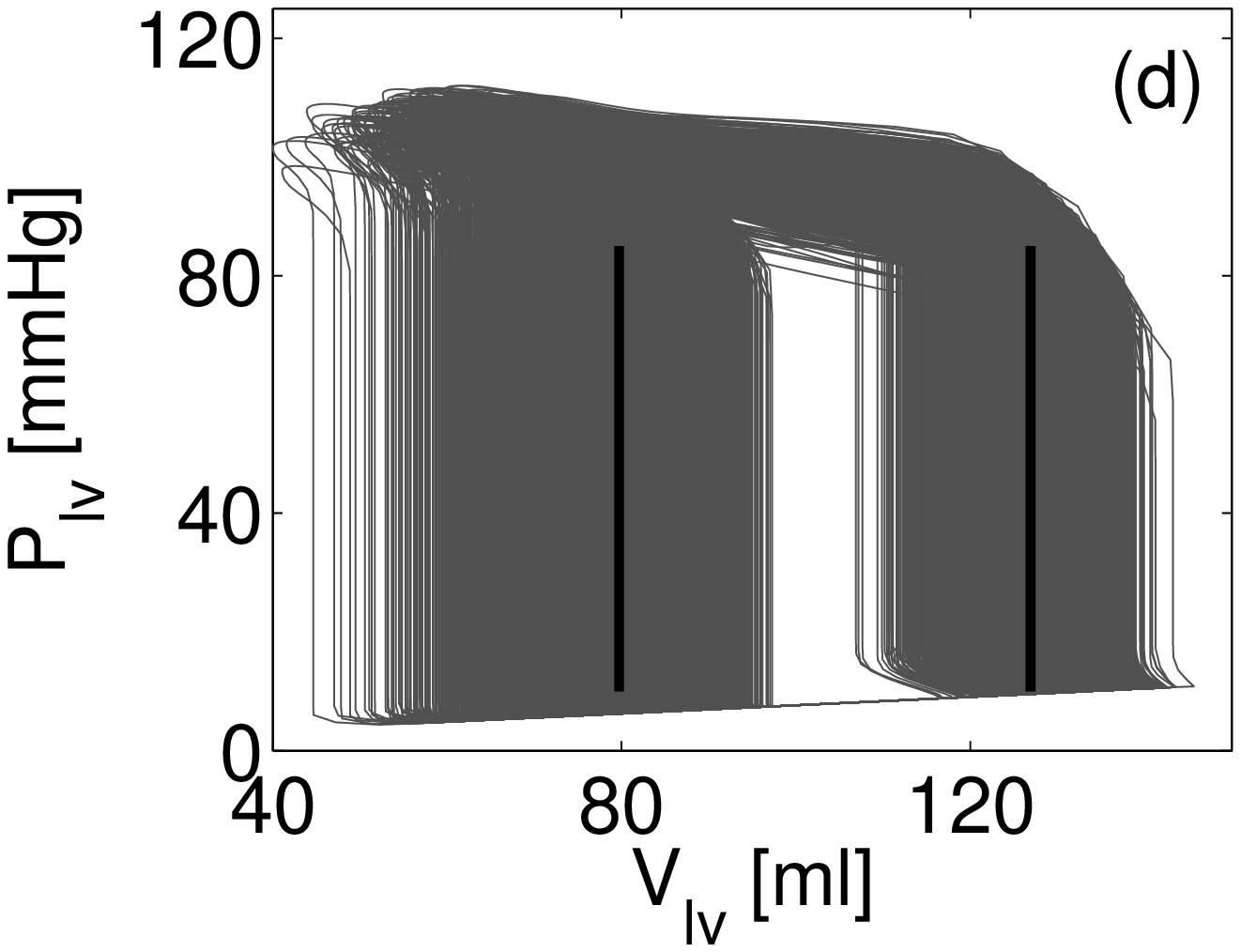}
\end{minipage}
\begin{minipage}[]{0.5\columnwidth}
\includegraphics[width=\columnwidth]{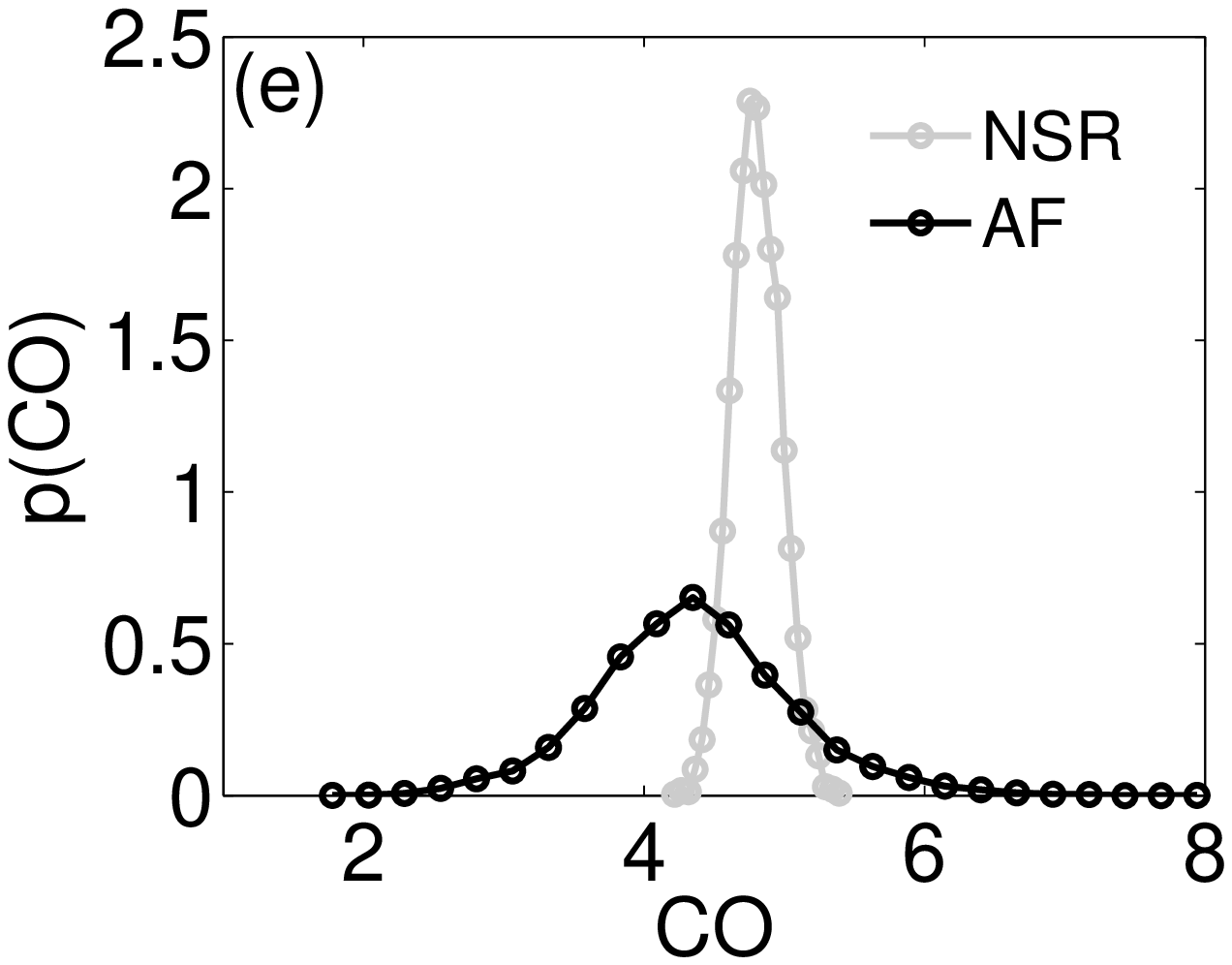}
\end{minipage}
\hspace{-0.3cm}
\begin{minipage}[]{0.5\columnwidth}
\includegraphics[width=\columnwidth]{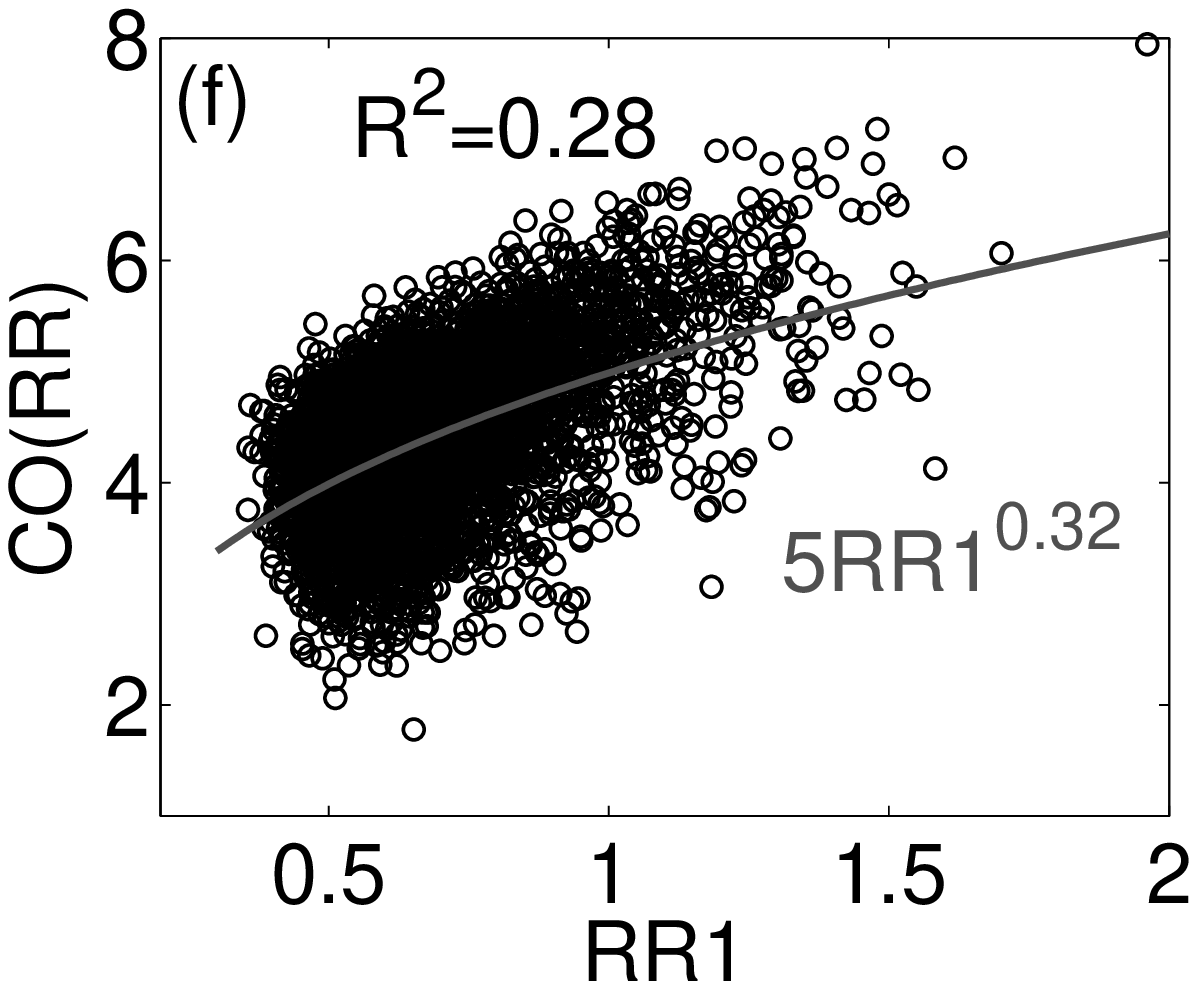}
\end{minipage}
\caption{(Top) Stroke Work: (a) PDFs; (b) AF stroke work, $SW(RR)$, as a function of the preceding heartbeat, $RR1$. (Middle) Pressure-volume loops of 5000 cycles: (c) NSR, (d) AF. Vertical lines represent the mean end-systolic and end-diastolic LV volumes. (Bottom) Cardiac Output: (e) PDFs; (f) AF cardiac output, $CO(RR)$, as a function of the preceding heartbeat, $RR1$. (b)-(f) Power-law fittings of the data and the coefficients of determination, $R^2$, are included. Light: NSR, dark: AF.}
\label{SW_CO}
\end{figure}

\noindent Available results on stroke work show a decrease during atrial fibrillation [B20,B24]. The relative decrease with respect to the normal rhythm is about $14\%$ [B24]. The stroke work here measured show a mean value of $\mu$=$0.87$ J in the normal case, and $\mu$=$0.57$ J in the fibrillated case, leading to a relative decrease of about $34\%$ during AF. The standard deviation is about one order of magnitude larger when beats are irregular (NSR $\sigma$=$0.02$, AF $\sigma$=$0.14$). Stroke work slightly varies during NSR, while wide oscillations are possible during fibrillation events (Fig. \ref{SW_CO}a). A direct proportionality between $SW(RR)$ and the preceding heartbeat, $RR1$, was found [B12], and here ascertained as well (Fig. \ref{SW_CO}b).

\noindent Left ventricle pressure-volume loops of 5000 cycles are reported for both NSR and AF in Fig. \ref{SW_CO}, panels (c) and (d), respectively. Vertical lines represent the mean end-systolic and end-diastolic LV volumes. Reverse flow evidenced by the lower right corner and upper left corner of the PV-loop is due to heart valve modeling \cite{Korakianitis-a}, accounting for the influence of the blood pressure effect, the friction effect from the tissue, and from blood motion. The PV loop markedly reduces during AF, as already observed through the $SW$ and also reported by [B5], while PV values are more spread out during the 5000 fibrillated cycles.

The cardiac output, defined as the volume of blood being pumped by the left ventricle in the time interval of one minute, $CO$=$SV \times HR$, is probably the most evaluated variable to quantify to impact of atrial fibrillation and several data are present in literature. A great part of these deals with measures before and after cardioversion (or catheter ablation), showing a consistent recovery of cardiac output when the normal sinus rhythm is established again [B6-B11,B16,B18,B20, B21,B22,B24,B25,B28,B33]. The relative increase, in absence of other important pathologies, with respect to pre-treatment configuration (that is, fibrillated conditions) varies in the range $7-13\%$ [B6,B9,B11,B21,B24, B28], $18\%$ [B7], or more than $50\%$ [B18,B33]. Few results [B2,B14,B26,B29] encounter no meaningful changes in terms of cardiac output between normal and fibrillated heartbeats.

\noindent The present outcomes align with the findings in literature: for the cardiac output there is a relative increase, when passing from fibrillation to normal rhythm, which is around $9.5\%$ (NSR $\mu$=$4.80$, AF $\mu$=$4.38$). The decrease of mean cardiac output during AF is accompanied with an increase of the standard deviation (NSR $\sigma$=$0.17$, AF $\sigma$=$0.67$), leading to a wider PDF (see Fig. \ref{SW_CO}c). We can detect a positive correlation between the cardiac output, $CO(RR)$, and the preceding heartbeat, $RR1$ (see Fig. \ref{SW_CO}d). 

\noindent The contemporary decrease of stroke work and cardiac output confirms that during atrial fibrillation the left ventricle experiences a reduced efficiency and performance.

\section{Discussion}

To better understand which hemodynamic changes are dominated by the forced decrease of the LV contractility and which changes do occur due to the heart rate variability, a fibrillated simulation with normal LV contractility ($E_{lv,max}=2.5$ mmHg/ml constant) is performed. Results in terms of mean and standard deviation values are summarized in the fourth column of Table \ref{value_table}. Trying to extract synthetic information from this new simulation, we observe that the changes mainly controlled by the heart rate variability involve the following variables: ejection fraction, stroke volume, stroke work, pulmonary arterial pressure, LA volume, LV pressure, RV pressure and volume. In fact, for these parameters the trends of the variations occurring during AF with reduced contractility with respect to NSR are confirmed by the AF simulation with normal contractility. On the contrary, changes related to RA volume and pressure, LV volume, systemic arterial pressure and cardiac output, are mainly driven by the reduced contractility. For example, indeed, the reduced LV contractility is the main responsible of the cardiac output drop. When the contractility is normal, the accelerated HR makes the cardiac output even increase. It should be noted that dividing the contributions of reduced contractility and heart rate variation is not always straightforward: while during AF the reduced LV contractility enhances a general increase of the LV volumes, the reduction of quantities related to the LV volumes, such as SV and EF, is in large part imputable to the irregular beating.

This work represents a first attempt to quantify, through a stochastic modeling, the role of acute AF on the whole cardiovascular system, thereby leading to a double advantage. First, atrial fibrillation conditions have been analyzed without the presence of other pathologies (e.g., hypertension, atrial dilatation, mitral regurgitation), which can all concur to affect the hemodynamic response. In this regard, our findings about moderate systemic hypotension and left atrial enlargement should be interpreted as pure consequences of AF alone and not induced by other pathologies. In particular, the reduced LV contractility promotes a reduction of the systemic arterial pressure, while the atrial enlargement is first due to the heart rate variability.
Second, the main cardiac variables and hemodynamic parameters can all be obtained at the same time, while clinical studies usually focus only on a few of them at a time.

The present outcomes have been compared with more than thirty clinical measures regarding AF. Although literature frequently offers data which are in contrast one with the other (e.g., systemic arterial pressure), the overall agreement is quite remarkable.

\noindent Reduced cardiac output with correlated drop of ejection fraction and decreased amount of energy converted to work by the heart during blood pumping, as well as higher left atrial volume and pressure values are some of the most representative outcomes aligned with literature and here emerging during AF with respect to NSR. Keeping in mind the different nature of measurements and clinical conditions, we here quantitatively compare (in terms of mean values) the present results with the ranges of some specific data available in literature for the most commonly measured hemodynamic parameters:
\begin{itemize}
\item $P_{pvn}$: present results (NSR: $9.60$ mmHg, AF: $10.72$ mmHg). [B2, B6, B7, B24, B11] (NSR: [$8\div19$] mmHg, AF: [$12\div21$] mmHg). 
\item $SV$: present results (NSR: $63.84$ ml, AF: $47.21$ ml). [B14, B18, B21] (NSR: [$68\div70$] ml, AF: [$49\div62$] ml). 
\item $EF$: present results (NSR: $53.27\%$, AF: $37.12\%$). [B3, B6, B13, B32, B34, B35] (NSR: [$41\%\div60\%$], AF: [$12\%\div52\%$]). 
\item $CO$ present results (NSR: $4.80$ l/min, AF: $4.38$ l/min). [B6, B7, B11, B14, B18, B22, B29] (NSR: [$4.1\div6.2$] l/min, AF: [$3.8\div5.1$] l/min). 
\end{itemize}

\noindent As can be observed, the values predicted by the present model fall within the typical ranges observed by in vivo measurements.


\section{Limitations}

Limiting aspects of the study are the absence of short-term regulation effects of the baroreceptor mechanism, which can act in order to restore the hemodynamic parameters towards normal values. The present model only represents acute episodes of AF lasting less than 24 hours, while the anatomical remodeling due to the effects of AF in the long-term is neglected. For this reason, effects and underlying structural changes due to persistent or chronic AF are not taken into account here. In the meantime, a clear distinction between acute and chronic hemodynamic effects of AF in clinical measurements is not possible. Literature data are contemporarily affected by short and long time feedbacks and compared with our outcomes, accounting solely for the early mechanical response of the system to isolated AF events.

\noindent Moreover, here we have focused on the unimodal $RR$ distribution only, while a broader inspection of multimodal distributions could give a more complete overview of the cardiovascular response to AF.

Regarding some more specific aspects of the modeling, a few results on the right heart behavior do not fully agree with the in vivo scenario, and this is probably due to the fact that the reduced contractility of the right ventricle and the ventricular interaction (here both neglected) should be properly accounted for. In the present work, the pulsatility properties of the heart are modeled by means of the time-varying elastance. This approach only phenomenologically accounts for the local Starling mechanism. Moreover, we have exclusively focused on the relation between elastance, the reduced contractility due to AF and the beating interval, RR. Other existing dependencies (e.g., on the patient disease condition) are not considered. 

\section{Conclusions}

The main goal of understanding the global response of the cardiovascular system during paroxysmal AF has been achieved by means of a lumped-parameter approach, which has been carried out paying special care to the stochastic nature of the irregular heartbeats and the reduced contractility of the heart. Although some fine details as well as the spatial cardiovascular description are here missing, the present stochastic modeling turns out to be a synthetic and powerful tool for a deeper comprehension of the arrhythmia impact on the whole cardiovascular system.

The simulations here performed allow us to isolate single cause-effect relations during AF events, a thing which is not possible in real medical monitoring. For example, a major outcome of the present work is that the drops of systemic arterial pressure and cardiac output are entirely induced by the reduced ventricular contractility during AF. On the contrary, the decrease of the ejection fraction and the LA enlargement are primarily caused by the irregular heart rate. Moreover, as all the variables are computed over a temporal range which guarantees their statistically stationarity, the present modeling provides a rich and accurate statistical description of the cardiovascular dynamics, a task which is rarely accomplished by in vivo measurements. Furthermore, the current modeling can provide new information on hemodynamic parameters (such as, for example, right ventricle dynamics), which are difficult to measure and almost never treated in literature. The proposed approach can be exploited to predict the response to AF with the combined presence of altered cardiac conditions (e.g., left atrial appendage occlusion or asportation), therefore recovering a clinical framework which often occurs in medicine. Future work will also take into account the modeling response to real beating series for both NSR and AF.

\section*{Acknowledgements}

The authors are grateful to Yubing Shi for valuable help with the model settings, and Umberto Morbiducci for fruitful discussion of the results.



\newpage

\noindent \textbf{\Large{Supplementary Material Online\\ Resource 1}}

\normalsize

\section*{Mathematical Modeling}

\begin{figure}[b]
\centering
\includegraphics[width=0.6\columnwidth]{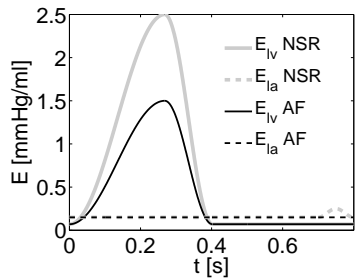}
\caption{Left heart elastances: (light) NSR, (dark) AF. Heartbeat in both cases is $RR$=$0.8$ s, for atrial fibrillation $RR1$=$RR2$. la = left atrium, lv = left ventricle.}
\label{elastance}
\end{figure}

The present lumped model, proposed by Korakianitis and Shi \cite{Korakianitis-a,Korakianitis-b}, extends the windkessel approach \cite{Westerhof,Ottesen} and consists of a network of compliances, $C$, resistances, $R$, and inductances, $L$, describing the pumping heart coupled to the systemic and pulmonary systems. All the four chambers of the pumping heart are described. The pulsatility properties are included by means of two pairs of time-varying elastance functions, one for the atria and one for the ventricles, which are then used in the constitutive equations (relating pressure, $P$, and volume, $V$). The left heart elastances in NSR are displayed in Fig. SM \ref{elastance} through the light curves. Changes to the elastance modeling due to AF (see Section 2.1 of the Main Text) are reported with dark color.

\subsection*{Equations and Numerical Scheme}

For the sake of simplicity, the equations are grouped into cardiac (with the four chambers) and circulatory (with systemic and pulmonary loops) sections.

\subsubsection*{Left atrium}

\begin{eqnarray}
\frac{d V_{la}}{dt} &=& Q_{pvn} - Q_{mi}, \nonumber \\
P_{la} &=& P_{la,un} + E_{la} (V_{la} - V_{la,un}), \\
Q_{mi} &=& \begin{cases} C Q_{mi} A R_{mi} \sqrt{P_{la} - P_{lv}}, \,\,\,\, \textmd{if} \,\,\,\, P_{la} \geq P_{lv}, \\
- C Q_{mi} A R_{mi} \sqrt{P_{lv} - P_{la}}, \,\,\,\, \textmd{if} \,\,\,\, P_{la} < P_{lv}, \nonumber
\end{cases}
\end{eqnarray}

\bigskip

\noindent where the subscript $un$ denotes the unstressed pressure and volume levels of each cardiovascular section.

\noindent The time-varying elastance is

\begin{equation}
E_{la}(t) = E_{la,min} + \frac{E_{la,max} - E_{la,min}}{2} e_a(t),
\end{equation}

\noindent and the atrium activation function is

\begin{eqnarray}
e_a(t) &=& \begin{cases} 0, \,\,\,\, \textmd{if} \,\,\,\, 0 \leq t \leq T_{ac}, \\
1 - \cos \left(\frac{t - T_{ac}}{RR - T_{ac}} 2 \pi \right), \,\,\,\, \textmd{if} \,\,\,\, T_{ac} < RR,\\
\end{cases}
\label{elastance_atrium}
\end{eqnarray}

\bigskip

\noindent where $E_{la,min}$ and $E_{la,max}$ are the minimum and maximum elastance values, respectively, while $T_{ac}$ is the beginning of atrial contraction.

The valve opening is decided by the angular position of the leaflets:

\begin{equation}
A R_{mi} = \frac{(1 - \cos \theta_{mi})^2}{(1 - \cos \theta_{max})^2}
\end{equation}

\noindent and the valve motion is governed by

\begin{eqnarray}
\frac{d^2\theta_{mi}}{dt^2} &=& \begin{cases} (P_{la} - P_{lv}) K_{p,mi} \cos \theta_{mi} - K_{f,mi} \frac{d \theta_{mi}}{dt} \\
\,\,\, + K_{b,mi} Q_{mi} \cos \theta_{mi} \\ \,\,\, - K_{v,mi} Q_{mi} \sin 2 \theta_{mi}, \,\,\,\,\,\,\, \textmd{if} \,\,\,\, Q_{mi} \geq 0, \\
(P_{la} - P_{lv}) K_{p,mi} \cos \theta_{mi} - K_{f,mi} \frac{d \theta_{mi}}{dt} \\
\,\,\, + K_{b,mi} Q_{mi} \cos \theta_{mi}, \,\,\,\,\,\,\, \textmd{if} \,\,\,\, Q_{mi} < 0,
\end{cases}
\end{eqnarray}

\bigskip

\subsubsection*{Left ventricle}

\begin{eqnarray}
\frac{d V_{lv}}{dt} &=& Q_{mi} - Q_{ao}, \nonumber \\
P_{lv} &=& P_{lv,un} + E_{lv} (V_{lv} - V_{lv,un}), \\
Q_{ao} &=& \begin{cases} C Q_{ao} A R_{ao} \sqrt{P_{lv} - P_{sas}}, \,\,\,\, \textmd{if} \,\,\,\, P_{lv} \geq P_{sas}, \\
- C Q_{ao} A R_{ao} \sqrt{P_{sas} - P_{lv}}, \,\,\,\, \textmd{if} \,\,\,\, P_{sas} < P_{lv}, \nonumber
\end{cases}
\end{eqnarray}

\bigskip

\noindent The time-varying elastance is

\begin{equation}
E_{lv}(t) = E_{lv,min} + \frac{E_{lv,max} - E_{lv,min}}{2} e_v(t),
\end{equation}

\noindent and the ventricle activation function is

\begin{eqnarray}
e_v(t) &=& \begin{cases} 1 - \cos \left( \frac{t}{T_{me}} \pi \right), \,\,\,\, \textmd{if} \,\,\,\, 0 \leq t < T_{me}, \\
1 + \cos \left(\frac{t - T_{me}}{T_{ce} - T_{me}} \pi \right), \,\, \textmd{if} \,\,\, T_{me} \leq t < T_{ce} \\
0, \,\,\,\, \textmd{if} \,\,\,\, T_{ce} \leq t < RR,
\end{cases}
\label{elastance_ventricle}
\end{eqnarray}

\bigskip

\noindent where $E_{lv,min}$ and $E_{lv,max}$ are the minimum and maximum elastance values, respectively. $T_{me}$ and $T_{ce}$ are the instants where the elastance reaches its maximum and constant values, respectively.

The valve opening is decided by the angular position of the leaflets:

\begin{equation}
A R_{ao} = \frac{(1 - \cos \theta_{ao})^2}{(1 - \cos \theta_{max})^2}
\end{equation}

\noindent and the valve motion is governed by

\begin{eqnarray}
\frac{d^2\theta_{ao}}{dt^2} &=& \begin{cases} (P_{lv} - P_{sas}) K_{p,ao} \cos \theta_{ao} - K_{f,ao} \frac{d \theta_{ao}}{dt} \\
\,\,\, + K_{b,ao} Q_{ao} \cos \theta_{ao}\\ \,\,\, - K_{v,ao} Q_{ao} \sin 2 \theta_{ao}, \,\,\,\,\,\,\, \textmd{if} \,\,\,\, Q_{ao} \geq 0, \\
(P_{lv} - P_{sas}) K_{p,ao} \cos \theta_{ao} - K_{f,ao} \frac{d \theta_{ao}}{dt} \\
\,\,\, + K_{b,ao} Q_{ao} \cos \theta_{ao}, \,\,\,\,\,\,\, \textmd{if} \,\,\,\, Q_{ao} < 0,
\end{cases}
\end{eqnarray}

\bigskip

\subsubsection*{Right atrium}

\begin{eqnarray}
\frac{d V_{ra}}{dt} &=& Q_{svn} - Q_{ti}, \nonumber \\
P_{ra} &=& P_{ra,un} + E_{ra} (V_{ra} - V_{ra,un}), \\
Q_{ti} &=& \begin{cases} C Q_{ti} A R_{ti} \sqrt{P_{ta} - P_{tv}}, \,\,\,\, \textmd{if} \,\,\,\, P_{ta} \geq P_{tv}, \\
- C Q_{ti} A R_{ti} \sqrt{P_{rv} - P_{ra}}, \,\,\,\, \textmd{if} \,\,\,\, P_{ra} < P_{rv}, \nonumber
\end{cases}
\end{eqnarray}

\bigskip

\noindent The time-varying elastance is

\begin{equation}
E_{ra}(t) = E_{ra,min} + \frac{E_{ra,max} - E_{ra,min}}{2} e_a(t),
\end{equation}

\noindent where the activation function is given by Eq. (\ref{elastance_atrium}), while $E_{ra,min}$ and $E_{ra,max}$ are the minimum and maximum elastance values, respectively.

The valve opening is decided by the angular position of the leaflets:

\begin{equation}
A R_{ti} = \frac{(1 - \cos \theta_{ti})^2}{(1 - \cos \theta_{max})^2}
\end{equation}

\noindent and the valve motion is governed by

\begin{eqnarray}
\frac{d^2\theta_{ti}}{dt^2} &=& \begin{cases} (P_{ra} - P_{rv}) K_{p,ti} \cos \theta_{ti} - K_{f,ti} \frac{d \theta_{ti}}{dt} \\
\,\,\, + K_{b,ti} Q_{ti} \cos \theta_{ti}\\ \,\,\, - K_{v,ti} Q_{ti} \sin 2 \theta_{ti}, \,\,\,\,\,\,\, \textmd{if} \,\,\,\, Q_{ti} \geq 0, \\
(P_{ra} - P_{rv}) K_{p,ti} \cos \theta_{ti} - K_{f,ti} \frac{d \theta_{ti}}{dt} \\
\,\,\, + K_{b,ti} Q_{ti} \cos \theta_{ti}, \,\,\,\,\,\,\, \textmd{if} \,\,\,\, Q_{ti} < 0,
\end{cases}
\end{eqnarray}

\bigskip

\subsubsection*{Right ventricle}

\begin{eqnarray}
\frac{d V_{rv}}{dt} &=& Q_{ti} - Q_{po}, \nonumber \\
P_{rv} &=& P_{rv,un} + E_{rv} (V_{rv} - V_{rv,un}), \\
Q_{po} &=& \begin{cases} C Q_{po} A R_{po} \sqrt{P_{rv} - P_{pas}}, \,\,\,\, \textmd{if} \,\,\,\, P_{rv} \geq P_{pas}, \\
- C Q_{po} A R_{po} \sqrt{P_{pas} - P_{rv}}, \,\,\,\, \textmd{if} \,\,\,\, P_{pas} < P_{rv}, \nonumber
\end{cases}
\end{eqnarray}

\bigskip

\noindent The time-varying elastance is

\begin{equation}
E_{rv}(t) = E_{rv,min} + \frac{E_{rv,max} - E_{rv,min}}{2} e_v(t),
\end{equation}

\noindent where the ventricle activation function is given by Eq. (\ref{elastance_ventricle}), while $E_{rv,min}$ and $E_{rv,max}$ are the minimum and maximum elastance values, respectively.

The valve opening is decided by the angular position of the leaflets:

\begin{equation}
A R_{po} = \frac{(1 - \cos \theta_{po})^2}{(1 - \cos \theta_{max})^2}
\end{equation}

\noindent and the valve motion is governed by

\begin{eqnarray}
\frac{d^2\theta_{po}}{dt^2} &=& \begin{cases} (P_{rv} - P_{pas}) K_{p,po} \cos \theta_{po} - K_{f,po} \frac{d \theta_{po}}{dt} \\
\,\,\, + K_{b,po} Q_{po} \cos \theta_{po}\\ \,\,\, - K_{v,po} Q_{po} \sin 2 \theta_{po}, \,\,\,\,\,\,\, \textmd{if} \,\,\,\, Q_{po} \geq 0, \\
(P_{rv} - P_{pas}) K_{p,po} \cos \theta_{po} - K_{f,po} \frac{d \theta_{po}}{dt} \\
\,\,\, + K_{b,po} Q_{po} \cos \theta_{po}, \,\,\,\,\,\,\, \textmd{if} \,\,\,\, Q_{po} < 0,
\end{cases}
\end{eqnarray}

\bigskip

\subsubsection*{Systemic Circuit}

\begin{eqnarray}
&& \begin{cases} \frac{d P_{sas}}{dt} = \frac{Q_{ao} - Q_{sas}}{C_{sas}}, \\
\frac{d Q_{sas}}{dt} = \frac{P_{sas} - P_{sat} - R_{sas} Q_{sas}}{L_{sas}}, \\
P_{sas} - P_{sas,un} = \frac{1}{C_{sas}} (V_{sas} - V_{sas,un}), \end{cases}
\end{eqnarray}

\bigskip

\begin{eqnarray}
&& \begin{cases} \frac{d P_{sat}}{dt} = \frac{Q_{sas} - Q_{sat}}{C_{sat}}, \\
\frac{d Q_{sat}}{dt} = \frac{P_{sat} - P_{svn} - (R_{sat} + R_{sar} + R_{scp}) Q_{sat}}{L_{sat}}, \\
P_{sat} - P_{sat,un} = \frac{1}{C_{sat}} (V_{sat} - V_{sat,un}), \end{cases}
\end{eqnarray}

\bigskip

\begin{eqnarray}
&& \begin{cases} \frac{d P_{svn}}{dt} = \frac{Q_{sat} - Q_{svn}}{C_{svn}}, \\
Q_{svn} = \frac{P_{svn} - P_{ra}}{R_{svn}}, \\
P_{svn} - P_{svn,un} = \frac{1}{C_{svn}} (V_{svn} - V_{svn,un}), \end{cases}
\end{eqnarray}

\bigskip

\subsubsection*{Pulmonary Circuit}

\begin{eqnarray}
&& \begin{cases} \frac{d P_{pas}}{dt} = \frac{Q_{po} - Q_{pas}}{C_{pas}}, \\
\frac{d Q_{pas}}{dt} = \frac{P_{pas} - P_{pat} - R_{pas} Q_{pas}}{L_{pas}}, \\
P_{pas} - P_{pas,un} = \frac{1}{C_{pas}} (V_{pas} - V_{pas,un})\\ \end{cases}
\end{eqnarray}

\bigskip

\begin{eqnarray}
&& \begin{cases} \frac{d P_{pat}}{dt} = \frac{Q_{pas} - Q_{pat}}{C_{pat}}, \\
\frac{d Q_{pat}}{dt} = \frac{P_{pat} - P_{pvn} - (R_{pat} + R_{par} + R_{pcp}) Q_{pat}}{L_{pat}}, \\
P_{pat} - P_{pat,un} = \frac{1}{C_{pat}} (V_{pat} - V_{pat,un}), \end{cases}
\end{eqnarray}

\bigskip

\begin{eqnarray}
&& \begin{cases} \frac{d P_{pvn}}{dt} = \frac{Q_{pat} - Q_{pvn}}{C_{pvn}}, \\
Q_{pvn} = \frac{P_{pvn} - P_{la}}{R_{pvn}}, \\
P_{pvn} - P_{pvn,un} = \frac{1}{C_{pvn}} (V_{pvn} - V_{pvn,un}). \end{cases}
\end{eqnarray}

\bigskip

The differential system is solved by means of a multistep adaptative solver for stiff problems, implemented by the \texttt{ode15s} Matlab function \cite{matlab}. The relative error tolerance, \texttt{RelTol}, is set equal to $10^{-10}$. The absolute error tolerance, \texttt{AbsTol}, is in general imposed equal to $10^{-6}$, except for equations involving aortic and pulmonary flows, $Q_{ao}$ and $Q_{po}$, where a more stringent tolerance ($10^{-10}$) is required to avoid numerical oscillations.

\noindent The variable order solver adopted is based on the numerical differentiation formulas (NDFs) and is chosen because is one of the most efficient and suitable routines for stiff problems. In fact, the cardiovascular differential system shows some stiffness features, that is the equations include some terms that can lead to rapid variation in the solutions. This aspect is particularly relevant for the valve dynamics, where sudden variations of the leaflet angular position occur when the valve opens and closes.

\subsection*{Initial Conditions and Cardiovascular Parameters}

Initial conditions, given in terms of pressures, volumes, flow rates and valve opening angles, are the same for both the normal sinus rhythm and the atrial fibrillation simulations. The total volume is taken as the average value for a healthy adult, $V_{tot}$=$5250$ ml. Flow rates and valve angles are initially considered as all the valves are closed and no flow is present. Initial pressures in the circulatory sections are given as typical values reached during a normal cardiac cycle. Volumes in the four chambers are obtained at $t$=$0$ subtracting from the total volume, $V_{tot}$, all the volume contributes at the generic vascular section $i$, using the constitutive relation $P_{i,t=0} - P_{i,un} = \frac{1}{C_i} (V_{i,t=0} - V_{i,un})$, where $P_{i,t=0}$ is imposed as previously said. Table SM \ref{ic_table} summarizes the adopted initial values. Cardiovascular parameters of the model are as well recalled in Tables from SM \ref{heart_table} to SM \ref{temporal_table}.

\begin{table}
  \centering
\begin{tabular}{|c|c|}
  \hline
  Variable & Value ($t=0$) \\
  \hline
  $V_{la,0}$ & 60 ml\\
  \hline
  $V_{lv,0}$ & 130 ml\\
  \hline
  $V_{ra,0}$ & 39 ml\\
  \hline
  $V_{rv,0}$ & 110 ml\\
  \hline
  $P_{sas,0}$ & 100 mmHg\\
  \hline
  $Q_{sas,0}$ & 0 ml/s\\
  \hline
  $P_{sat,0}$ & 100 mmHg\\
  \hline
  $Q_{sat,0}$ & 0 ml/s\\
  \hline
  $P_{svn,0}$ & 10 mmHg\\
  \hline
  $P_{pas,0}$ & 20 mmHg\\
  \hline
  $Q_{pas,0}$ & 0 ml/s\\
  \hline
  $P_{pat,0}$ & 20 mmHg\\
  \hline
  $Q_{pat,0}$ & 0 ml/s\\
  \hline
  $P_{pvn,0}$ & 10 mmHg\\
  \hline
  $\theta_{mi,0} = d \theta_{mi,0}/dt$ & 0 rad\\
  \hline
  $\theta_{ao,0} = d \theta_{ao,0}/dt$ & 0 rad\\
  \hline
  $\theta_{ti,0} = d \theta_{ti,0}/dt$ & 0 rad\\
  \hline
  $\theta_{po,0} = d \theta_{po,0}/dt$ & 0 rad\\
  \hline
  \end{tabular}
\caption{Initial conditions.} \label{ic_table}
\end{table}

\begin{table}
  \centering
\begin{tabular}{|c|c|}
  \hline
  Parameter & Value \\
  \hline
  $CQ_{ao}$ & 350 ml/(s mmHg$^{0.5}$)\\
  \hline
  $CQ_{mi}$ & 400 ml/(s mmHg$^{0.5}$)\\
  \hline
  $E_{lv,max}$ & 2.5 mmHg/ml\\
  \hline
  $E_{lv,min}$ & 0.07 mmHg/ml\\
  \hline
  $P_{lv,un}$ & 1 mmHg\\
  \hline
  $V_{lv,un}$ & 5 ml\\
  \hline
  $E_{la,max}$ & 0.25 mmHg/ml\\
  \hline
  $E_{la,min}$ & 0.15 mmHg/ml\\
  \hline
  $P_{la,un}$ & 1 mmHg\\
  \hline
  $V_{la,un}$ & 4 ml\\
  \hline
  $CQ_{po}$ & 350 ml/(s mmHg$^{0.5}$)\\
  \hline
  $CQ_{ti}$ & 400 ml/(s mmHg$^{0.5}$)\\
  \hline
  $E_{rv,max}$ & 1.15 mmHg/ml\\
  \hline
  $E_{rv,min}$ & 0.07 mmHg/ml\\
  \hline
  $P_{rv,un}$ & 1 mmHg\\
  \hline
  $V_{rv,un}$ & 10 ml\\
  \hline
  $E_{ra,max}$ & 0.25 mmHg/ml\\
  \hline
  $E_{ra,min}$ & 0.15 mmHg/ml\\
  \hline
  $P_{ra,un}$ & 1 mmHg\\
  \hline
  $V_{ra,un}$ & 4 ml\\
  \hline
\end{tabular}
\caption{Heart parameters.} \label{heart_table}
\end{table}

\begin{table}
  \centering
\begin{tabular}{|c|c|}
  \hline
  Parameter & Value \\
  \hline
  $C_{sas}$ & 0.08 ml/mmHg\\
  \hline
  $R_{sas}$ & 0.003 mmHg s/ml\\
  \hline
  $L_{sas}$ & 0.000062 mmHg s$^2$/ml\\
  \hline
  $P_{sas,un}$ & 1 mmHg \\
  \hline
  $V_{sas,un}$ & 25 ml \\
  \hline
  $C_{sat}$ & 1.6 ml/mmHg\\
  \hline
  $R_{sat}$ & 0.05 mmHg s/ml\\
  \hline
  $L_{sat}$ & 0.0017 mmHg s$^2$/ml\\
  \hline
  $P_{sat,un}$ & 1 mmHg \\
  \hline
  $V_{sat,un}$ & 775 ml \\
  \hline
  $R_{sar}$ & 0.5 mmHg s/ml\\
  \hline
  $R_{scp}$ & 0.52 mmHg s/ml\\
  \hline
  $R_{svn}$ & 0.075 mmHg s/ml\\
  \hline
  $C_{svn}$ & 20.5 ml/mmHg\\
  \hline
  $P_{svn,un}$ & 1 mmHg \\
  \hline
  $V_{svn,un}$ & 3000 ml \\
  \hline
\end{tabular}
\caption{Systemic circulation parameters.} \label{systemic_circulation_table}
\end{table}

\begin{table}
  \centering
\begin{tabular}{|c|c|}
  \hline
  Parameter & Value \\
  \hline
  $C_{pas}$ & 0.18 ml/mmHg\\
  \hline
  $R_{pas}$ & 0.002 mmHg s/ml\\
  \hline
  $L_{pas}$ & 0.000052 mmHg s$^2$/ml\\
  \hline
  $P_{pas,un}$ & 1 mmHg \\
  \hline
  $V_{pas,un}$ & 25 ml \\
  \hline
  $C_{pat}$ & 3.8 ml/mmHg\\
  \hline
  $R_{pat}$ & 0.01 mmHg s/ml\\
  \hline
  $L_{pat}$ & 0.0017 mmHg s$^2$/ml\\
  \hline
  $P_{pat,un}$ & 1 mmHg \\
  \hline
  $V_{pat,un}$ & 175 ml \\
  \hline
  $R_{par}$ & 0.05 mmHg s/ml\\
  \hline
  $R_{pcp}$ & 0.07 mmHg s/ml\\
  \hline
  $R_{pvn}$ & 0.006 mmHg s/ml\\
  \hline
  $C_{pvn}$ & 20.5 ml/mmHg\\
  \hline
  $P_{pvn,un}$ & 1 mmHg \\
  \hline
  $V_{pvn,un}$ & 300 ml \\
  \hline
\end{tabular}
\caption{Pulmonary circulation parameters.} \label{pulmonary_circulation_table}
\end{table}

\begin{table}
  \centering
\begin{tabular}{|c|c|}
  \hline
  Parameter & Value \\
  \hline
  $K_{p,mi}$, $K_{p,ao}$, $K_{p,ti}$, $K_{p,po}$ & 5500 ml/mmHg\\
  \hline
  $K_{f,mi}$, $K_{f,ao}$, $K_{f,ti}$, $K_{f,po}$ & 50 s$^{-1}$\\
  \hline
  $K_{b,mi}$, $K_{b,ao}$, $K_{b,ti}$, $K_{b,po}$ & 2 rad/(s ml)\\
  \hline
  $K_{v,mi}$, $K_{v,ti}$ & 3.5 rad/(s ml)\\
  \hline
  $K_{v,ao}$, $K_{v,po}$ & 7 rad/(s ml)\\
  \hline
  $\theta_{max}$ & 5/12 $\pi$ rad\\
  \hline
\end{tabular}
\caption{Valve dynamics parameters.} \label{valve_table}
\end{table}

\begin{table}
  \centering
\begin{tabular}{|c|c|}
  \hline
  Parameter & Value \\
  \hline
  $T_{ac}$ & 0.875 $RR$ s\\
  \hline
  $T_{me}$ & 0.3 $\sqrt{RR}$ s\\
  \hline
  $T_{ce}$ & 3/2 $T_{me}$ s\\
  \hline
\end{tabular}
\caption{Temporal parameters. Activation times are those typically introduced in the time-varying elastance models  \cite{Korakianitis-a,Korakianitis-b,Ruiz,Ottesen}, and are here settled considering that, at $RR$=$0.8$ s, the ventricular systole lasts about $0.3$ s, while the atrial systole length is about $0.1$ s \cite{Guyton}.}
\label{temporal_table}
\end{table}

\section*{Sensitivity analysis of the AF simulations}

To test the sensitivity of our results, we have also analyzed the cardiovascular response with two other $RR$ series (PDFs are depicted in Fig. SM \ref{RR_distributions_series} and compared with the reference fibrillation distribution in black):

\begin{itemize}
\item EMG distribution: $\mu$=$0.5$ s, $\sigma$=$0.17$ s, $cv$=$0.35$ (red);
\item EMG distribution: $\mu$=$0.67$ s, $\sigma$=$0.22$ s, $cv$=$0.34$ (green).
\end{itemize}

\noindent Although the coefficients of variation, $cv$, of both these distributions are higher than what usually observed during atrial fibrillation ($cv$=$0.24$), each distribution emphasizes, one at a time, the two main fibrillated beating features already present in the reference fibrillated distribution: a reduced mean heartbeat (EMG red distribution with $\mu$=$0.5$ s), and a higher $RR$ variability (EMG green distribution with $\sigma$=$0.22$ s). The reference fibrillated outcomes described in the Main Text (EMG black distribution: $\mu$=$0.67$ s, $\sigma$=$0.17$ s, $cv$=$0.26$) are added in black for comparison.

\begin{figure}
\centering
\includegraphics[width=0.6\columnwidth]{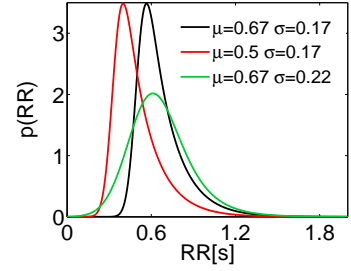}
\caption{AF: $RR$ distributions used for the sensitivity analysis of the results.}
\label{RR_distributions_series}
\end{figure}

\begin{figure}
\begin{minipage}[]{0.5\columnwidth}
\includegraphics[width=\columnwidth]{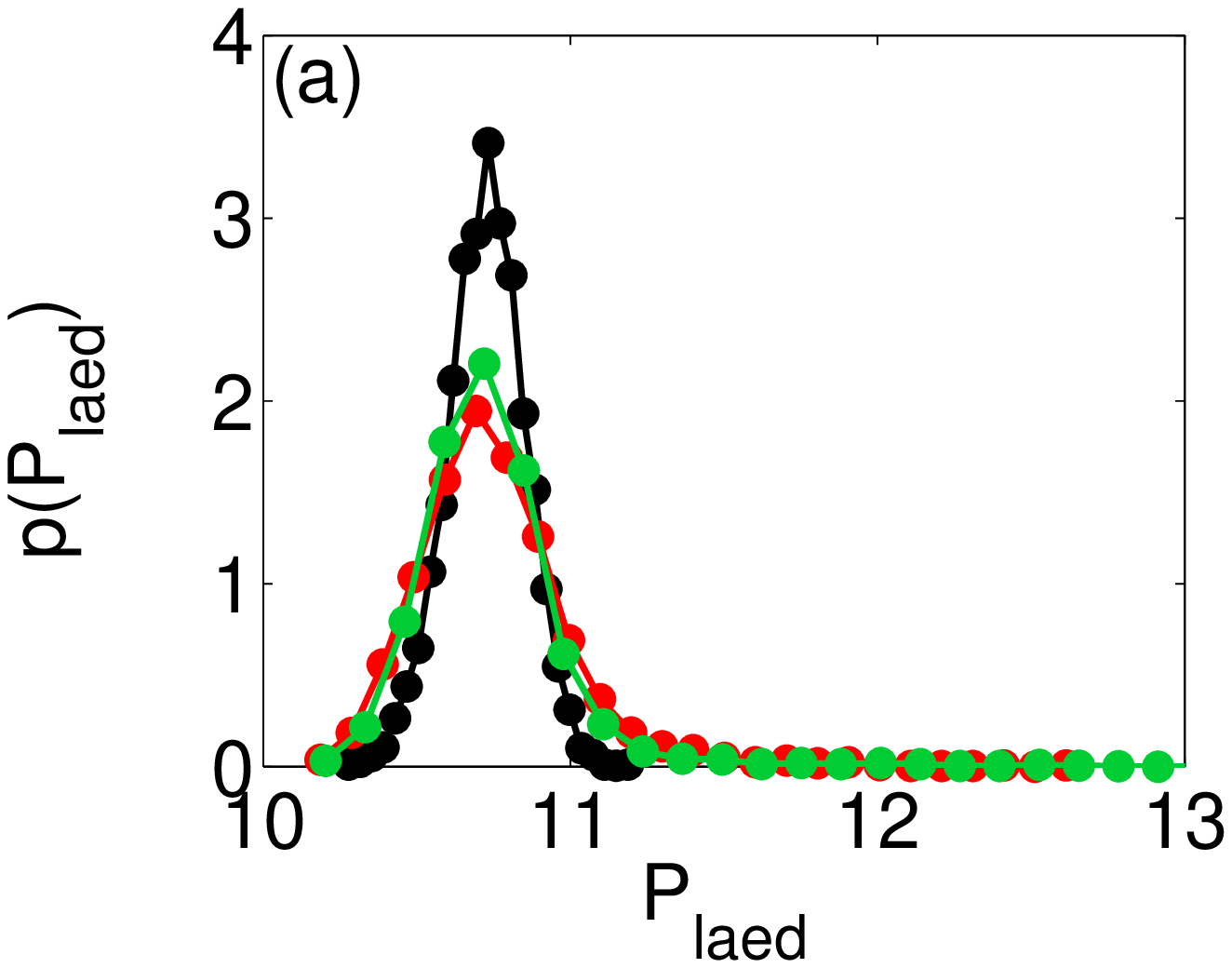}
\end{minipage}
\hspace{-0.3cm}
\begin{minipage}[]{0.5\columnwidth}
\includegraphics[width=\columnwidth]{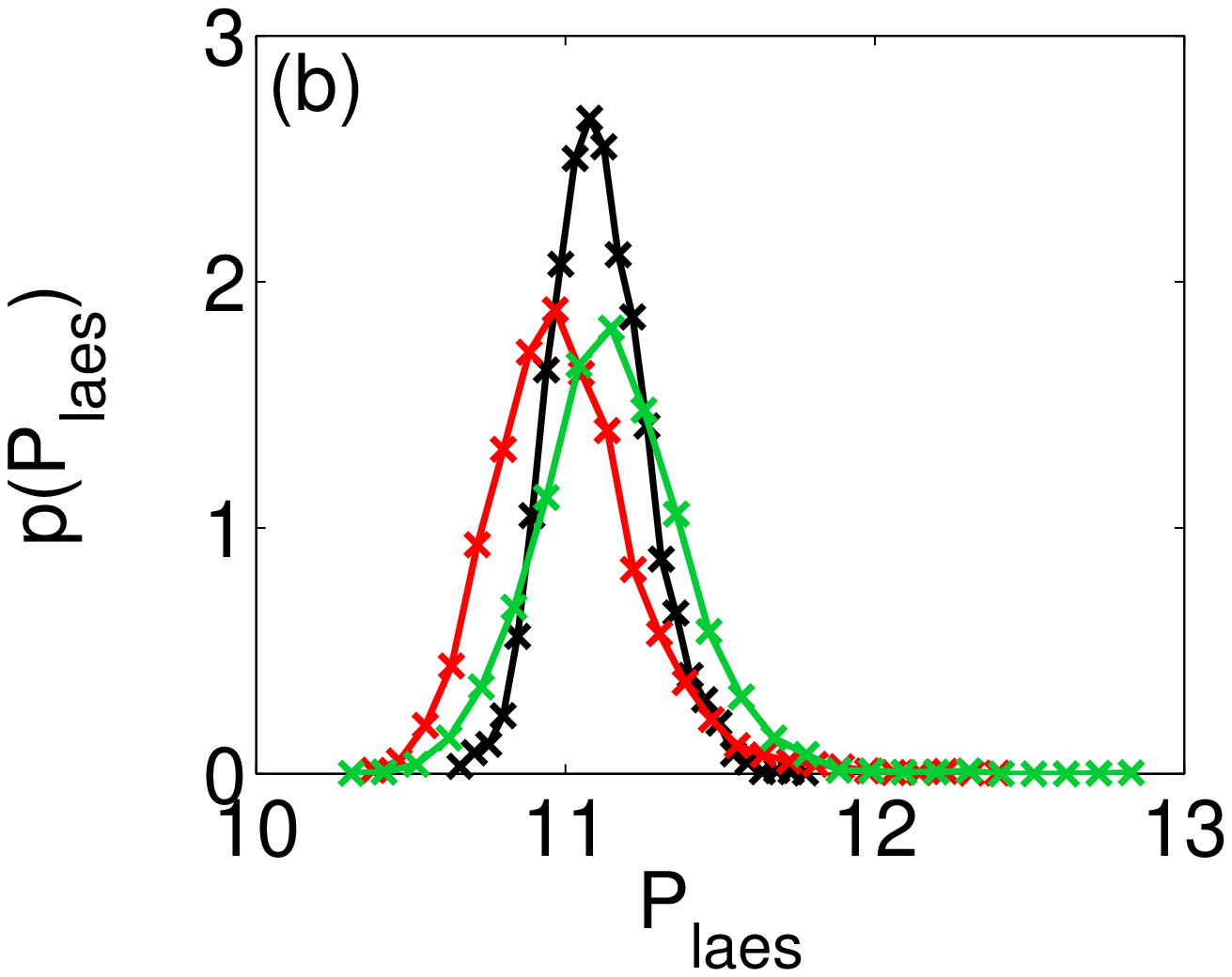}
\end{minipage}
\begin{minipage}[]{0.5\columnwidth}
\includegraphics[width=\columnwidth]{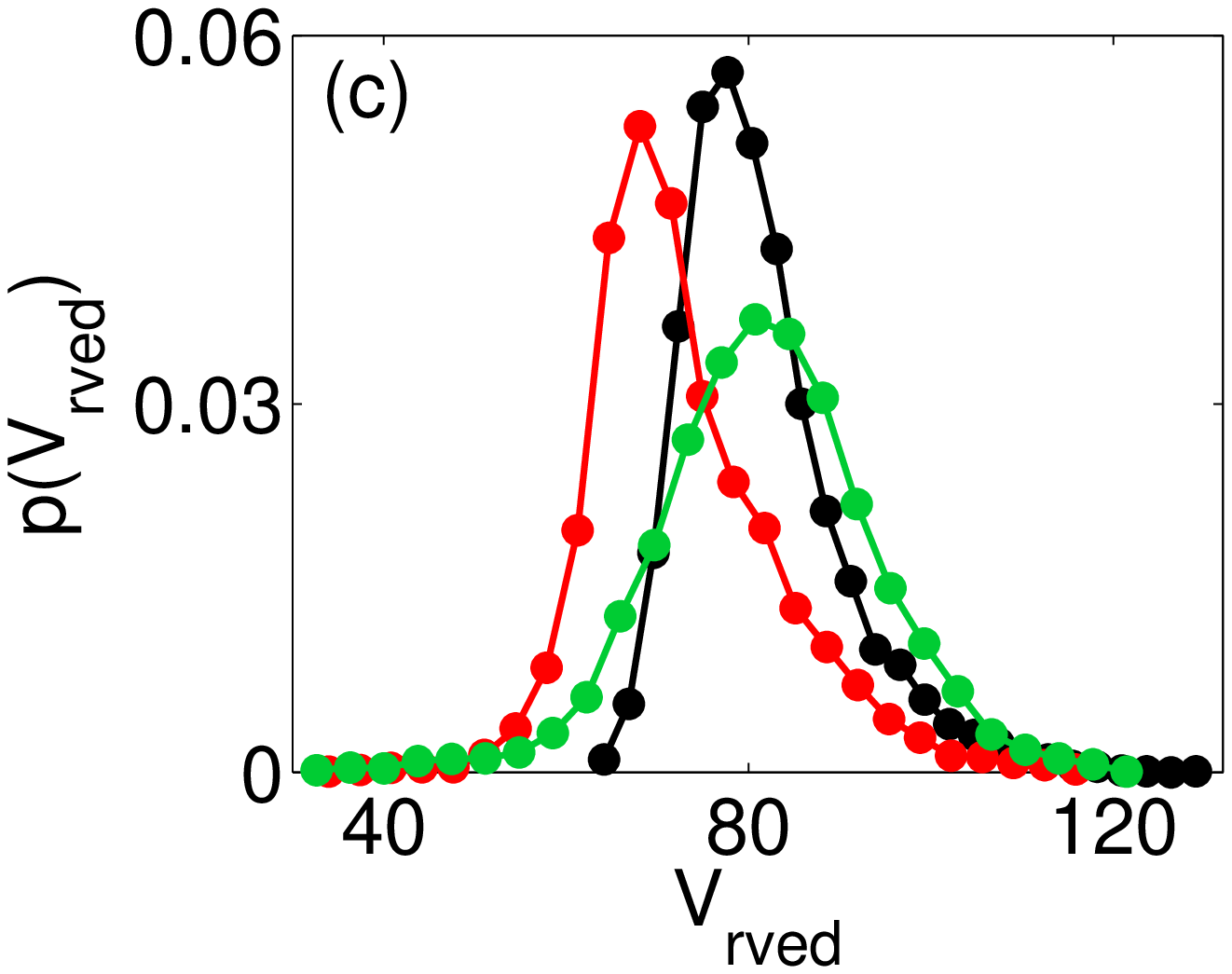}
\end{minipage}
\hspace{-0.3cm}
\begin{minipage}[]{0.5\columnwidth}
\includegraphics[width=\columnwidth]{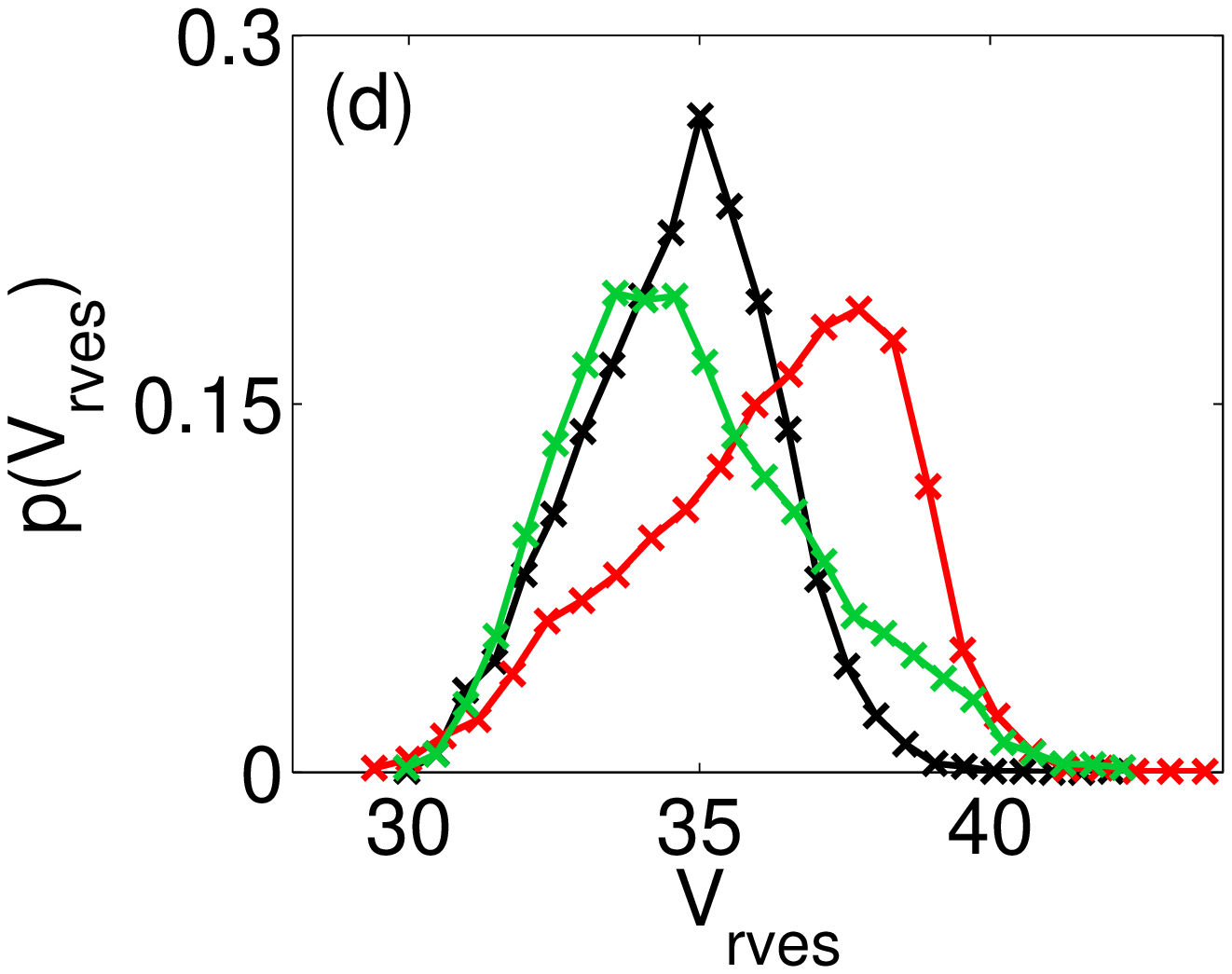}
\end{minipage}
\begin{minipage}[]{0.5\columnwidth}
\includegraphics[width=\columnwidth]{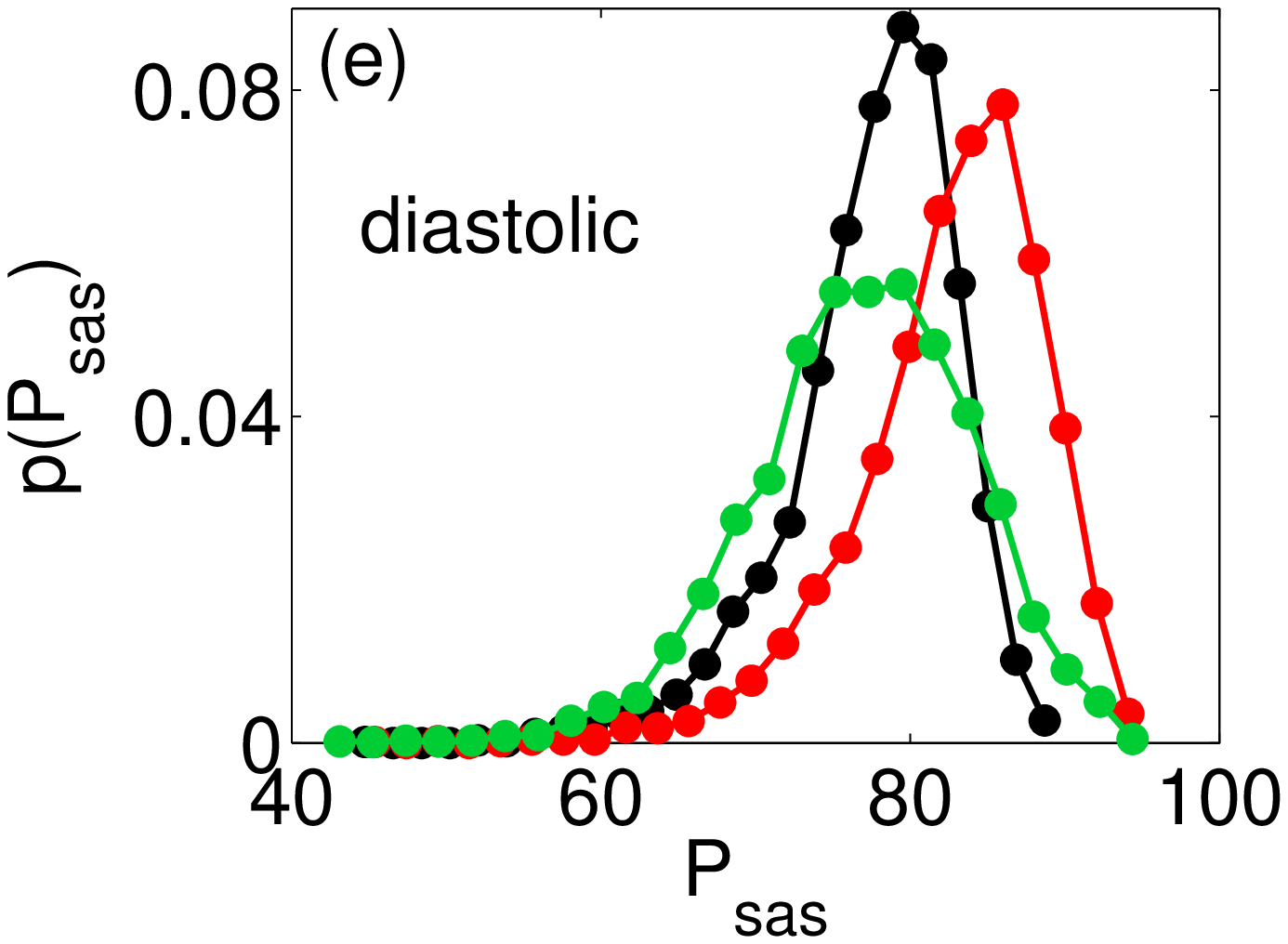}
\end{minipage}
\hspace{-0.3cm}
\begin{minipage}[]{0.5\columnwidth}
\includegraphics[width=\columnwidth]{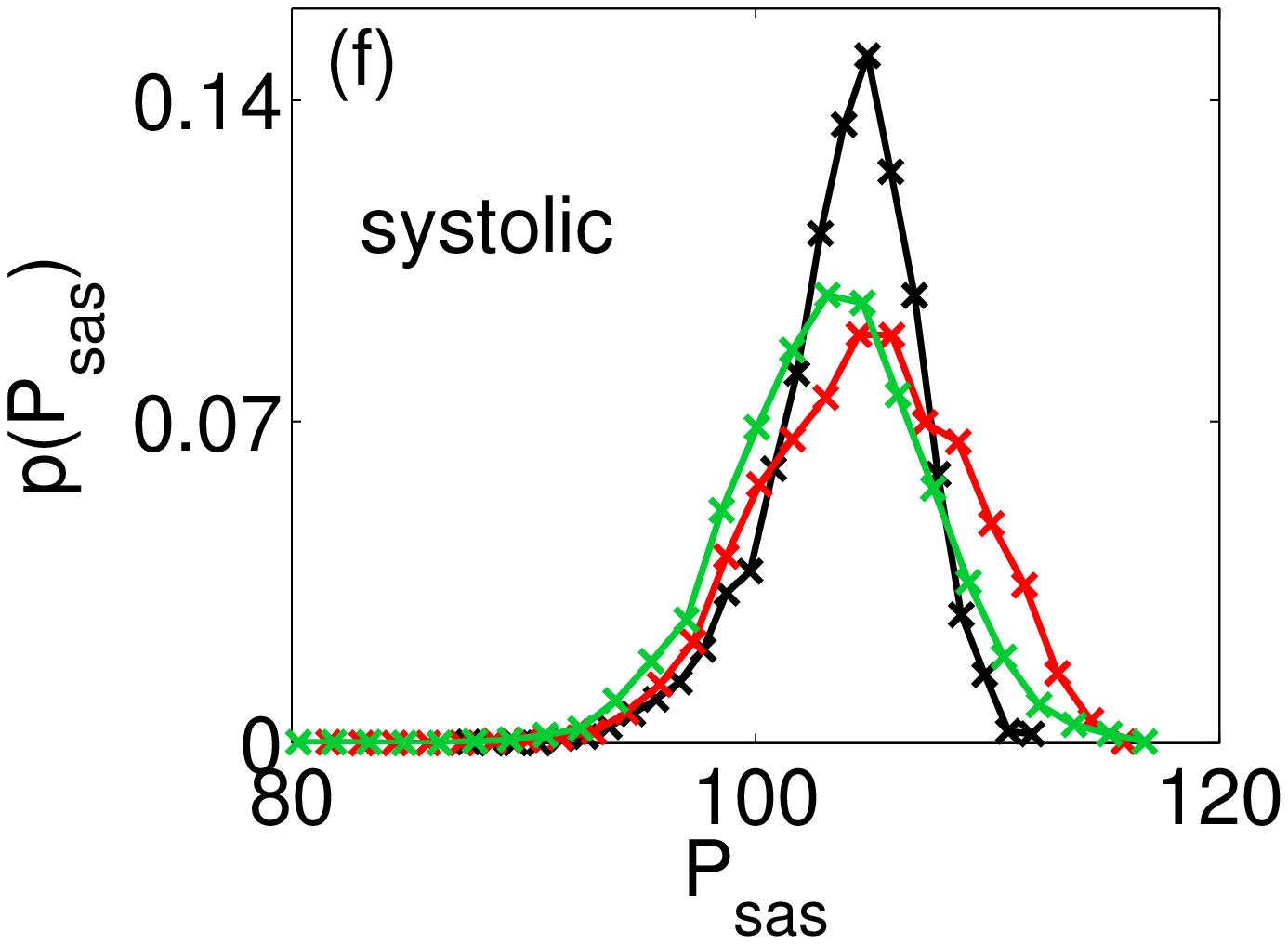}
\end{minipage}
\begin{minipage}[]{0.5\columnwidth}
\includegraphics[width=\columnwidth]{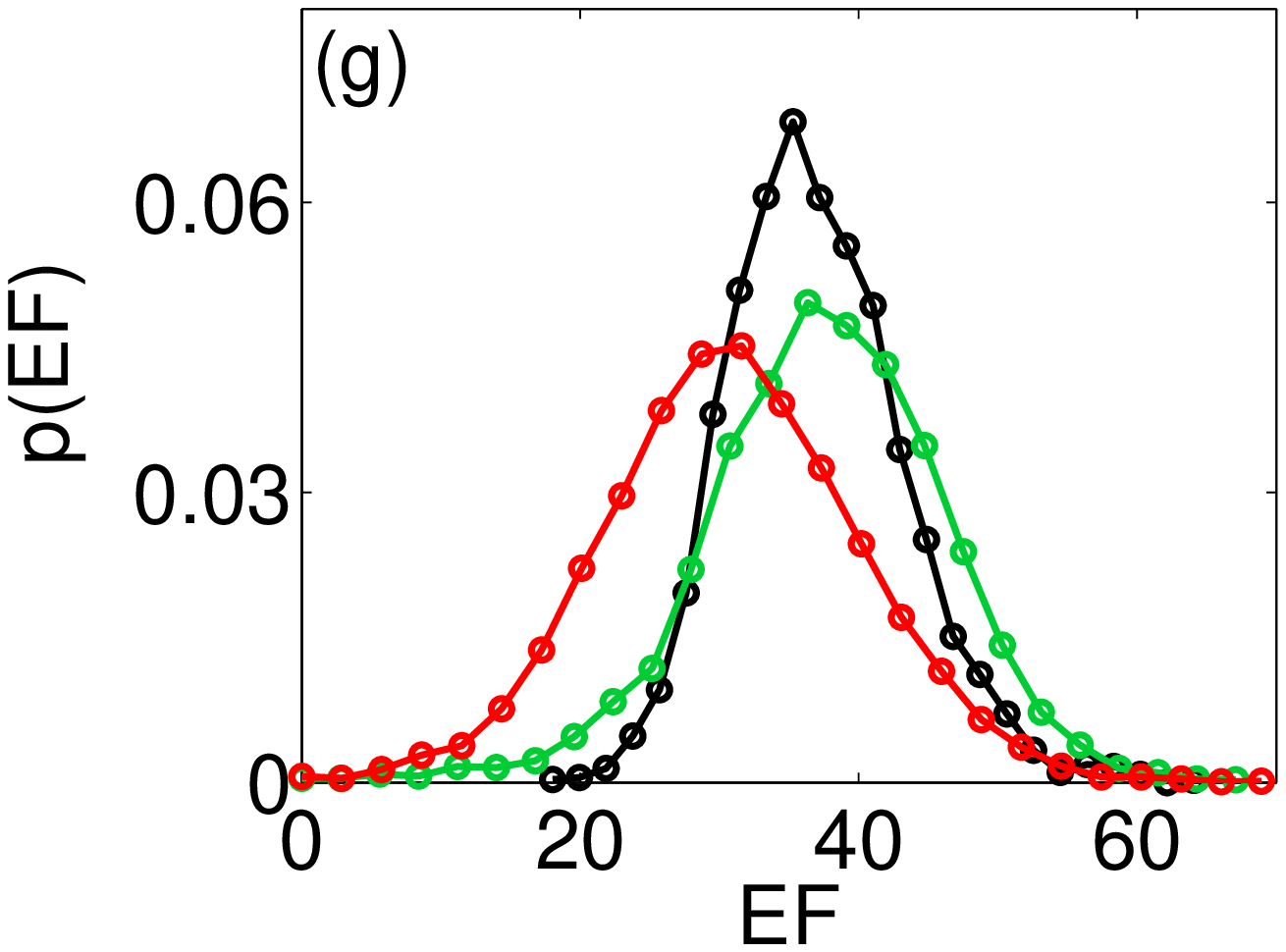}
\end{minipage}
\hspace{-0.3cm}
\begin{minipage}[]{0.5\columnwidth}
\includegraphics[width=\columnwidth]{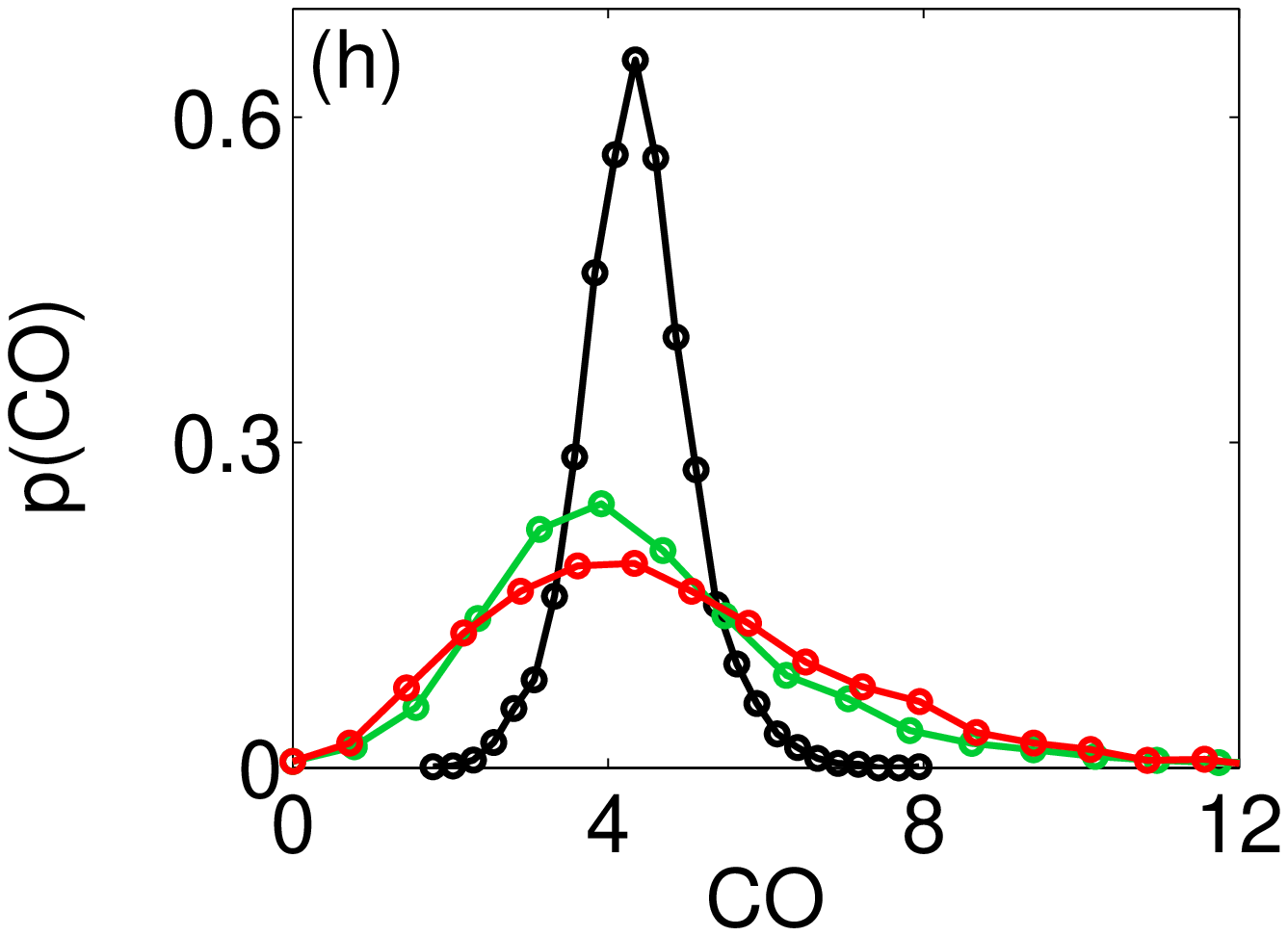}
\end{minipage}
\caption{PDFs of cardiovascular variables during AF. (black) reference EMG distribution: $\mu$=$0.67$ s, $\sigma$=$0.17$ s; (red) EMG distribution: $\mu$=$0.5$ s, $\sigma$=$0.17$ s; (green) EMG distribution: $\mu$=$0.67$ s, $\sigma$=$0.22$ s. (a-b) Left atrium pressure: (a) end-diastolic, (b) end-systolic; (c-d) Right ventricle volume: (c) end-diastolic, (d) end-systolic; (e-f) Systemic arterial pressure: (e) diastolic, (f) systolic; (g) Ejection fraction; (h) Cardiac output.}
\label{pdf_comparison}
\end{figure}

We here include the PDFs of some cardiovascular variables: $P_{la}$, $V_{rv}$, $P_{sas}$, $EF$, $CO$. Both end-diastolic and end-systolic left atrial pressures show more pronounced right tails for the green and red curves, while mean values remain practically unvaried with respect to the black reference curve (Fig. SM \ref{pdf_comparison}, panels a and b). Concerning the right ventricular volumes, the red distribution tends to lower the end-diastolic mean value and increase the end-systolic mean value, with respect to the black distribution. The green distribution does not induce substantial differences, apart from a more marked left tail for the end-diastolic values (Fig. SM \ref{pdf_comparison}, panels c and d). Systolic arterial pressure displays the same behavior in the three cases, while only the red distribution shifts the diastolic pressure towards higher values (Fig. SM \ref{pdf_comparison}, panels e and f). Regarding the ejection fraction (Fig. SM \ref{pdf_comparison}g), the red distribution accentuates a bit more the behavior revealed by the reference fibrillated distribution, rather than the green distribution does. In particular, the red distribution leads to a mean $EF$ of about $31\%$, while the mean $EF$ values for the black and green distributions are the same ($37\%$). Both green and red distributions furthermore heighten the cardiac output variability ($\sigma$=$2.1$), with respect to the reference distribution ($\sigma$=$0.67$), see Fig. SM \ref{pdf_comparison}h.

\noindent Despite of the specific values, which are sensitive to the choice of the $RR$ distribution adopted, outcomes obtained by means of the two fibrillated (red and green) distributions confirm, and in some cases even strengthen, the general behavior observed adopting the black distribution. Therefore, results achieved with the reference fibrillated distribution can be regarded as a typical response of the present model in fibrillated conditions.

\newpage

\noindent \textbf{\Large{Supplementary Material Online\\ Resource 2}}

\normalsize

\section*{AF references compared with the present stochastic modeling}

\noindent [B1] Agner BR, Akeson P, Linde JJ, Jensen GB, Dixen U (2013) Assessment of left atrial volume by magnetic resonance in patients with permanent atrial fibrillation. The short-axis method vs the single plane area-length method. The Open Cardiovascular Imaging Journal 4: 4-10

\noindent [B2] Alboni P, Scarfo S, Fuca G, Paparella N, Yannacopulu P (1995) Hemodynamics of idopathic paroxysmal atrial-fibrillation. Pacing Clin Electrophysiol 18: 980-985

\noindent [B3] Anter E, Jessup M, Callans DJ (2009) Atrial fibrillation and heart failure: treatment considerations for a dual epidemic. Circulation 119: 2516-2525

\noindent [B4] Braunwald E, Frye R, Aygen M, Gilbert J (1960) Studies on Straling's law of the heart. III. Observations in patients with mitral
stenosis and atrial fibrillation on the relationships between left ventricular end diastolic segment length, filling pressure, and the characteristics of ventricular contraction. J Clin Invest 39: 1874-84

\noindent [B5] Brookes CIO, White PA, Staples M, Oldershaw PJ, Redington AN, Collins PD, Noble MIM (1998) Myocardial contractility is not constant during spontaneous atrial fibrillation in patients. Circulation 98: 1762-1768

\noindent [B6] Chirillo F, Brunazzi MC, Barbiero M, Giavarina D, Pasqualini M, Franceschini-Grisolia E et al (1997) Estimating mean pulmonary wedge pressure in patients with chronic atrial fibrillation from transthoracic doppler indexes of mitral and pulmonary venous flow velocity. J Am Coll Cardiol 30: 19-26

\noindent [B7] Clark DM, Plumb VJ, Epstein AE, Kay GN (1997) Hemodynamic effects of an irregular sequence of ventricular cycle lengths during atrial fibrillation. J Am Coll Cardiol 30: 1034-1045

\noindent [B8] Conover MB (2003) Understanding Electrocardiography. Elsevier Health Sciences, pp 507

\noindent [B9] Corliss RJ, McKenna DH, Crumpton CW, Rowe GG (1968) Hemodynamic effects after conversion of arrhythmias. J Clin Invest 47: 1774-1786

\noindent [B10] Chugh A, Pelosi F, Morady F (2008) Atrial fibrillation and atrial flutter. In: Eagle KA, Baliga RR (eds) Practical Cardiology: Evaluation and Treatment of Common Cardiovascular Disorders. Wolters Kluwer health, pp 313-326

\noindent [B11] Daoud EG, Weiss R, Bahu M, Knight BP, Bogun F, Goyal R et al (1996) Effect of an irregular ventricular rhythm on cardiac output. Am J Cardiol 78: 1433-1436

\noindent [B12] Dodge HT, Kirkham FT, King CV (1957) Ventricular dynamics in atrial fibrillation. Circulation 15: 335-347

\noindent [B13] Gentlesk PJ, Sauer WH, Gerstenfeld EP, Lin D, Dixit S, Zado E et al (2007) Reversal of left ventricular dysfunction following ablation of atrial fibrillation. J Cardiovasc Electr 18: 9-14

\noindent [B14] Giglioli C, Nesti M, Cecchi E, Landi D, Chiostri M., Gensini GF et al (2013) Hemodynamic effects in patients with atrial fibrillation submitted to electrical cardioversion. Int J Cardiol 168: 4447-4450

\noindent [B15] Gosselink ATM, Blanksma PK, Crijns HJGM, Van Gelder IC, De Kam P, Hillege HL et al (1995) Left ventricular beat-to-beat performance in atrial fibrillation: contribution of Frank-Starling mechanism after short rather than long RR intervals. J Am Coll Cardiol 26: 1516-1521

\noindent [B16] Graettinger JS, Carleton RA, Muenster JJ (1964) Circulatory consequences of change in cardiac rhythm produced in patients by transthoracic direct-current shock. J Clin Invest 43: 2290-302

\noindent [B17] Greenfield JC, Harley A, Thompson HK, Wallace AG (1968) Pressure-flow studies in man during atrial fibrillation. J Clin Invest 47: 2411-2421

\noindent [B18] Halmos PB, Patterson GC (1965) Effect of atrial fibrillation on cardiac output. Brit Heart J 27: 719-723

\noindent [B19] Kaliujnaya VS, Kalyuzhny SI (2005) The assessment of blood pressure in atrial fibrillation. Comput Cardiol 32: 287-290

\noindent [B20] Khaja F, Parker JO (1972) Hemodynamic effects of cardioversion in chronic atrial fibrillation. Arch Intern Med 129: 433-440

\noindent [B21] Killip T, Baer RA (1966) Hemodynamic effects after reversion from atrial fibrillation to sinus rhythm by precordial shock. J Clin Invest 45: 658-671

\noindent [B22] Morris JJ, Entman M, North WC, Kong Y, McIntosh H (1965) The changes in cardiac output with reversion of atrial fibrillation to sinus rhythm. Circulation 31: 670-678

\noindent [B23] Muntinga HJ, Gosselink ATM, Blanksma PK, De Kam PJ, Van Der Wall EE, Crijns HJGM (1999) Left ventricular beat to beat performance in atrial fibrillation: dependence on contractility, preload, and afterload. Heart 82: 575-580

\noindent [B24] Orlando JR, Van Herick R, Aronow WS, Olson HG. Hemodynamics and echocardiograms before and after cardioversion of atrial fibrillation to normal sinus rhythm. Chest 76: 521-526

\noindent [B25] Prystowsky EN, Benson Jr DW, Fuster V, Hart RG, Kay GN, Myerburg RJ et al (1996) Management of patients with atrial fibrillation. Circulation 93: 1262-1277

\noindent [B26] Resnekov L, McDonald L (1971) Electroversion of lone atrial fibrillation and flutter including haemodynamic studies at rest and on exercise. Brit Heart J 33: 339-350

\noindent [B27] Rottlaender D, Motloch LJ, Schmidt D, Reda S, Larbig R, Wolnyet M et al (2012) Clinical impact of atrial fibrillation in patients with pulmonary hypertension. Plos One 7: e33902

\noindent [B28] Samet P (1973) Hemodynamic sequelae of cardiac arrhythmias. Circulation 47:399-407

\noindent [B29] Shapiro W, Klein G (1968) Alterations in cardiac function immediately following electrical conversion of atrial fibrillation to normal sinus rhythm. Circulation 38: 1074-84

\noindent [B30] Sievers B, Kirchberg S, Addo M, Bakan A, Brandts B, Trappe HJ (2004) Assessment of left atrial volumes in sinus rhythm and atrial fibrillation using the biplane area-length method and cardiovascular magnetic resonance imaging with TrueFISP. J Cardiov Magn Reson 6: 855-863

\noindent [B31] Suarez GS, Lampert S, Ravid S, Lown B (1991) Changes in left atrial size in patients with lone atrial fibrillation. Clin Cardiol 14: 652-656

\noindent [B32] Therkelsen SK, Groenning BA, Svendsen JH, Jensen GB (2006) Atrial and ventricular volume and function evaluated by magnetic resonance imaging in patients with persistent atrial fibrillation before and after cardioversion. Am J Cardiol 97: 1213-1219

\noindent [B33] Upshaw CB (1997) Hemodynamic changes after cardioversion of chronic atrial fibrillation. Arch Intern Med 157: 1070-1076

\noindent [B34] Wozakowska-Kaplon B (2005) Changes in left atrial size in patients with persistent atrial fibrillation: a prospective echocardiographic study with a 5-year follow-up period. Int J Cardiol 101: 47-52

\noindent [B35] Wylie JV, Peters DC, Essebag V, Manning WJ, Josephson ME, Hauser TH (2008) Left atrial function and scar after catheter ablation of atrial fibrillation. Heart Rhythm 5: 656-662


\begin{thebibliography}{}



\bibitem{Alpert} Alpert JS (2000) Atrial fibrillation: a growth industry in the 21st century. Eur Heart J 21: 1207-1208


\bibitem{Benjamin} Benjamin EJ, Wolf PA, D'Agostino RB, Silbershatz H, Kannel WB, Levy D (1998) Impact of atrial fibrillation on the risk of death: the Framingham heart study. Circulation 98: 946-952


stenosis and atrial fibrillation on the relationships between left ventricular end diastolic segment length, filling pressure, and the characteristics of ventricular contraction. J Clin Invest 39: 1874-84



\bibitem{Cha} Cha YM, Redfield MM, Shen WK, Gersh BJ (2004) Atrial fibrillation and ventricular dysfunction: a vicious electromechanical cycle. Circulation 109: 2839-2843












\bibitem{Fuster} Fuster V, Ryden LE, Cannom DS, Crijns HJ, Curtis AB, Ellenbogen KA et al (2006) ACC/AHA/ESC 2006 Guidelines for the management of patients with atrial fibrillation. Circulation 114: e257-354









\bibitem{Haddad} Haddad F, Hunt SA, Rosenthal DN, Murphy DJ (2008) Right ventricular function in cardiovascular disease, Part I: anatomy, physiology, aging, and functional assessment of the right ventricle. Circulation 117: 1436-1448



\bibitem{Hayano} Hayano J, Yamasaki F, Sakata S, Okada A, Mukai S, Fujinami T (1997) Spectral characteristics of ventricular response to atrial fibrillation. Am J Physiol Heart Circ Physiol 273: H2811-H2816


\bibitem{Hennig} Hennig T, Maass P, Hayano J, Heinrichs S (2006) Exponential distribution of long heart beat intervals during atrial fibrillation and their relevance for white noise behaviour in power spectrum. J Biol Phys 32: 383-392



\bibitem{Kannel} Kannel WB, Wolf PA, Benjamin EJ, Levy D (1998) Prevalence, incidence, prognosis, and predisposing conditions for atrial
fibrillation: population-based estimates. Am J Cardiol 82: 2N-9N



\bibitem{Kobayashi} Kobayashi M, Musha T (1982) 1/f fluctuation of heartbeat period. IEEE T Bio-Med Eng 29: 456-457

\bibitem{Korakianitis-a} Korakianitis T, Shi Y (2006) Numerical simulation of cardiovascular dynamics with healthy and diseased heart valves. J Biomech 39: 1964-1982

\bibitem{Korakianitis-b} Korakianitis T, Shi Y (2006) A concentrated parameter model for the human cardiovascular system including heart valve dynamics and atrioventricular interaction. Med Eng Phys 28: 613–628

\bibitem{Krahn} Krahn AD, Manfreda J, Tate RD, Mathewson FA, Cuddy TE (1995) The natural history of atrial fibrillation: incidence, risk factors, and prognosis in the Manitoba follow-up study. Am J Med 98: 476-484

\bibitem{Lloyd-Jones} Lloyd-Jones D, Adams RJ, Brown TM, Carnethon M, Dai S, De Simone G et al (2010) Heart disease and stroke statistics-2010 update a report from the American Heart Association. Circulation 121: E46-E215

\bibitem{Magnani} Magnani JW, Rienstra M, Lin H, Sinner MF, Lubitz SA, McManus DD et al (2011) Atrial fibrillation: current knowledge and future directions in epidemiology and genomics. Circulation 124: 1982-1993

\bibitem{Mielniczuk} Mielniczuk LM, Lamas GA, Flaker GC, Mitchell G, Smith SC, Gersh BJ et al (2007) Left ventricular end-diastolic pressure and risk of subsequent heart failure in patients following an acute myocardial infarction. Congest Heart Fail 13: 209-214

\bibitem{Osranek} Osranek M, Bursi F, Bailey KR, Grossardt BR, Brown RD, Kopecky SL et al (2005)  Left atrial volume predicts cardiovascular events in patients originally diagnosed with lone atrial fibrillation: three-decade follow-up. Eur Heart J 26: 2556-2561







\bibitem{Pikkujamsa} Pikkujamsa SM, Makikallio TH, Airaksinen KEJ, Huikuri HV (2001) Determinants and interindividual variation of R-R interval dynamics in healthy middle-aged subjects. Am J Physiol Heart Circ Physiol 280: H1400-H1406



\bibitem{Rokas} Rokas S, Gaitanidou S, Chatzidou S, Pamboucas C, Achtipis D, Stamatelopoulos S (2001) Atrioventricular node modification in patients with chronic atrial fibrillation: role of morphology of RR interval variation. Circulation 103: 2942-2948





\bibitem{Sanfilippo} Sanfilippo AJ, Abascal VM, Sheehan M, Oertel LB, Harrigan P, Hughes RA et al (1990) Atrial enlargement as a consequence of atrial fibrillation A prospective echocardiographic study. Circulation 82: 792-797

\bibitem{Sosnowski} Sosnowski M, Korzeniowska B, Macfarlane P, Tendera M (2011) Relationship between R-R interval variation and left ventricular function in sinus rhythm and atrial fibrillation as estimated by means of heart rate variability fraction. Cardiol J 18: 538-545




\bibitem{Tanabe} Tanabe M, Onishi K, Dohi K, Kitamura T, Ito M, Nobori T et al (2006) Assessment of left ventricular systolic function in patients with chronic atrial fibrillation and dilated cardiomyopathy using the ratio of preceding to prepreceding R-R intervals. Int J Cardiol 108: 197-201

\bibitem{Tateno} Tateno K, Glass L (2001) Automatic detection of atrial fibrillation using the coefficient of variation and density histograms of RR and $\Delta$RR intervals. Med Biol Eng Comput 39: 664-671




\bibitem{Tsang} Tsang TSM, Barnes ME, Bailey KR, Leibson CL, Montgomery SC, Takemoto Y et al (2001) Left atrial volume: important risk marker of incident atrial fibrillation in 1655 older men and women. Mayo Clin Proc 76: 467-475


\bibitem{Verdecchia} Verdecchia P, Reboldi GP, Gattobigio R, Bentivoglio M, Borgioni C, Angeli F et al (2003) Atrial fibrillation in hypertension: predictors and outcome. Hypertension 41: 218-223


\bibitem{Weismuller} Weism$\ddot{\textmd{u}}$ller P, Kratz C, Brandts B, Kattenbeck K, Trappe HJ, Ranke C (2001) AV nodal pathways in the R-R interval histogram of the 24-hour monitoring ECG in patients with atrial fibrillation. Ann Noninvas Electro 6: 285-289



\end{thebibliography}

\begin{thebibliography}{}





\bibitem{Guyton} Guyton AC, Hall JE (2006) Textbook of Medical Physiology. Elsevier Saunders, pp 1116

\bibitem{Korakianitis-a} Korakianitis T, Shi Y (2006) Numerical simulation of cardiovascular dynamics with healthy and diseased heart valves. J Biomech 39: 1964-1982

\bibitem{Korakianitis-b} Korakianitis T, Shi Y (2006) A concentrated parameter model for the human cardiovascular system including heart valve dynamics and atrioventricular interaction. Med Eng Phys 28: 613–628

\bibitem{matlab} MATLAB version 7.9.0 (2009) Natick, Massachusetts: The MathWorks Inc

\bibitem{Ottesen} Ottesen JT, Olufsen MS, Larsen JK (2004) Applied Mathematical Models in Human Physiology. SIAM, pp 298

\bibitem{Ruiz} Ruiz P, Amin Rezaienia M, Rahideh A, Keeble TR, Rothman MT, Korakianitis T (2013) In vitro cardiovascular system emulator (bioreactor) for the simulation of normal and diseased conditions with and without mechanical circulatory support. Artif Organs 37: 549-560

\bibitem{Westerhof} Westerhof N, Stergiopulos N, Noble MIM (2010) Snapshots of Hemodynamics. Springer, pp 286



\end{thebibliography}
\end{document}